\documentclass[fleqn,usenatbib]{mnras}

\usepackage{newtxtext,newtxmath}
\usepackage[T1]{fontenc}
\usepackage{ae,aecompl}
\usepackage{xcolor}
\usepackage{graphicx}	% Including figure files
\usepackage{amsmath}	% Advanced maths commands
\usepackage{pdflscape}	% Landscape pages
\usepackage{soul}

\newcommand{\OIII}{[O\textsc{iii}]$_{\lambda5007}$\ }
\newcommand{\OIIIdoublet}{[O\textsc{iii}]\ }
\newcommand{\OIIIdoubletfull}{[O\textsc{iii}]$_{\lambda\lambda4959,5007}$\ }
\newcommand{\OII}{[O\textsc{ii}]\ }
\newcommand{\OIIdoubletfull}{[O\textsc{ii}]$_{\lambda\lambda3727,29}$\ }
\newcommand{\Ha}{H$\alpha$\ }
\newcommand{\Hb}{H$\beta$\ }
\title[WISP X FORS2]{Emission-line galaxies at $z\sim1$ from near-IR HST Slitless Spectroscopy: metallicities, star formation rates and redshift confirmations from VLT/FORS2 spectroscopy}

\author[K. Boyett et al.]{K. Boyett,$^{1,2,3}$\thanks{E-mail: Kit.Boyett@unimelb.edu.au}  %0000-0003-4109-304X
A. J. Bunker,$^{3}$  %0000-0002-8651-9879
J. Chevallard,$^{3}$  %0000-0002-7636-0534
A. Battisti,$^{2,4}$  %0000-0003-4569-2285
A. L. Henry,$^{5}$ \newauthor  %0000-0002-6586-4446
S. Wilkins,$^{6,7}$  %0000-0003-3903-6935
M. A. Malkan,$^{8}$  %0000-0001-6919-1237
J. Caruana,$^{7,9}$  %0000-0002-6089-0768
H. Atek,$^{10}$  %0000-0002-7570-0824
I. Baronchelli,$^{11, 12}$  %0000-0003-0556-2929
\newauthor
J. Colbert,$^{13}$  %0000-0001-6482-3020
Y. S. Dai,$^{14}$  %0000-0002-7928-416X
Jonathan P. Gardner,$^{15}$  %0000-0003-2098-9568
M. Rafelski,$^{5,16}$ %0000-0002-9946-4731
C. Scarlata,$^{17}$  %0000-0002-9136-8876
\newauthor
H. I. Teplitz,$^{13}$  %0000-0002-7064-5424
X. Wang,$^{18,19,20}$  %0000-0002-9373-3865
\\
$^{1}$School of Physics, University of Melbourne, Parkville 3010, VIC, Australia\\
$^{2}$ARC Centre of Excellence for All Sky Astrophysics in 3 Dimensions (ASTRO 3D), Australia\\
$^{3}$Department of Physics, University of Oxford, Denys Wilkinson Building, Keble Road, Oxford OX1 3RH, UK\\
$^{4}$Research School of Astronomy and Astrophysics, Australian National University, Cotter Road, Weston Creek, ACT 2611, Australia\\
$^{5}$Space Telescope Science Institute, 3700 San Martin Dr., Baltimore, MD 21218, USA \\
$^{6}$Astronomy Centre,University of Sussex, Falmer, Brighton BN1 9QH, UK\\
$^{7}$Institute of Space Sciences and Astronomy, University of Malta, Msida MSD2080, Malta\\
$^{8}$Department of Physics and Astronomy, University of California, Los Angeles, 430 Portola Plaza, Los Angeles, CA 90095, USA\\
$^{9}$Department of Physics, University of Malta, Msida MSD2080, Malta\\
$^{10}$Institut d'Astrophysique de Paris, CNRS, Sorbonne Universit\'e, 98bis Boulevard Arago, 75014, Paris, France\\
$^{11}$INAF - Istituto di Radioastronomia, via Gobetti 101, 40129 Bologna, Italy\\
$^{12}$Italian ALMA Regional Centre, via Gobetti 101, 40129 Bologna, Italy\\
$^{13}$IPAC, Mail Code 314-6, California Institute of Technology, 1200 E. California Blvd., Pasadena CA, 91125, USA\\
$^{14}$Chinese Academy of Sciences South America Center for Astronomy, National Astronomical Observatories of China, \\ Chinese Academy of Sciences, Beijing 100101, China\\
$^{15}$Astrophysics Science Division, NASA Goddard Space Flight Center, 8800 Greenbelt Rd, Greenbelt, MD 20771, USA\\
$^{16}$Department of Physics and Astronomy, Johns Hopkins University, Baltimore, MD 21218, USA\\
$^{17}$Minnesota Institute for Astrophysics, University of Minnesota, 116 Church St SE, Minneapolis, MN 55455, USA\\
$^{18}$School of Astronomy and Space Science, University of Chinese Academy of Sciences (UCAS), Beijing 100049, China\\
$^{19}$National Astronomical Observatories, Chinese Academy of Sciences, Beijing 100101, China\\
$^{20}$Institute for Frontiers in Astronomy and Astrophysics, Beijing Normal University,  Beijing 102206, China
}

\date{Accepted XXX. Received YYY; in original form ZZZ}

\pubyear{2024}

\begin{document}
\label{firstpage}
\pagerange{\pageref{firstpage}--\pageref{lastpage}}
\maketitle

% Abstract of the paper
\begin{abstract}
We follow up emission line galaxies identified through the near-infrared slitless HST/WFC3 WISP survey with VLT/FORS2 optical spectroscopy.  
Over 4 WISP fields, we targetted 85 of 138 line emission objects at $0.4<z<2$ identified in WFC3 spectroscopy. Half the galaxies are fainter than $H_{AB}=24$\,mag, and would not have been included in many well-known surveys based on broad-band magnitude selection. We confirm 95\% of the initial WFC3 grism redshifts in the 38 cases where we detect lines in FORS2 spectroscopy. However, for targets which exhibited a single emission line in WFC3, up to 65\% at $z<1.28$ did not have expected emission lines detected in FORS2 and hence may be spurious (although this false-detection rate improves to 33\% using the latest public WISP emission line catalogue). From the Balmer decrement the extinction of the WISP galaxies is consistent with $A($H$\alpha)=1$\,mag. From SED fits to multi-band photometry including Spitzer $3.6\,\mu$m, we find a median stellar mass of $\log_{10}(M_\star/M_{\odot})=8.94$. Our emission-line-selected galaxies tend to lie above the star-forming main sequence (i.e.\ higher specific star formation rates). Using [O\textsc{iii}], [O\textsc{ii}] and H$\beta$ lines to derive gas-phase metallicities, we find typically sub-solar metallicities, decreasing with redshift. Our WISP galaxies lie below the $z=0$ mass-metallicity relation, and galaxies with higher star formation rates tend to have lower metallicity. Finally, we find a strong increase with redshift of the H$\alpha$ rest-frame equivalent width in this emission-line selected sample, with higher $EW_0$ galaxies having larger [O\textsc{iii}]/H$\beta$ and O32 ratios on average, suggesting lower metallicity or higher ionisation parameter in these extreme emission line galaxies. 
\end{abstract}

% Select between one and six entries from the list of approved keywords.
% Don't make up new ones.

%%%%%%%%%%%%%%%%%%%%%%%%%%%%%%%%%%%%%%%%%%%%%%%%%%

%%%%%%%%%%%%%%%%% BODY OF PAPER %%%%%%%%%%%%%%%%%%
\section{Introduction}

In this paper we study galaxies seen in H$\alpha$ emission across \lq\lq cosmic noon" ($z \sim 0.5-2$), the era when the star formation rate (SFR) comoving density peaks \citep{Madau14}.
In recent years there has been significant effort in slitless spectroscopic surveys from HST, such as the WFC3 Infrared Spectroscopic Parallel (WISP) Survey in the near-IR (\citealt{Atek10, Battisti24}), and currently this is being greatly expanded with ongoing space-based Wide Field Slitless Spectroscopy (hereafter, WFSS) such as {\em JWST}-NIRISS \citep[e.g.,][]{Boyett22B} and NIRCam \citep[e.g.,][]{Oesch23arXiv, Kashino23}, and Euclid \citep[e.g.][]{Laureijs11}. WFSS selects directly on emission line strength (which is often a good proxy for the star formation rate), and is sensitive to high equivalent width sources. These extreme emission line objects, with potentially high specific star formation rates, may well be missed in traditional multi-object spectroscopic surveys which are based on broad-band photometric selections. 

However, many WISP spectra contain only single emission lines (some with low signal-to-noise) where line designation may be in error. To address this concern, we analyse VLT/FORS2 spectroscopic optical follow-up on selected WISP galaxies. A key goal is redshift confirmation, affirming WISP line identifications and improving redshift measurements with higher spectral resolution. The optical spectroscopy also enables the study of line ratio diagnostics, to measure the gas-phase metallicity of the WISP galaxies and to identify potential Active Galactic Nuclei (AGN). 

In this work we describe the reduction and analysis of the VLT/FORS2 spectroscopy, and combine this with the WFC3 spectra and photometry across a wide wavelength range (including Spitzer 3.6\,$\mu$m imaging) to study the nature and evolution of emission line galaxies at high redshift.

The outline of this paper is as follows. We discuss the HST/WFC3 and VLT/FORS2 spectroscopic observations, along with the target selection, data reduction and ancillary photometry in Section \ref{sec:obs}. 
We identify the FORS2 detected emissions lines and from these determine redshifts in Sections \ref{sec:identification} and \ref{sec:redshift_valid}. We use the measured line emission and photometry to investigate AGN contamination (Section \ref{sec:AGN}), star formation rates and stellar masses (Section \ref{sec:SFR}), metallicity (Section \ref{sec:line_diagnostics}), and a sub-sample of galaxies with extreme equivalent widths (EWs, Section\ref{sec:HaEW}).
Where applicable, we use a standard $\Lambda$CDM cosmology with parameters \textit{H}$_{0}=70$ km/s/Mpc, $\Omega_{m}=$0.3, and $\Omega_{\wedge}=$0.7. All magnitudes are in the AB system \citep{Oke83}. When quoted, all EWs are in the rest-frame. 

\section{Observations}\label{sec:obs}

The initial results of the WFC3 Infrared Spectroscopic Parallel survey (see Section \ref{sec:WISPS-section}) are reported in \citet{Atek10} with subsequent data releases being made available through MAST\footnote{\label{footnote:MAST}The latest data release (v6.2) is available on the Milukski archive for space telescopes page \url{https://archive.stsci.edu/prepds/wisp/}} as the survey expanded \citep{Battisti24}. In this Section we report the optical spectroscopy follow-up performed at the VLT on 4 selected WISP fields (Par:309, 64, 62 and 236), providing complementary 0.5 to 0.9\,$\mu$m wavelength coverage, ideal for rest-optical emission lines (e.g., [O\textsc{ii}]$\lambda\lambda3727,29$, H$\beta$ and [O\textsc{iii}]$\lambda\lambda4959, 5007$) in our anticipated redshift range, and offering up to an order of magnitude greater spectral resolution relative to the WFC3 grisms. 
First we will briefly describe the WISP HST/WFC3 WFSS, then we will detail follow-up target selection, observing strategy, and spectroscopic data reduction. Finally we will discuss the complementary broadband photometry.

\subsection{HST/WFC3 wide field slitless spectroscopy}
\label{sec:WISPS-section}
The WISP survey (\citealt{Atek10, Battisti24}) is a pure parallel program on the Hubble Space Telescope, using the WFC3 G102 and G141 infrared grisms to detect emission line galaxies through WFSS from 0.7 to 1.8$\mu m$ over $\sim1600$ arcmin$^2$ of blank fields. Sensitive to H$\alpha$ emission across a broad redshift range, $0.5 < z < 1.6$, the WISP survey detects emission line galaxies whose continuum may otherwise be too faint to appear in standard broadband wide field photometric surveys. Selecting on emission line strength rather than broadband luminosity is ideal for targetting low mass systems exhibiting high specific star formation rates (sSFR) which are expected to be abundant above $z>1$  \citep[e.g.,][]{VanDerWel11}
but are below the broadband photometric threshold for selection in  typical spectroscopic surveys (see Section \ref{sec:HaEW}). 
We do note, however, that our sample from WISP is essentially emission-line-flux limited, equivalent to a star-formation based selection rather than one based on a stellar-mass limited sample (i.e., the WISP survey probes the nature and evolution of the star-forming population with redshift). We will discuss the possible biases introduced by such a selection in Section~\ref{sec:HaEW}.
WFSS offers the advantage over broad-band pre-selection for spectroscopic follow-up in that we are selecting directly on the quantity of interest (the star formation rate, indicated by the H$\alpha$ flux), hence reducing any observational biases in a survey of star forming galaxies (SFGs). \citet{Atek10} and \citet{Battisti24} provide the data reduction method and detail the identification and flux measurements of the observed emission lines ([O\textsc{ii}]$\lambda\lambda3727,29$, H$\beta$, [O\textsc{iii}]$\lambda\lambda4959,5007$, H$\alpha$, [S\textsc{ii}]$\lambda\lambda6716,30$) from the WISP HST/WFC3 slitless spectra, with accompanying redshift estimates. 

\subsubsection{WFSS emission line catalogues}

A publicly available WISPS emission line catalogue (v6.2)\textsuperscript{\ref{footnote:MAST}} is presented in \citet{Battisti24}. The slitmasks for the VLT/FORS2 observations (see Section \ref{sec:FORS2}) were based on an earlier version of the catalogue which had a lower emission line S/N threshold and less thorough visual inspection of emission line candidates, and hence had a larger number of potential targets. We note that the v6.2 catalogue does not cover one of our four target fields with FORS2 (Par62) where 13 galaxies were assigned slits. For the 72 galaxies we targetted in the other three fields, many are not included in the later v6.2 catalogue (53/72) due to its higher S/N cut, although as we discuss in Section~\ref{sec:identification} some of these galaxies were subsequently spectroscopically confirmed in our FORS2 observations. In column v6.2\_par\_obj of Table~\ref{tab:photometry} we show the matching ID in the v6.2 catalogue for the 19 galaxies which we targetted with FORS2 which are included in the v6.2 catalogue in \citet{Battisti24}.

We note that when we match the original emission line catalogue, from which the slitmask design was based, to the v6.2 catalogue; only 34 of the 114 original candidates are included in v6.2 (we exclude Par62 from the matching as we know that this field was not included in v6.2).
From these 34 candidates, the reported redshift is consistent between both catalogues for 24 candidates but has changed significantly in 11 cases due to the identification of the detected emission lines in the WFSS being updated (see Appendix \ref{App:catalogue_redshift}). 

\subsection{VLT/FORS2 Observation and Target Selection} 
\label{sec:FORS2}
Between July and August 2014, follow-up optical spectroscopic observations were conducted for a sub-sample of WISP galaxies using the FOcal Reducer/low dispersion Spectrograph 2 (FORS2, \citealt{FORS2_1998}), a multi-object spectrometer on the Very Large Telescope UT1 (Antu), as part of programme 093.A-0893 (P.I A. Bunker).
FORS2 uses two red-sensitive MIT/LL $2k\times 4k$ CCDs, with a native pixel scale of $0.125''$\,pixel$^{-1}$. To reduce the readnoise we binned the pixels $2\times 2$ (hereafter we refer to the binned pixels, 0.25$''$), which still Nyquist-sampled the seeing and spectral resolution. The observations are detailed in Table~\ref{Tab:Obs_Dates}, and were taken at low airmass (1.0--1.35)
and in good seeing ($\sim0.6''$ FWHM). We used a slit width of 1$''$ across all of the masks. The 600RI grism (with a central wavelength of 6800\,\AA) was used in combination with the GG435 order-blocking filter. Each slit mask was observed twice, each with a 1400\,s exposure. The mask was nodded by 3$''$ between the two exposures to place the targets first in the upper half of their slit and then in the lower ($\pm1.5''$ from the centre of the slit), in order to perform local background subtraction (discussed in Section \ref{sec:data_reduction}). Acquisition images of the field were taken in the $R$-band before the slitmask was moved into the focal plane. The field-of-view of FORS2 is $6.8\times 6.8$\,arcmin$^2$. Since our WISP fields were smaller (the HST/WFC\,3 detector is only 2.2\,arcmin across) we placed all our targets in just one of the two FORS2 detectors. 
 
We chose 4 WISP survey fields to observe which were equatorial or in the Southern hemisphere with RAs appropriate to the observing season.
Two of the WISP fields (Par309 and Par236) were observed twice with different slitmask designs to accommodate the larger number of WISP emission line targets in these fields, 
and as part of these masks designs three targets appeared on both masks (and these are indicated in our Tables \ref{tab:ancillary} - \ref{tab:derived_properties} as \lq\lq double-galaxy").
The emission line catalogue from the WISP HST/WFC3 spectroscopy (Section~\ref{sec:WISPS-section}) has between 24 and 48 candidate galaxies per field, based on the early catalogues used to select the targets for FORS2 spectroscopy (not all of which are included in the latest v6.2 public catalogue).
We could not accommodate all these candidates in each field on a single FORS2 slitmask, since we wanted a minimum slitlet length of 8$''$ to facilitate sky subtraction by nodding at two positions along the slit. Over the 2.2$'$ HST/WFC3 field we could allocate at most $\sim 15$ targets, and we gave each galaxy a score based on the WISP line emission, considering the observed H$\alpha$\footnote{\label{note1}If H$\alpha$ was not detected within the wavelength coverage (e.g., at too high a redshift, $z>1.6$) the [O\textsc{iii}] doublet was used instead.}  emission line equivalent width, signal-to-noise (S/N) and the determined redshift. Specifically, the scores were to favour:
\begin{itemize}
    \item{a high signal-to-noise ratio in the WISP WFC3 spectroscopy from our early catalogues, with $S/N>10$ scoring 3, $5<S/N<10$ scoring 2, and $S/N<5$ scoring 1;}
    \item{a high observed-frame equivalent width emission line sources, with H$\alpha$\textsuperscript{\ref{note1}} $\rm{EW}>300$\,\AA\ scoring 3, $100<$EW$<300$\,\AA\ scoring 2, and $\rm{EW}<100$\,\AA\ scoring 1. }
\end{itemize}

The scores for observed equivalent width and S/N were multiplied together to obtain an overall score between 1 and 9, and if the redshift based on the WFC3 WFSS lay beyond $z=1.28$ we down-weight the overall score by a factor of 4, since few bright emission lines would fall in our FORS2 wavelength range at higher redshifts. In the early WISP survey catalogues used to for the FORS2 mask design about 20-30\% of galaxies per field lie at these higher redshifts (and 36\% lie at $z>1.28$ in the subsequent v6.2 catalogue for the entire WISP survey). Beyond this redshift, 15 galaxies are selected for follow-up to check if conflicting emission lines appear (8 of which are in the v6.2 catalogue, we note that the redshift of 2/8 has changed between the catalogues and these 2 are now in the redshift range where we expect to observe corresponding line emission in FORS2); this can provide confirmation on whether or not the HST/WFC3 grism detected lines were identified correctly.

Based on these overall scores we used the UCSC-LRIS mask design software\footnote{\url{https://www.ucolick.org/~phillips/lris/}}, developed by D.\, Phillips and collaborators for LRIS on the Keck telescope, to maximise the sum of the overall scores of the objects placed on the slitmask, optimising the centre and position angle for this. Of 138 galaxies in the initial WISP survey emission line catalogue used in the slitmask design for these 4 fields, 85 have slits placed over them (a completeness of $62\%$).
In Figure \ref{fig:Hmag_dist}, we present the distribution of H-band magnitudes for our FORS2 sample.
In Figure \ref{fig:Mv_dist} we show the distribution of the rest-frame $V$-band absolute magnitudes ($\sim5500$\,\AA), M$_{V}$, for the WISP galaxies within the sample that have observed photometric coverage (observed-frame F110W-F160W), and we note the wide range of galaxy luminosities in our emission line selected sample. We determine M$_{V}$ using the best available spectroscopic redshift, adopting the FORS2 spectroscopic redshift when available (see Section \ref{sec:redshift_valid}), otherwise using the WFC3 spectroscopic redshift.
We will discuss the redshift distribution in Section \ref{sec:redshift_valid} with the distribution of spectroscopic redshifts presented in Figure \ref{fig:red_dist}. 

\begin{table*}
\begin{tabular}{ccccccc}
Observation  & RA & Dec &Date       & Airmass & Seeing   \\
mask & (Deg) & (Deg) & & & \\
\hline
 \hline
309\_1          &324.79799 & -38.419686 & 2014-08-03 & 1.0655  & 0.55$''$   \\
309\_2          &324.79799 & -38.419686 & 2014-08-03 & 1.0615  & 0.6$''$    \\
64\_2           & 219.36889 & -1.8316702& 2014-07-18 & 1.2465  & 0.6$''$    \\
62\_1           & 195.318858 & -0.03223 & 2014-07-18 & 1.374   & 0.575$''$  \\
236\_1          & 231.20360& 0.41788568& 2014-07-01 & 1.129   & 0.675$''$   \\
236\_2          & 231.20360& 0.41788568& 2014-07-29 & 1.3515  & 0.625$''$   \\ \hline
std\_star       & 28.710728 & -27.47780 & 2014-07-25 & 1.33    & 0.625$''$                                
\end{tabular}
\caption{Observation conditions for our 6 FORS2 masks and our standard star.}
\label{Tab:Obs_Dates}
\end{table*}

\subsection{FORS2 data reduction}
\label{sec:data_reduction}
The European Southern Observatory (ESO) provide a reduction pipeline for the FORS2 instrument - EsoReflex \citep{Freudling13}. However, this pipeline reduces individual exposures and does not combine the nodded observations at different positions that we took. Hence we perform our own data reduction, using standard techniques in the 
Image Reduction and Analysis Facility (IRAF). We now briefly outline the process. 

Each exposure first had the bias removed using the overscan regions using the {\tt colbias} package. For each slitmask we averaged several spectroscopic flat fields taken during the day with the same instrumental configuration. Then we normalized the flat field for each slitlet by dividing by the spectral shape of the flat field lamp (obtained by collapsing the 2D spectrum spatially, along the long axis of the slit), leaving the pixel-to-pixel sensitivity variations. We then flat fielded our science integrations for each 2D slitlet through division by the appropriate flat field. 

We performed cosmic ray rejection through two methods. We first used {\tt LACOSMIC} \citep{LACOSMIC} on the individual 2D science spectra for each slitlet. We also used {\tt crreject} within the IRAF {\tt imcombine} package, rejecting outliers at the $4\sigma$ level when the two spatially-offset spectra were aligned (we used the CCD gain of 1.43e/DN and detector read noise of 2.9\,electrons so that the Poisson noise was correctly calculated). The cosmic rays found were added to a mask, which also included bad pixels and regions not illuminated by the slitlet. 

Finally, we subtracted one nod position from the other, and this ``$A-B$'' spectrum had the sky emission subtracted out although residuals due to time variation were still present. We combined the 2D ``$A-B$'' spectrum with the inverted ``$B-A$'' applying a spatial offset to reverse the effect of the nod along the slit, so that the spectrum of the target galaxy added at the same location. Bad pixels, unilluminated areas and cosmic rays were excluded in this combination. 

The resulting 2D combined spectra for each slitlet (i.e. targetted galaxy) had not yet been corrected for spatial curvature, or wavelength and flux calibrated. However, at this point each pixel was independent (no interpolation had yet been applied), so the noise could be measured directly and was found to agree well with the expected Poisson noise (with the readout noise added in quadrature) derived from the spectrum of the sky emission. 

The spectra of each object is then extracted using the {\tt apextract.apall} package in IRAF. For the brighter objects, where the spectral continuum was easily seen, we traced how the position of the spectrum varied with wavelength across the detector. For fainter objects, where no continuum or just individual emission lines were visible, we used the edge of the slit (which was readily seen in the flat field exposures) to map the spatial distortion. 
We use a 1$''$ extraction width, corresponding to 4 pixels centred on the target, in combination with the 1$''$-width slits this creates a 1$''\times$1$''$ extraction aperture.
Some sources were spatially extended, so we also performed a second extraction over a wider 10-pixel (2.5$''$) height along the slit. In both cases we applied background subtraction using the \texttt{IRAF.background} package to remove any residual sky emission which remained after the differencing of the offset exposures, by subtracting the clipped average counts in pixels between 6 and 10 pixels from the object position (which avoided self-subtracting the object flux, and also avoided the ends of the slitlets). For each extraction width we also generated a 1D noise spectrum from the Poisson noise model of the object and sky counts (taking into account the gain of the detector). 

Wavelength calibration for each slitmask was done using NeArHgHe arc lamps observed during the day with the same instrument configuration. We extracted the arc spectrum for each slitlet using the same trace as the object, but without the background subtraction. We fit a cubic mapping of pixel to wavelength using about 50 arc lines with a scatter of 0.3\,\AA, and calculated the average scale to be 1.62\,\AA\,pix$^{-1}$. The wavelength calibration was checked against sky emission lines in our spectra to confirm there had been no shift in the grism central wavelength. We then produced two versions of the wavelength-calibrated spectrum, the first involving interpolation onto a uniform wavelength scale of 1\,\AA\,pix$^{-1}$, and the other preserving the original pixels but allocating a wavelength to each pixel on a non-linear scale (so that the pixels are independent). For slitlets close to the centre of the CCD, the wavelength coverage was $5150-8470$\,\AA , although slits placed on targets towards the edge of the field (offset along the wavelength axis) can have the wavelength coverage shifted by up to $\pm$300\,\AA .
We computed the spectral resolution from the FWHM of arclines and unblended sky lines, and found this to be 5.6\,\AA, equivalent to a resolving power of $R=\lambda/\Delta\lambda_{FWHM}=1200$ at our central wavelength. However, we note that for objects which do not fill the 1$''$ slit the spectral resolution will be better. This spectral resolution is far greater than the $R=210$ and $R=130$ that the G102 and G141 HST/WFC3 grism achieve at 1.1\,$\mu$m and 1.4\,$\mu$m respectively.  

As part of the FORS2 observing program four standard stars were obtained to flux calibrate the science images, but these were not necessarily observed on the same nights as their target masks. 
Only one of the standard stars (LTT1020) was taken in similar seeing to our slitmask observations, so we used this to determine our flux calibration. The extraction and wavelength calibration were performed as described above. We determined the conversion of counts to flux density by comparison with the reference values for our standard star \citep{Hamuy_1992, Hamuy_1994}. For each slitmask we had added about 4 additional brighter compact/point sources as a check of the spectro-photometry. These objects have photometry from the Sloan Digital Sky Survey \citep[SDSS DR12,][]{Fukugita:96} or the VLT Survey Telescope ATLAS \citep[VHS-ATLAS DR3,][]{Shanks:15}. The literature flux in the $r$ and $i$ bands (within our FORS2 wavelength coverage) was compared to the FORS2 spectrum continuum flux (convolved with the $r$- or $i$-band filter transmission function) averaged across the same wavelength range. The FORS2 observed flux consistently falls short of the anticipated flux based on the flux calibration from the single standard star. The discrepancy is seen to be $30\%$ in the $r$-band and $15\%$ in the $i$-band. 
The origin of this flux discrepancy may be due to a number of factors: different seeing; some slight mis-alignment in the mask acquisition; telescope tracking drift or changes in atmospheric transparency. We applied the appropriate scaling factor (for the mask and wavelength) to the emission line fluxes from our FORS2 spectroscopy. 

The difference in correction factors between the $r$-band and the $i$-band regions of the spectra does not have a large impact on the spectral diagnostics we wish to explore. For example a difference of 15$\%$ in the [O\textsc{ii}]/H$\beta$ emission line flux ratio would result in an offset of 0.06 dex in the $x$-axis in Figure \ref{fig:OIII/I/Hb} (which shows the metallicity tracks in [O\textsc{iii}]/H$\beta$ vs. [O\textsc{ii}]/H$\beta$), which is comparable to the error bars and also the bin size of the SDSS comparison dataset. The [O\textsc{iii}]/H$\beta$ ratio is unaffected by the correction factor since these lines are close in wavelength to each other. A larger potential impact is on the H$\beta$ flux, and hence the Balmer decrement -- perturbing the Balmer decrement H$\alpha$/H$\beta$ by $15\%$ changes the reddening of A(H$\alpha$) by 0.04 mag.

Finally, for the three galaxies that were each assigned to two masks, allowing them to be observed twice, the 1D wavelength and flux calibrated spectrum from each mask are averaged together to achieve greater S/N. We note that in these 3 cases, the measured continuum flux density and emission lines flux ([O\textsc{ii}] was detected in \texttt{309\_1\_9:309\_2\_2}\footnote{The FORS2 ID is constructed as the combination of the WISP field, the slitmask and the slit the target was assigned to. e.g., 309\_1\_9 was the galaxy assigned to slit 9 on the first slitmask targetting WISP field 309.} and \texttt{236\_1\_2:236\_2\_5}, while no lines were detected in \texttt{236\_1\_9:236\_2\_7}) between the two observations are consistent, supporting consistent flux calibration between the different masks.

\subsection{Multi-band photometry}
In addition to the HST WFC3 grism slitless and VLT/FORS2 spectroscopy, these four WISP fields (62, 64, 236 $\&$ 309) have been observed with ground- and space-based telescopes to provide complementary imaging (described below). However, not all fields have coverage in all filters and the imaging available to each WISP field is given in Table \ref{tab:ancillary}. We will briefly describe the imaging and photometric analysis.  

As part of the WISP survey (\citealt{Atek10, Battisti24}) HST/WFC3 WFSS was obtained in combination with direct imaging in typically two HST/WFC3 broadband filters, F110W and one of either F140W or F160W (apart from field 236 which only had F140W obtained). The near-infrared bands are observed to provide reference images for wavelength calibration of the WFC3 WFSS. As part of the WISP program, HST/UVIS direct imaging was observed for certain WISP fields; this included F475X and F600LP for field 64, and F814W for field 309.  Here multiple exposures of the reduced imaging were drizzled into a final science image that retains the HST/WFC3 native 0.128$''$ pixel scale. Due to the variable number of HST orbits that each WISP field was observed for, the exposure times for each direct imaging observation vary and we measure $5\sigma$ limit depth in a 0.4$''$ radius aperture in the range 26.05 to 24.74; full details are given in Table \ref{tab:ancillary}. 

\begin{figure}
    \centering
    \includegraphics[width = \columnwidth]{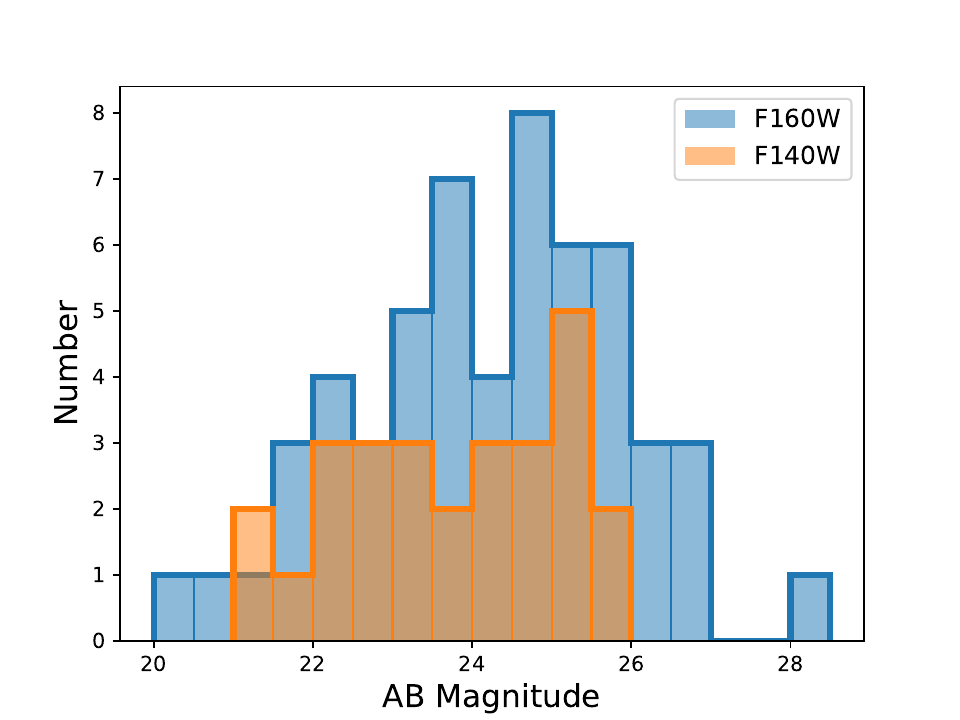}
    \caption{The $H$-band (F160W, blue) and $JH$-band (F140W, orange) AB magnitude distribution for the 85 objects in our sample. $JH$-band is given instead of the $H$-band in WISP field 236 due to the availability of imaging; see Table \ref{tab:ancillary}. 51$\%$ of galaxies have magnitude fainter than AB=24\,mag.}
    \label{fig:Hmag_dist}
\end{figure}

\begin{figure}
    \centering
    \includegraphics[width = \columnwidth]{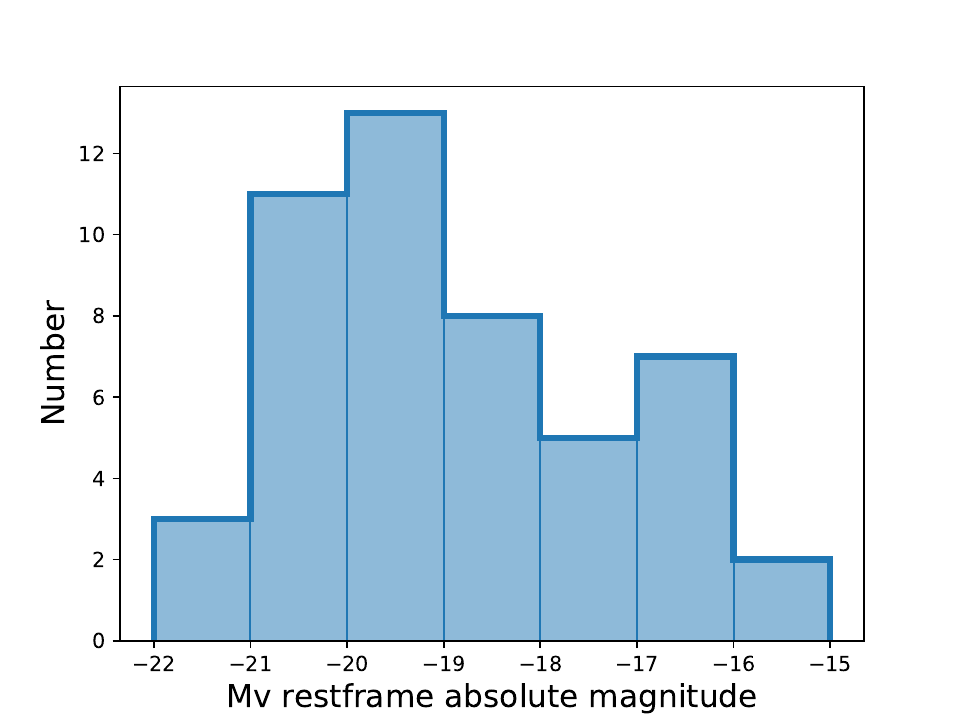}
    \caption{The distribution of the rest-frame $V$-band absolute magnitudes ($\sim$5500\,\AA) for the sub-set of targetted WISP galaxies with observed photometric broad-band coverage (observed-frame F110W-F160W). Absolute magnitudes are determined using the FORS2 spectroscopic redshift when available, otherwise using the WISPS WFC3 spectroscopic redshift.}
    \label{fig:Mv_dist}
\end{figure}

To complement the HST imaging, the WISP program obtained follow-up ground-based Palomar/WIYN Sloan $u$, $g$, $r$, $i$ and Spitzer/IRAC 3.6$\,\mu$m observations for WISP fields 62, 64 and 309. The IRAC imaging is drizzled to a 0.6$''$ pixel scale (1.2$''$ pixel native scale) and has $5\sigma$ 1.2$''$ radius aperture depths in the range 23.35 to 23.58. The AB magnitude distribution for the F160W \lq\lq H-band" (F140W \lq\lq JH-band" for WISP field 236) is given in Figure \ref{fig:Hmag_dist}. 

From the available imaging, we determine the photometry with \texttt{IRAF.phot} using apertures fixed at the location of the emission line galaxies for each WISP target to provide complementary analysis to existing SExtractor \citep{Bertin&Arnouts96} derived photometry from the WISP team (see \citealt{Atek10, Battisti24} for discussion of SExtractor parameter details). The SExtractor photometry was measured by training on each filter image separately (i.e., without a reference image), which meant galaxies would not have a flux measurement in a particular band if they were too faint to be detected in that image. In these cases our application of fixed-position apertures allows targets with faint continuum emission to have photometric measurements obtained even when they were undetected above the SExtractor specified thresholds. 
Across the HST direct imaging, 0.4$''$-radius apertures are laid on the coordinates determined from the F160W or F140W reference image of each WISP target. For our aperture photometry, we apply an aperture correction determined from point sources to account for the flux falling outside of the aperture and to return an approximate total magnitude (which is appropriate for compact sources, but will underestimate the total flux if the source is significantly extended).
Due to the lower resolution of the Spitzer/IRAC imaging we measure total fluxes using 1.2$''$-radius apertures 
and apply a
0.7 magnitude aperture correction from \citet{Eyles05} appropriate for a compact source. The lower resolution of Spitzer/IRAC additionally means that for a small number of WISP targets the measured flux within an aperture is confused with the contribution of close neighbours. In each of these cases we utilise the WISP team photometry from \citet{Battisti24} which models the contribution of multiple sources to reduce the confusion.   
For the majority of objects, consistency is found between the photometry derived from both the SExtractor and fixed-aperture photometric analysis methods. However, disparity in the measured photometry is found in extended sources, when the target extends beyond the fixed aperture. For extended sources, the variable radius employed by SExtractor recovers a more reliable total flux estimate. For faint sources, which were not necessarily detected above the required SExtractor thresholds (see \citealt{Atek10} for details), fixed aperture photometry provides reliable measurements using coordinates matched to the HST/WFC3 reference images. Therefore, upon inspection of each galaxy, the preferred choice of photometric method is dependent on whether the galaxy is considered to be extended in comparison to the PSF of the images and the size of the fixed aperture.  

\begin{table*}
\begin{center}
\begin{tabular}{cccccc}
Filter $_{(*)}$      & $\lambda_{\rm{central}}$ ($\mu$m)    & Par-309 & Par-236 & Par-62 & Par-64 \\ \hline
\hline
u $_{(1)}$     &      0.383 & $25.83$ & - & - & $25.97$\\
g $_{(1)}$ &    0.487  & $25.59$ & - & - & $25.97$ \\
F475X $_{(3)}$  & 0.490         & -    & -    & -   & 25.39  \\
r $_{(1)}$  &    0.625  & $25.14$ & - & - & - \\
F600LP $_{(3)}$ & 0.719         & -    & -    & -   & 24.88  \\
i $_{(2)}$  & 0.768     & - & - & - & $25.18$ \\
F814W $_{(3)}$   & 0.806       & 25.17   &  -   &  -  & -   \\
F110W $_{(3)}$ & 1.152        & 26.05   &  -   & 25.38  & 25.95  \\
F140W $_{(3)}$ & 1.392         & -    & 25.29   &  24.74  & -   \\
F160W $_{(3)}$ & 1.540          & 25.13   & -    & -  & 24.97  \\
IRAC $3.6 \mu$m $_{(4)}$&  3.557           & 23.58   & -    & 23.58  & 23.35 
\end{tabular}
\caption{HST UVIS, WFC3 and ground-based Sloan-like ancillary photometric data (taken by the WISP collaboration) available for each of the four WISPS fields observed. The $5\sigma$ 0.4$''$-radius aperture depths are given for each, with the IRAC given for a larger 1.2$''$-radius aperture due to the lower resolution. 
$*$ Imaging from: $1)$ Magellan-Megacam; $2)$ WIYN-Mini Mosaic; $3)$ HST; $4)$ Spitzer/IRAC}
\label{tab:ancillary}
\end{center}
\end{table*}

\begin{table*}
\begin{tabular}{lllllllll}
ID\_fors$^a$ & RA [deg] & DEC [deg] & orig\_par\_obj$^b$ & v6.2\_par\_obj$^c$ & $z$\_fors & $z$\_wisp\_orig & $z$\_v6.2 & flag$^d$ \\ \hline
64\_2\_10 
&219.365463
&-1.828888
&64\_206
&--
&0.6588$\pm$0.0001
&0.660$\pm$0.001
&--
&2
\\
309\_1\_16 
&324.797699
&-38.433235
&309\_40
&309\_40
&0.8123$\pm$0.0001
&0.8109$\pm$0.0003
&0.8156$\pm$0.0004
&0\\
...
\end{tabular}
\caption{Identification of targets across various catalogues. $^a)$ Throughout this work we refer to targets from the WISP field they were observed in (e.g. par 309) and the FORS2 mask and slit they were assigned to (e.g., \texttt{309\_1\_16}). $^b)$ Each target also has an original identification based on the early emission line catalogue in that field (e.g. par 309, object 40). $^c)$ if a target appears in the v6.2 catalogue we also provide the identification from \citet{Battisti24}.  
$^d)$ Finally, we provide a WFC3 single emission line flag ($2=$ single emission line reported, $1=$ multiple lines reported but only one
$\geq3\sigma$, $0=$ multiple $\geq3\sigma$ lines). Available for the full sample as a machine readable table.
}
\label{tab:identification}
\end{table*}

\begin{table*}
\begin{tabular}{llllllllllll}
ID\_fors &  u & g & F475X & r & F600LP & i & F814W & F110W & F140W & F160W & IRAC \\ \hline
64\_2\_10 
&0.18$\pm$0.03
&0.19$\pm$0.03
&0.27$\pm$0.05
&-
&0.45$\pm$0.08
&-
&-
&0.51$\pm$0.03 
&-
&0.54$\pm$0.07
&1.46$\pm$0.33
\\
309\_1\_16 
&2.77$\pm$0.04
&3.09$\pm$0.05
&-
&4.42$\pm$0.08
&-
&-
&4.57$\pm$0.13 
&5.12$\pm$0.08
&- 
&5.02$\pm$0.25 
&7.94$\pm$0.19
  \\
...
\end{tabular}
\caption{Ancillary photometry ($\mu$Jy), available for the full sample as a machine readable table.}
\label{tab:photometry}
\end{table*}

\section{Results}
The combination of HST/WFC3 slitless grism and VLT/FORS2 provides spectroscopic coverage over a broad range from the observed-optical to near-infrared, with complementary multi-band photometry constraining the broadband continuum from the observed-optical into the mid-infrared. This breadth of data provides coverage of key rest-optical emission lines over our redshift range (e.g., [O\textsc{ii}]$\lambda\lambda3727,29$, H$\beta$, [O\textsc{iii}]$\lambda\lambda4959,5007$ and H$\alpha$) which we will use to assess the validity of HST/WFC3 grism WISP line identifications and redshifts, correct WISP H$\alpha$ derived SFR by estimating the dust extinction, and utilise line diagnostics and photometry-derived stellar masses to assess the gas-phase metallicity and interstellar medium (ISM) conditions for our galaxies. 

\subsection{Emission line identification and flux measurement}
\label{sec:identification}
With the 2D and 1D FORS2 spectrum for each galaxy extracted, we visually inspect each to identify all significant emission. Visual inspection establishes 38 of 85 unique WISP targets display FORS2 emission lines (the breakdown by field and mask is given in Table \ref{table:FORS2-detections}). We note that this FORS2 emission line detection rate is low and we emphasise 
that many of the single line emitters followed-up were low S/N in the WISP grism spectrum and FORS2 is being used to test the reality of these (see discussion in Section \ref{sec:non-detect}). As a preliminary check before a more sophisticated redshift determination, a simple overlay is created to mark the locations of expected [O\textsc{ii}], H$\beta$ and [O\textsc{iii}] emission lines on the FORS2 2D spectra based on the WISP spectroscopic redshift estimates for each galaxy. This provides immediate visual confirmation that 36  galaxies have observed emission lines consistent with the expected redshift, whereas 2 galaxies show emission lines inconsistent with the WISP redshift (see Section \ref{sec:redshift_valid} and Appendix \ref{App:comparison_redshift} for details). The FORS2 observed emission lines are typically the \OIIdoubletfull doublet and often H$\beta + $ \OIIIdoubletfull. We will discuss redshift validation further in Section \ref{sec:redshift_valid}.

The remaining 47 targets without obvious emission lines in the FORS2 spectrum can be further separated into two categories. First, those where no emission lines were predicted to lie in the FORS2 wavelength range (at $z>1.28$) based on the WISP redshift estimates, which comprised 15 of the 85 galaxies targetted, all 15 of which had no lines detected in our FORS2 spectroscopy. Here, the lack of contradictory line detections in FORS2 supports the WISP redshift estimate and line identification by ruling out alternative redshift/line-identification solutions that would have placed emission in the FORS2 wavelength coverage (although we caution this does not rule out the possibility that a single-line detection in the HST/WFC3 WFSS is spurious). Secondly, targets where we had expected emission lines to fall within the FORS2 wavelength range (70 of our targetted sample, $z<1.28$) but no detections were made (32/70), either due to the emission line flux being too faint to be detected or due to misidentification of the WISP emission line (or the WISP emission line being spurious), which we will discuss further in Section \ref{sec:non-detect}. 

The 1D science and noise spectra of the 38 targets with visually identified FORS2 emission line detections are processed through the Penalized Pixel-Fitting spectral fitting package \citep[PPXF,][]{Cappellari:2017} to fit a continuum and emission line model to the spectra and to obtain emission line flux measurements. Here the galaxy spectra are brought to the rest-frame using the HST/WFC3 grism WISP redshift estimate. The galaxy continuum in these emission line selected targets is usually non-dominant and a fourth order additive polynomial is used to model the continuum; the individual emission lines are modelled as either a single gaussian or a pair in the case of the [O\textsc{ii}]$\lambda\lambda3727,29$ doublet. The flux ratio between this pair is tied to constraints set from atomic physics \citep{Osterbrock_06}. Any velocity offset exhibited by the emission lines allows the HST/WFC3 WFSS redshift estimate to be refined by the FORS2 spectroscopy. The emission line fluxes are given in Table \ref{tab:emission_flux}.

\subsection{Redshift validation}
\label{sec:redshift_valid}
When a galaxy in a spectroscopic survey only has a single emission line detected, determination of the line identification and redshift is ambiguous. One goal of our optical spectroscopy follow-up is to validate the WISP redshifts, especially for the 47/85 galaxies that only had a single emission line detection in the HST/WFC3 WFSS.

Visual inspection of the 2D and 1D FORS2 spectra identified 38 of the 85 unique galaxies with significant line emission ($>3\sigma$). Of these 38, 36 galaxies had FORS2 emission line detections that supported their WISP designated spectroscopic redshift, whilst 2 galaxies had FORS2 significant line emission that was inconsistent with the expected redshift. Hence, in the sub-set of cases where there was significant line detection in FORS2 (38/85, 45\% of targetted objects) the tentative redshift from WFC3 was confirmed 95\% of the time (36/38). Where FORS2 was able to detect emission lines, there was a very high success rate in confirming that the WISP redshifts are accurate, although due to the lower WFC3 
WFSS spectral resolution compared to the FORS2, the FORS2 redshift is more precise. 

Of the 38 galaxies with significant FORS2 line emission, 17 were cases when the WISP redshift was based on only a single line detection (most commonly attributed to H$\alpha$ and in one case [O\textsc{iii}]). Here, the reliability of the WISP spectroscopic redshift is less robust than cases when multiple emission lines were detected, which could further constrain the redshift. The 2 galaxies with FORS2 detections inconsistent with the expected WISP spectroscopic redshift both had only single line detections in WFC3, and present catastrophic WISP misidentification.
From the detection of multiple FORS2 emission lines we determine new line identifications and redshifts for these 2 galaxies. 
FORS2 ID \texttt{236\_2\_12} had an initial tentative WISP spectroscopic redshift of $z\sim0.36$ based on a single emission line assumed to be H$\alpha$. However, FORS2 spectroscopy identified line emission of H$\beta$, [O\textsc{iii}] and H$\alpha$ consistent with $z=0.12$. With no standard emission line expected at the WISP line wavelength (rest-frame 0.89$\mu$m based on the FORS2 spectroscopic redshift), the WISP survey line detection (S/N = 2.9) is determined to be spurious and may have been caused by a cosmic ray or some detector artefact coincident with the dispersed slitless spectrum of the galaxy. FORS2 ID \texttt{62\_1\_13} similarly had a single line detection attributed to H$\alpha$ at $z\sim0.62$ however, a clear FORS2 [O\textsc{ii}] doublet is identified which determines a $z=1.13$ spectroscopic redshift. The WISP detection (S/N = 6.1) lies coincident with a rest-frame wavelength of 4990\,\AA\ based on the FORS2 redshift, which we identify as a being consistent with the blended [O\textsc{iii}]$\lambda\lambda4959,5007$ doublet. This is a case of a single detected WISP HST/WFC3 emission line being real, but mis-attributed to H$\alpha$ rather than [O\textsc{iii}]. Of the single-line emitters from WISP/WFC3 for which we detect lines in FORS2, in 10\% (2/17) of the cases the original single-line in WFC3 must be [O\textsc{iii}]$\lambda\lambda4959,5007$, with 90\% being  H$\alpha$.

\begin{figure*}
    \centering
    \includegraphics[width=\textwidth]{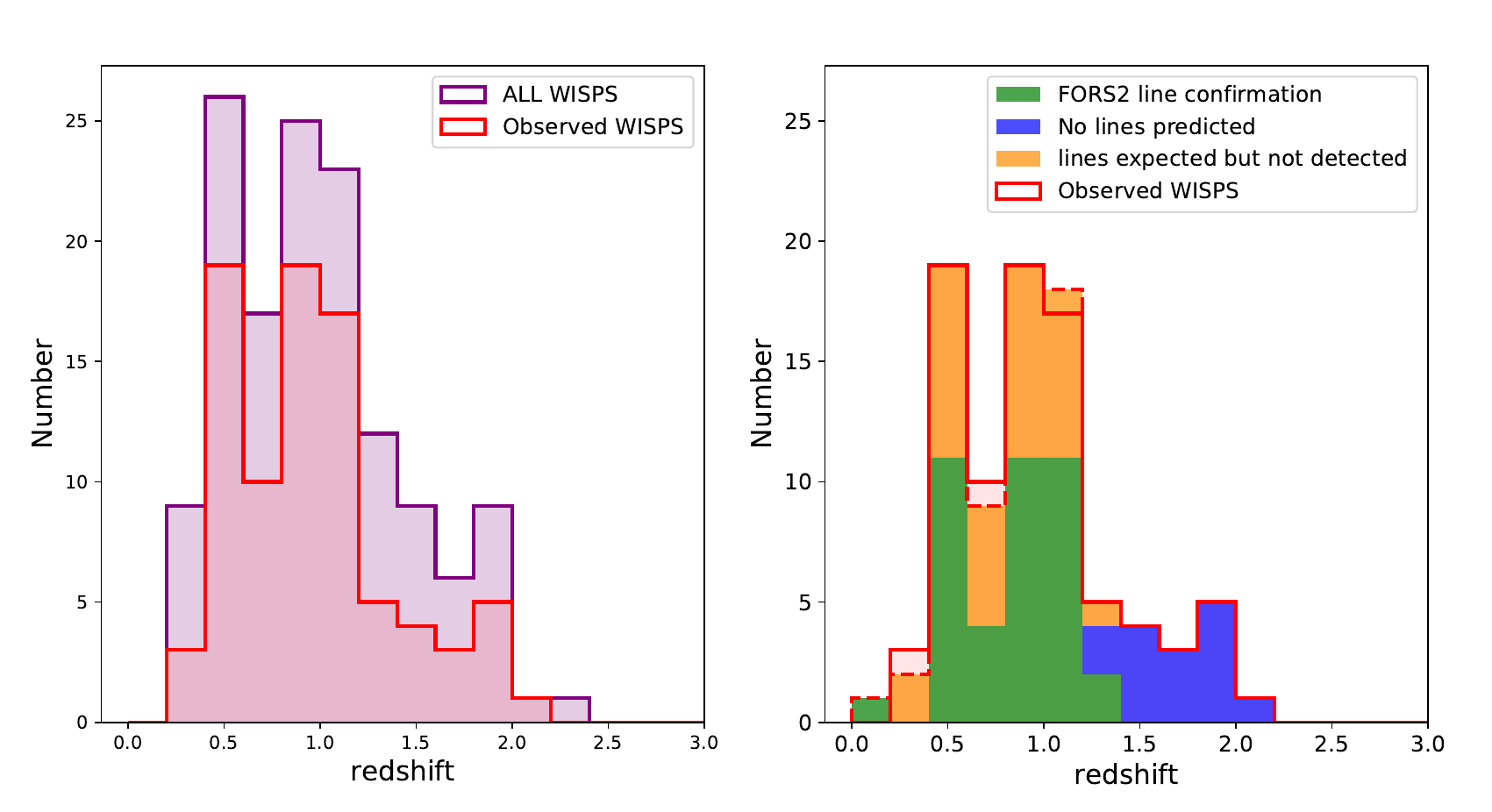}
    \caption{Left: The HST/WFC3 WFSS redshift distribution of the 85 galaxies (red), out of the 138 WISP emission line detected sources (purple), that we follow up with FORS2 spectroscopy. Right: The updated redshift distribution based on the FORS2 follow-up (solid red, with updates shown in dashed). 38 galaxies (green) had FORS2 detected emission lines allowing their spectroscopic redshifts to be validated or corrected. 32 galaxies (orange) are predicted to exhibit emission lines within the FORS2 wavelength coverage but none are detected in FORS2 at S/N$=3\sigma$ detection threshold. 15 galaxies (blue) were not predicted to exhibit emission lines within the FORS2 wavelength coverage and the lack of contradictory lines help rule out lower redshift solutions.}
    \label{fig:red_dist}
\end{figure*}

\begin{figure*}
    \centering
    \includegraphics[width=\textwidth]{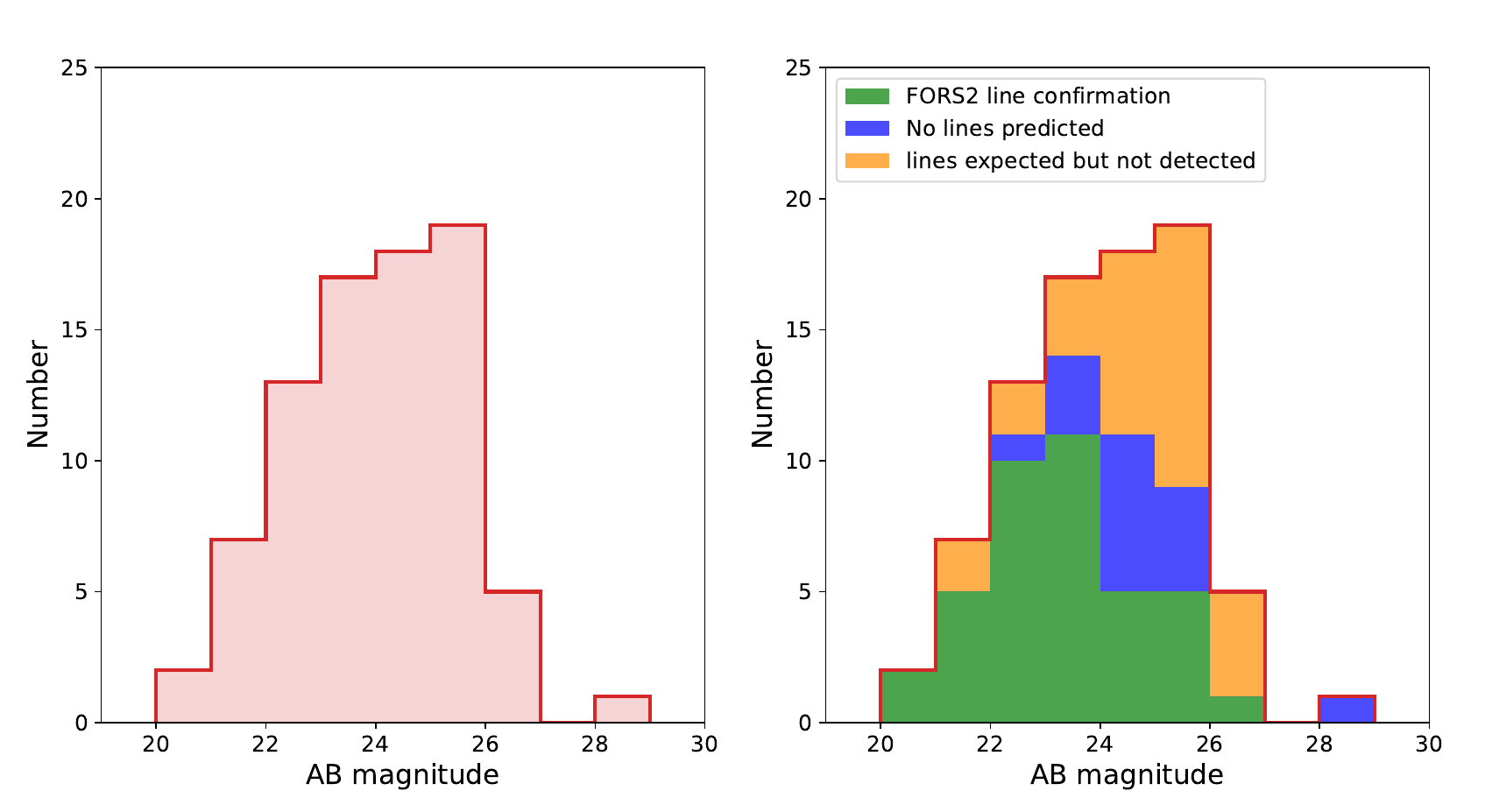}
    \caption{Left: The Near-IR AB magnitude distribution of the 85 galaxies in our sample (red), using either the $H$-band or the $JH$-band depending on available imaging. Right: The near-IR AB magnitude distribution coloured by the same redshift validation as in Figure \ref{fig:red_dist}. Galaxies (green) that had FORS2-detected emission lines are typically brighter than average, but still cover a broad magnitude range in F160W/F140W of 20$<AB<27$\,mag.}
    \label{fig:red_dist_mag}
\end{figure*}

In Figure~\ref{fig:red_dist} we first present the WISP spectroscopic redshift distribution (left panel) for the complete WISP survey sample of galaxies in the four target fields and the sub-sample that were followed up with our optical spectroscopy. In the right panel of Figure \ref{fig:red_dist}, we present the updated redshift distribution, considering the FORS2 spectroscopy, identifying how robustly each sub-group has its redshift validated. 
In Appendix \ref{App:comparison_redshift} we compare the original catalogue redshift with the updated redshifts in Figure \ref{fig:spectroscopic_redshift}.
In Figure \ref{fig:red_dist_mag}, we present the Near-IR magnitude distribution, again identifying how robustly each sub-group has its redshift validated. 

\subsubsection{Undetected emission lines}
\label{sec:non-detect}
Given that a significant fraction of our sample 
did not display emission lines in their FORS2 spectroscopy, despite the sufficient wavelength coverage (32/70 at $z<1.28$), we now consider whether this was expected. 
We note that of these 32, 26 were single line detections in the WFC3 spectroscopy and of the remaining 6, only 3 had multiple lines detected at $>3\sigma$.
In turn we now examine whether the expected signal-to-noise of the three strongest rest-optical emission lines ([O\textsc{ii}], H$\beta$ and [O\textsc{iii}]) would be below the $3\sigma$ detection threshold of FORS2. Hence, whether these non-detections can be attributed to the line emission being too faint or to the HST/WFC3 WFSS emission lines being misidentified or spurious.

To determine the expected observed flux for each undetected emission line, we will utilise the line ratio between the average detected emission line flux of H$\alpha$ and the average detected flux of the line in question, for the sub-sample where both lines were detected. In this calculation we remove two sources which were spatially extended compared to the FORS2 slit width, as these would artificially increase the measured line ratio due to slit losses. We require detections in both emission lines (H$\alpha$ and the line in question) and a detection of at least one line in the FORS2 spectroscopy to confirm the redshift. Within our sample there are 34 galaxies that meet this criterion for detection of both H$\alpha$ and [O\textsc{ii}] lines. The H$\alpha$ to [O\textsc{ii}] line flux ratio of the average observed line fluxes is $3.6\pm0.7$. For H$\beta$, there are 15 galaxies that meet these criteria with a measured observed H$\alpha$ to H$\beta$ line flux ratio to of $6.0\pm1.0$ (this is consistent with the Balmer decrement presented in Section \ref{sec:balmer} where we also consider the [N\textsc{ii}] contribution and stellar absorption). For [O\textsc{iii}]$\lambda\lambda4959,5007$, 24 galaxies meet these criteria with a measured observed line flux ratio to H$\alpha$ of $1.5\pm0.2$. 

To calculate the expected signal-to-noise of each undetected line, the WISP measured H$\alpha$ flux for each galaxy is divided by the appropriate line flux ratio (see above) to estimate the corresponding observed flux. Then the associated noise spectrum for each galaxy is used to estimate the noise over a 500$\,$km\,s$^{-1}$ aperture centred on the expected wavelength, determined from the WISP grism spectroscopic redshift. Together these provide an estimate of the expected S/N for each of the emission lines that were undetected in FORS2.

Out of the 32 galaxies without FORS2 detections but with WFC3 emission-line-determined redshifts that predict emission lines to fall in the FORS2 coverage, 29 were expected to exhibit [O\textsc{ii}] with a S/N estimate ranging between 4 and 43 (the remaining 3 galaxies were at redshifts or slit-mask locations where [O\textsc{ii}] was not in the spectral coverage). The 17 galaxies expected to exhibit [O\textsc{iii}] had estimated S/N ranging between 5 and 57. Finally for H$\beta$, the same 17 galaxies are estimated to exhibit a S/N in the range of 2 to 21 for the observed line ratio, or up to a S/N between 5 to 44 for the intrinsic line ratio of H$\alpha$/H$\beta$ (2.86 from \citealt{Osterbrock89}). Therefore, we would have expected to observe significant line emission in the FORS2 spectra for the majority of these objects, following the typical line ratios of the FORS2 detected galaxies. We note that 26 of these 32 undetected galaxies had WISP redshifts based on a single-emission line detection in WFC3. Hence, for the majority of these galaxies we believe the WFC3 emission line may have been spurious (e.g., a mistaken artefact within the grism imaging) or have been mis-identified such that no lines should have been expected to fall in the FORS2 coverage. 

As a check, we stack the FORS2 spectra for the galaxies without FORS2 emission line detections despite expectation. The stacking is performed by averaging the linearised 1D FORS2 spectrum of each object. Due to the lower spectral resolution of the WFC3 grism the redshift precision means the expected location of the non-detected emission line may lie over a broad pixel range ($\sim 30$\AA\, due to the precision of the redshift that could be constrained from the WFC3 grism). We examine evidence for a flux excess over a $\pm15$\AA\, wavelength range centred on the expected position, after subtracting a fit to the continuum emission. We find no evidence at more than $2\sigma$ for a flux excess in the stacked [O\textsc{ii}], [O\textsc{iii}] or H$\beta$ spectra. We note that due to the redshift uncertainty in WFC3 being much larger than that due to the FORS2 spectral resolution, the undetected emission line may fall over a large wavelength (and pixel) range and this increases the noise in the stack and can mask the benefits of improving the S/N of the  individual spectra by stacking the data. This stacking result supports that the WFC3 detected single-line emission of these galaxies may not be genuine, or that the line identification is incorrect, or the object in the direct-image associated with the spectrum (required to set the wavelength zeropoint) was mis-identified.

Within our complete sample there are 43 single emission line galaxies in WFC3 where we expect lines to fall in our FORS2 coverage ($z<1.28$), with 15 (35$\%$) having their redshift confirmed. The inverse of this sets a maximum potential false-detection rate of galaxies identified in the WFC3 WFSS with only a single emission line to $<65$$\%$. 
This is a upper limit on the false detection rate (i.e., spurious emission lines, false-positives) because it includes cases when the WFSS line may be real but the emission line rest-wavelength is mis-identified as well as cases when the WFSS line may be spurious.
We find that this upper limit to the single emission line false-detection rate is roughly in line with that determined by \citet{Baronchelli20, Baronchelli21} of about $\sim50\%$ using both supervised and unsupervised machine learning models trained on the complete WISPS datasets. It is clear that in WFSS surveys the factions of false detections among the single lines is far from marginal.

\subsubsection{Confirmation rates in the v6.2 public catalogue}
\label{sec:v6.2_confirmation_rates}
Finally, when we restrict our sample to only those that appeared in the recent v6.2 emission line catalogue, which had a higher S/N requirement, we find a much higher success rate for redshift confirmation. Of 19 galaxies in v6.2 which we target, 13 have redshifts in the v6.2 emission line catalogue where emission lines should be observable with FORS2. We confirm FORS2 line detections in 11 out of 13, giving a confirmation rate of 85$\%$. The 6 galaxies at redshifts $z>1.28$, where no lines are anticipated to fall in FORS2 wavelength coverage, show no emission lines in FORS2 as expected. Of the 19 galaxies targetted from the v6.2 catalogue, 5 had a single emission line in WFC3. Of these, 2 have confirmed redshifts from FORS2, and 2 have WISP redshifts beyond the range where we would expect to see line emission in the FORS2 coverage ($z>1.28$), so of those single-emission line  galaxies in the v6.2 catalogue expected to have emission lines in FORS2 we have a 67\% (2/3) confirmation rate compared to 35\% for the larger preliminary catalogue used initially to choose targets (15 confirmed single emission line galaxies out of 43 at $z<1.28$). 
The remaining 1 single-emission-line galaxy out of 5 from the v6.2 catalogue is likely to be spurious (a potential false-detection rate of 33\%, compared with 65\% for the original early catalogue, when we consider galaxies at $z<1.28$). Here the detection rate is clearly shown to improve with higher S/N thresholds in WFSS. However, we do note that there are 25 galaxies with FORS2 line detections which confirm the WISPS spectroscopic redshift that do not appear in v6.2 (see Table \ref{tab:identification}), of which 12 were single line emitters. 

\subsection{Emission line flux corrections}
In this sub-section we will consider the necessary corrections for the observed emission line fluxes to recover line luminosities. The emission line fluxes are presented in Table \ref{tab:emission_flux}, and are available electronically\footnote{All the tables presented in this work are available 
at \url{https://github.com/Kitboyett/Boyett_et_al_2024}}.

\begin{table*}
\begin{tabular}{ccccccccccccc}
ID\_fors & redshift$^a$ & [O\textsc{ii}]$^{\rm{b}}$ & H$\beta^{\rm{b}}$ & [O\textsc{iii}]$^{\rm{b}}$ & H$\alpha^{\rm{b,c}}$ & [S\textsc{ii}]$^{\rm{b}}$ & [O\textsc{ii}]$^{\rm{b,d}}$ & H$\beta^{\rm{b,d,e}}$ & [O\textsc{iii}]$^{\rm{b,d}}$ & H$\alpha^{\rm{b,c,d,e}}$ & [S\textsc{ii}]$^{\rm{b,d}}$ & $L_{\rm H\alpha}^{\rm b,c}$ \\ \hline
64\_2\_10 
&0.6588$\pm$0.0001 
&2.9$\pm$0.2 
&1.7$\pm$0.1 
&9.1$\pm$0.3
&7$\pm$1
& -
&15$\pm$1.2  
&7.1$\pm$0.3 
&31$\pm$1 
&18$\pm$4 
& -
& 12$\pm$2
\\
309\_1\_16 
&0.8123$\pm$0.0001
&43$\pm$7 
&15$\pm$3
&92$\pm$4
&56$\pm$2
&5$\pm$1
&220$\pm$37 
&61$\pm$14
&320$\pm$13
&160$\pm$5  
&13$\pm$3
&180$\pm$6
\\
...
\end{tabular}
\caption{Emission line measurements (fluxes: $10^{-17}$erg/s/cm$^2$, Luminosity $10^{40}$erg/s) for our full sample measured from FORS2 or taken from the early WISP emission line catalogue. $a)$ Spectroscopic redshift from FORS2 detected line emission when available, otherwise set to the WFC3 redshift. $b)$. Line measurement corrected for seeing. $c)$. H$\alpha$ measurements corrected for [N \textsc{ii}] contribution. $d)$.
Line measurements corrected for reddening assuming $A_{\rm{H}\alpha}=1$ mag and the \citet{Calzetti00} extinction law. $e)$ H$\alpha$ and H$\beta$ corrected for Balmer absorption, adopting 2 and 3\AA\, EW corrections respectively from \citet{Dom_nguez_2013}. Available as a machine readable table.}
\label{tab:emission_flux}
\end{table*}

\subsubsection{[NII] contribution correction for H$\alpha$ flux}
\label{sec:NIIcorr}

The low spectral resolution of the WFC\,3 WFSS means the H$\alpha$, [N\textsc{ii}]$\lambda6548$ and [N\textsc{ii}]$\lambda6584$ emission lines are blended and the WISP flux measurement therefore provides an upper limit to the H$\alpha$ emission. To obtain an estimate for the contribution from the two [N\textsc{ii}] emission lines (which have a theoretical flux ratio of 3:1 between [N\textsc{ii}]$\lambda6584$ and [N\textsc{ii}]$\lambda6548$, \citealt{Osterbrock_06}), the [N\textsc{ii}]$\lambda6584$ over H$\alpha$ ratio is estimated from the star-forming abundance sequence in the redshift-dependent BPT diagnostic diagram from \citet[equation 5]{Kewley_13(mean)}. The star-forming abundance sequence provides a [N\textsc{ii}]$\lambda6584$/H$\alpha$ flux ratio as a function of the [O\textsc{iii}]$\lambda5007$/H$\beta$ flux ratio under the assumption that our sample consist of star forming galaxies. The mean value for the [N\textsc{ii}]$\lambda\lambda6548,6584$ contribution to the total H$\alpha$  + [N\textsc{ii}]$\lambda\lambda6548,6584$ flux is $0.19 \pm 0.10$ for the sub-sample of 19 galaxies which had [O\textsc{iii}]$\lambda5007$, H$\beta$ and H$\alpha$ detections, exhibiting a mean redshift of $z = 0.69 \pm 0.34$. We also consider a plausible upper limit on the [N\textsc{ii}]$\lambda$(6548 + 6584) (hereafter [N\textsc{ii}]) contribution to H$\alpha$  + [N\textsc{ii}] flux for this sub-sample, which we determine to be $0.26 \pm 0.10$, following the 0.1 dex upper boundary on the star forming abundance sequence provided by \citet{Kewley_13(upper)}. For the remaining galaxies in the sample, the H$\alpha$ flux is corrected for the [N\textsc{ii}] contribution using the mean from this sub-sample (where [N\textsc{ii}] contributes a fraction 0.19 to the blended flux). Our correction provides an updated estimate to the [N\textsc{ii}] contributions adopted in previous WISP collaboration papers \citep{Atek10, Atek_11, Dom_nguez_2013, Colbert13, Masters14} which lie in the range $4 - 20\%$, dependent on the H$\alpha$ equivalent width and stellar mass of the galaxy. Our value for the [N\textsc{ii}] contribution is marginally larger than some previous estimates, due to the availability of a redshift-dependent BPT star-forming abundance sequence. Higher [N\textsc{ii}] contributions are found at higher redshifts which may be due to a larger ionisation parameter, evolution of metallicity or a different N/O abundance ratio \citep{Faisst18}, whereas the previous computations were all reliant on $z \sim 0$ galaxy diagnostics.  

\begin{table*}
\centering
\begin{tabular}{ccccc}
field $\&$ & Fors2 line  & no FORS2 & no FORS2    &  total  \\

mask &  detections &  detections (high-$z$) & detections ($<S/N$) & targets  \\ \hline
\hline
309\_1           & 11    & 1                        & 5                       & 17                   \\
309\_2           & 3     & 7                        & 4                       & 14              \\
62\_1            & 6     & 2                        & 5                       & 13                 \\
64\_2            & 5     & 4                        & 4                       & 13                   \\
236\_1           & 8     & 0                        & 8                       & 16                  \\
236\_2           & 5     & 1                        & 6                       & 12                 \\ \hline
Totals           & 38    & 15                       & 32                      & 85
\end{tabular}
\caption{FORS2 follow-up detection rate split between each of the six observed mask configurations.  }
\label{table:FORS2-detections}
\end{table*}

\subsubsection{Extinction correction and the Balmer Decrement}
\label{sec:balmer}

\begin{figure*}
    \centering
    \includegraphics[width=\textwidth]{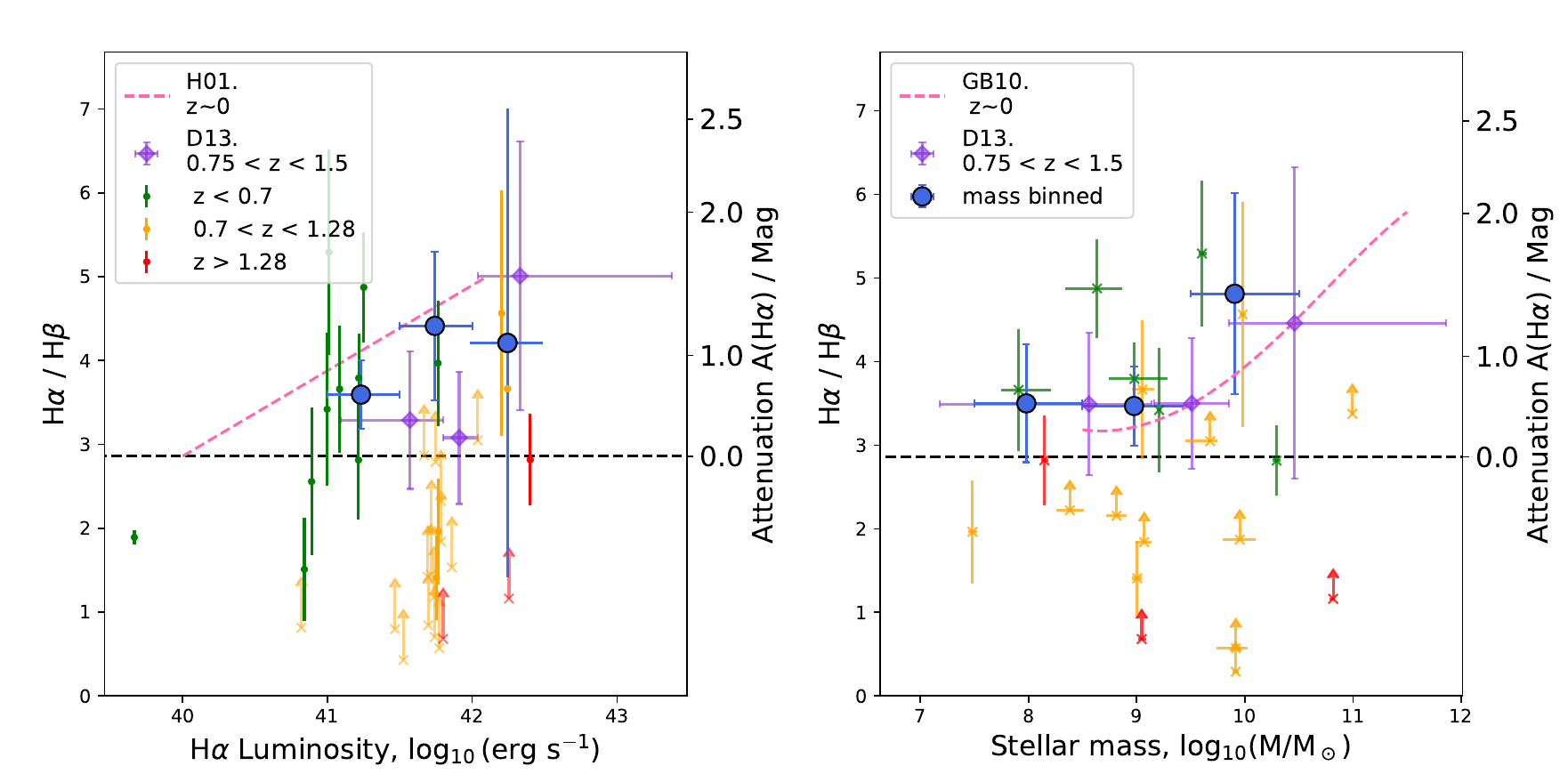}
    \caption{Balmer decrement relation with H$\alpha$ luminosity (left) and stellar mass (right). Balmer lines corrected for seeing and [N\textsc{ii}] contribution, we additionally correct the Balmer decrement for stellar absorption. 
    In both panels our detections are spilt into low (green, $z< 0.7$), medium (orange, $0.7 < z < 1.28$) and high (red, $z > 1.28$) redshift bins, with galaxies that have H$\beta$ detections below $2\sigma$ plotted as lower limits on H$\alpha$/H$\beta$. Two galaxies are removed which had artefacts around the H$\beta$ emission line in the spectroscopy. 
    In purple diamonds, we present the WISP sample from \citet{Dom_nguez_2013} ($0.75\leq z\leq 1.5$).
    Left: Binned data points (blue) show the luminosity ranges $\log_{10}(L_{\rm H\alpha} / \rm{erg\,s}^{-1})$ of $41.0-41.5$, $41.5-42.0$ and $42.0-42.5$. 
    Pink dashed line shows the binned SDSS ($z\sim0$) sample from \citet{Hopkins01}.
    Right: Binned data points (blue) show the stellar mass ranges $\log_{10}($M$_\star$/M$_\odot)$ of $7.4-8.5$, $8.5-9.5$ and $9.5-10.5$. 
    Pink dashed line shows the binned SDSS ($z\sim0$) sample from \citet{GarnBest2010}.
    }
    \label{fig:balmer_dec_ha_lum}
\end{figure*}

Observation of the H$\alpha$ emission line in the WISP survey allows galaxy properties including SFR to be measured. However, measurements of the SFR based on H$\alpha$ luminosity alone may be underestimated due to dust attenuation, which we wish to correct for.
For the majority of our sample we have wavelength coverage of the H$\alpha$ emission line in the HST/WFC3 WFSS and also the H$\beta$ emission line, which falls in our FORS2 wavelength coverage at $z\lesssim0.7$, and in the G102 WFC3 grism at $0.7<z<1.4$. 
We assume the intrinsic flux ratio between the H$\alpha$ and H$\beta$ emission lines (the Balmer decrement) is $f($H$\alpha)/f($H$\beta)=2.86$, set by atomic physics for Case B recombination for a temperature T$=10^4 K$ and an electron density $n_e = 10^2$ cm$^{-3}$ \citep{Osterbrock_06}.

The flux ratio observed in galaxies is typically larger than this due to wavelength dependent differential extinction, with the shorter-wavelength H$\beta$ line typically more attenuated.
In this section, we will use the observed Balmer decrement from galaxies in our sample along with a Calzetti extinction law \citep{Calzetti00} to determine and correct for the dust extinction.

We note that since the H$\beta$ emission line flux is typically more than three times weaker than H$\alpha$ it is frequently undetected (see Section \ref{sec:non-detect}). This happens even in the cases where we confirm the putative WISP redshift from a single emission line in the slitless WFC3 WISP spectrum with another line (typically [O\textsc{ii}] or [O\textsc{iii}]) in the follow-up FORS2 spectroscopy.

To determine the Balmer decrement within our sample the measured H$\alpha$ and H$\beta$ emission line fluxes must be corrected for the effects of [N\textsc{ii}] contribution to the measured H$\alpha$ flux and the effect of stellar absorption for both lines.
First, the H$\alpha$ flux is corrected for contamination from the [N\textsc{ii}] doublet, as described in Section~\ref{sec:NIIcorr}.
Secondly, we correct for the intrinsic Balmer line absorption that occurs in the photospheres of stars in these galaxies. The strength of this absorption depends on the stellar population, and we follow \citet{Dom_nguez_2013} who study the dust properties of WISP galaxies and adopt their correction values for the emission line rest-frame equivalent width of 3\,\AA\ for H$\beta$ and 2\,\AA\ for H$\alpha$, which \citet{Dom_nguez_2013} determine for their lowest stellar mass sub-sample
$M_\star\approx 10^{8.5}\,M_{\odot}$ (which is consistent with the mean stellar mass of our sample, described in Section \ref{sec:stellar_mass}). For this correction we assume a continuum slope flat in $f_\nu$ appropriate for ongoing star formation in the absence of significant reddening \citep[e.g.,][]{Wilkins11}.

The Balmer decrement is measured for the sub-sample of galaxies that have detections $>2\sigma$ in both H$\alpha$ and H$\beta$ emission lines. For galaxies with a robust redshift (i.e., confirmed through a FORS2 line detection or multiple emission line detections in the WISP survey) but with a H$\beta$ signal-to-noise below our detection threshold, a $2\sigma$ lower limit to the Balmer decrement is derived using the 2$\sigma$ upper limit to the H$\beta$ flux and we plot these limits for individual galaxies in Figure \ref{fig:balmer_dec_ha_lum}. We also stack the data in bins of H$\alpha$ luminosity by adding the measured line fluxes from different galaxies (even if undetected in individual cases) to get the average Balmer decrement for a sub-sample of galaxies. We do not consider three galaxies that clearly have extended morphology beyond the FORS2 1$''$ slit-width, where the slit-losses would lead to an overestimate of the Balmer decrement (these are noted in Table \ref{tab:emission_flux}\footnote{FORS2 ID: \texttt{236\_1\_10}, \texttt{62\_1\_16} and \texttt{309\_1\_5}}). Together the average Balmer decrement for the sub-sample of galaxies with robust redshifts is $4.08 \pm 0.45$. In Figure \ref{fig:balmer_dec_ha_lum}, the Balmer decrement measurements and lower limits are plotted against their H$\alpha$ luminosity (corrected for [N\textsc{ii}] contribution). The individual galaxies are colour coded by redshift and are also binned (closed blue circles) into three H$\alpha$ luminosity regimes (log$_{10}(L_{\rm H\alpha} / \rm{erg\,s}^{-1}$) = 41.0-41.5, 41.5-42.0 and 42.0-42.5). 
Two galaxies are removed from the $42.0 < \log_{10}(L_{\rm H\alpha} / \rm{erg\,s}^{-1}) < 42.5$ bin, on account of their WFC3 spectra exhibiting strong negative counts around H$\beta$ due to data reduction artefacts\footnote{FORS2 ID: \texttt{64\_2\_15} and \texttt{236\_2\_16}}. The binned data lie slightly below the local Balmer decrement - H$\alpha$ luminosity relation from \citet{Hopkins01} and are in broad agreement with the moderate redshift ($0.75 < z < 1.5$) observations made by \citet{Dom_nguez_2013}.

When splitting our stacks into bins of H$\alpha$ luminosity, we note a trend that galaxies with higher H$\alpha$ emission line luminosity tend to have higher Balmer ratios and hence more dust extinction. However, we note that our lowest luminosity bin ($\log_{10}(L_{\rm H\alpha} / \rm{erg\,s}^{-1})=41.0-41.5$) is almost exclusively comprised of galaxies at $z<0.7$ with H$\beta$ falling in the FORS2 spectroscopy, but the higher luminosity bins are mainly the higher redshift galaxies with H$\beta$ from the WFC3 WFSS. Hence the trend may either be due to line luminosity, or instead a trend with redshift. Many models predict that dust should increase with time (i.e.\ decrease with redshift), although how the dust extinction in star forming regions evolves is less clear (e.g., \citealt{Zavala21} find the contribution from dust obscured star formation to the total remains flat up to $z\sim2$).
Our Balmer decrement results, binned in H$\alpha$ luminosity, are slightly below the $z\approx 0$ work of \citet{Hopkins01} from SDSS, who also fit an extinction dependence on H$\alpha$ luminosity. However, to guard against a possible selection effect with luminosity (in that it is easier to detect the less-obscured emission lines), we also plot the observed Balmer decrement as a function of stellar mass (right panel of Figure \ref{fig:balmer_dec_ha_lum}, see Section \ref{sec:stellar_mass} for mass measurement). As also found by \citet{Battisti22} using a larger WISP sample with WFC3 spectroscopy alone, our measured Balmer decrements stacked in stellar mass bins are in approximate agreement with the mass:reddening relation seen in low-redshift galaxies in SDSS \citep{GarnBest2010}, where we consider the SDSS fibre-region stellar mass (with $\log M_\star^{tot}\approx \log M_\star^{fib}+0.5$) as used in  \citet{Battisti22}. Our results are consistent with  the scenario where reddening increases with stellar mass (as also found in WISP by \citealt{Dom_nguez_2013}), although we note that our small sample size and large error bars mean that the three bins are also consistent with uniform reddening.

Although we find tentative evidence for a correlation between H$\alpha$ luminosity and extinction, we also note that each luminosity bin is consistent within $1\sigma$ with an extinction $A(\rm H\alpha) = 1$\,mag, consistent with the findings of \citet{Sobral_2009}, corresponding to an observed Balmer decrement of 4.07.
We note that the average Balmer decrement, stacking our full sample of galaxies with robust redshifts, is $4.08 \pm 0.45$.  In order to correct the emission line fluxes, later used in line diagnostic analysis (Section \ref{sec:line_diagnostics}), we assume a fixed extinction corresponding to this standard value of $A($H$\alpha) = 1$\,mag. For a \citet{Calzetti00} extinction law this corresponds to $A$([O\textsc{ii}]) $\sim 1.77$\,mag and $A$(H$\beta),\,\,\, A$([O\textsc{iii}]) $\sim 1.36$\,mag for the other nebular emission lines. We note that the continuum might be affected by a different level of attenuation than the nebular lines \citep{Calzetti00}, and we consider this further in Section~\ref{sec:SFR-uv}.

\subsection{Identifying potential AGN}
\label{sec:AGN}
\begin{figure}
    \centering
    \includegraphics[width=\columnwidth]{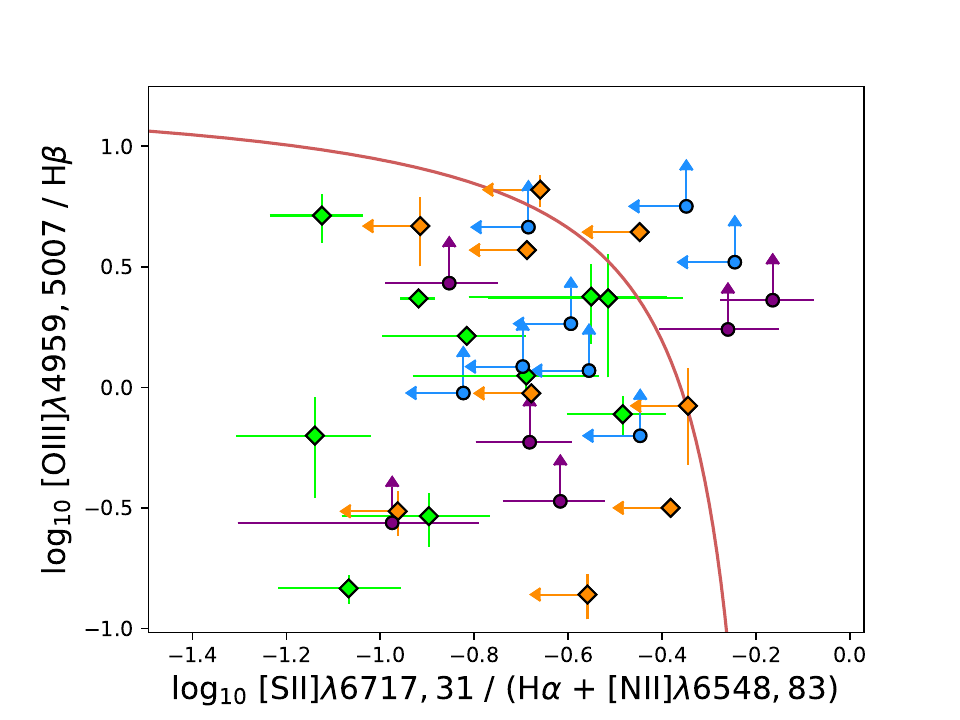}
    \includegraphics[width=\columnwidth]{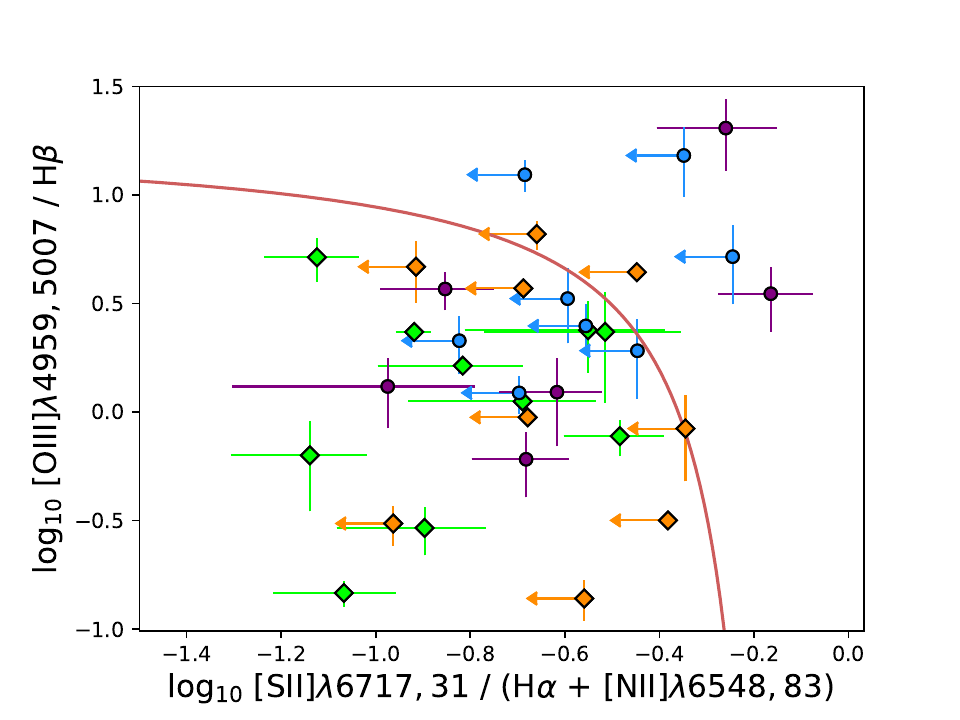}
    \caption{The [S\textsc{ii}]-modified BPT diagram to identify AGN candidates (which lie in the region above the curve), using the \citet{Henry21} separation curve (red).
    In the two panels we treat non-detections ($ \geq 2 \sigma$) of H$\beta$ differently. Top: Galaxies without H$\beta$ detections are treated as $ 2\sigma$ lower limits on the flux ratios. Bottom: Galaxies without H$\beta$ detections are plotted using $f$(H$\alpha$)/4.07 (that consistent with a H$\alpha$ attenuation of 1 magnitude).
    Galaxies with H$\beta$ detections are shown in diamonds (green, or orange when [S\textsc{ii}] is undetected), whilst galaxies without detections in H$\beta$ are shown in circles (blue, or purple when [S\textsc{ii}] is undetected). Galaxies without [S\textsc{ii}] detections are set to $2\sigma$ [S\textsc{ii}] upper limits for the flux ratios.}
    \label{fig:BPT}
\end{figure}

Emission line galaxies can be powered by star formation and/or AGN activity, and although we do not have X-ray data by which to select AGN, we can still address the relative contribution by looking at the line ratios of the emission lines which relate to the hardness of the ionising spectrum. A BPT diagram \citep{Balwin_1981, Veilleux87} traditionally uses [N\textsc{ii}] and [O\textsc{iii}] forbidden lines in ratio with their counterpart (close in wavelength) recombination lines H$\alpha$ and H$\beta$ to discriminate between star formation and AGN as the source of photoionisation. However, for our dataset, [N\textsc{ii}] is unusable in the low spectral resolution WFC\,3 grism as it is blended with H$\alpha$. Hence we use  [S\textsc{ii}]$\lambda\lambda6717,6731$ (hereafter [S\textsc{ii}]) as an alternative diagnostic to [N\textsc{ii}] \citep[e.g.,][]{Kewley:2006}.
A soft ionising spectrum indicating a star forming radiation field consists of low [O\textsc{iii}]$\lambda5007$/H$\beta$ and [S\textsc{ii}]$\lambda\lambda6717,31$/(H$\alpha$+[N\textsc{ii}]) line flux ratios whilst Seyfert galaxies and LINERs with harder ionising spectra occupy the high ratio region.
Our galaxies are plotted on the [S\textsc{ii}]-BPT diagnostic diagram in Figure \ref{fig:BPT}, and we overlay the $z\sim0$ AGN/star-formation separation line diagnostic from \citet{Henry21} which is the theoretical maximum for SFG calculated using Cloudy (v17; \citealt{Ferland17}) models (see also \citealt{Henry18}).

We have 35 galaxies where we cover all the lines in the [S\textsc{ii}]-BPT diagram. This breaks down to 11 galaxies with ($>2\sigma$) detections in each of H$\alpha$, [O\textsc{iii}], [S\textsc{ii}], H$\beta$. There are 10 galaxies that have only [S\textsc{ii}] undetected and 6 galaxies with only H$\beta$ undetected. The remaining 8 are undetected in both [S\textsc{ii}] and H$\beta$. We plot 2$\sigma$ limits on the flux ratios in the upper panel of Figure \ref{fig:BPT}. In the lower panel of Figure \ref{fig:BPT} we replace the limits on undetected H$\beta$ with an inferred flux based on the H$\alpha$ flux, assuming a case B flux ratio and an assumed dust attenuation of $A(H\alpha)=1$\,magnitude with a \citet{Calzetti00} extinction law (Section~\ref{sec:balmer}), corresponding to $f($H$\alpha) /f($H$\beta)=4.07$ (where the H$\alpha$ flux has had the estimated [N\textsc{ii}] removed, and both H$\alpha$ and H$\beta$ have been corrected for stellar absorption, see Sections \ref{sec:NIIcorr} and \ref{sec:balmer} respectively).

From this sub-sample, 29 of the 35 galaxies on our [S\textsc{ii}]-BPT diagram lie below the SFG/AGN demarcation line at $z\sim0$, although it should be noted that 10 of these 29 have lower limits on H$\beta$ that mean they are conceivably still consistent with the AGN region (this falls to 3 if we use H$\beta$ fluxes inferred from the measured H$\alpha$ emission line), whilst 19 (54\%) are confirmed as inconsistent with being AGN.
The remaining 6 galaxies lie above the diagnostic demarcation line in the region consistent with being dominated by AGN, although 4 are consistent with the star forming region within the uncertainties. We therefore determine that 33 of the 35 galaxies (94\%) are consistent with being star forming galaxies (either falling below the demarcation line for $z\sim 0$ within the error bars, or with limits consistent with the star forming region).
Of the galaxies above the dividing line, inspection of the F160W $H$-band WFC\,3 images reveals that the two inconsistent with being SFGs appear to be spatially unresolved (FORS2 ID: \texttt{309\_1\_2} \& \texttt{309\_1\_6}).
We note that \texttt{309\_1\_6} also exhibited [Ne \textsc{iii}]$\lambda3869$ emission, further evidence of a hard radiation field (although we note that 2 additional galaxies that are consistent with being SFGs also had [Ne \textsc{iii}] detected).

In recent years it has been suggested that there is evolution with redshift in the dividing line between SFG and AGN in the BPT diagram, in the sense that the demarcation rises with increasing redshift, and this is particularly prominent in the [N\textsc{ii}]-based BPT \citep[e.g.,][who note that this evolution may also be due to a larger electron density or a harder ionising radiation field]{Kewley_13(mean)}. However, \citet{Shapley19} study the [S\textsc{ii}]-based BPT diagram for a sample of galaxies from the MOSDEF survey at $z\sim$1.5-2.3 (slightly higher redshift on average than in our sample), and find that the evolution is less extreme for [S\textsc{ii}] than when using [N\textsc{ii}], which they argue is attributable to two competing effects: a less significant contribution from Diffuse Ionised Gas (DIG) leading to a lower [S\textsc{ii}]/H$\alpha$ ratio at higher-redshift, and an increased ionisation parameter which pushes this line ratio in the opposite direction (higher [S\textsc{ii}]/H$\alpha$). Hence, the net result is minimal overall evolution in the [S\textsc{ii}]-BPT diagram, and we adopt the $z\sim0$ dividing line as our AGN criterion.

\begin{figure}
    \centering
    \includegraphics[width = \columnwidth]{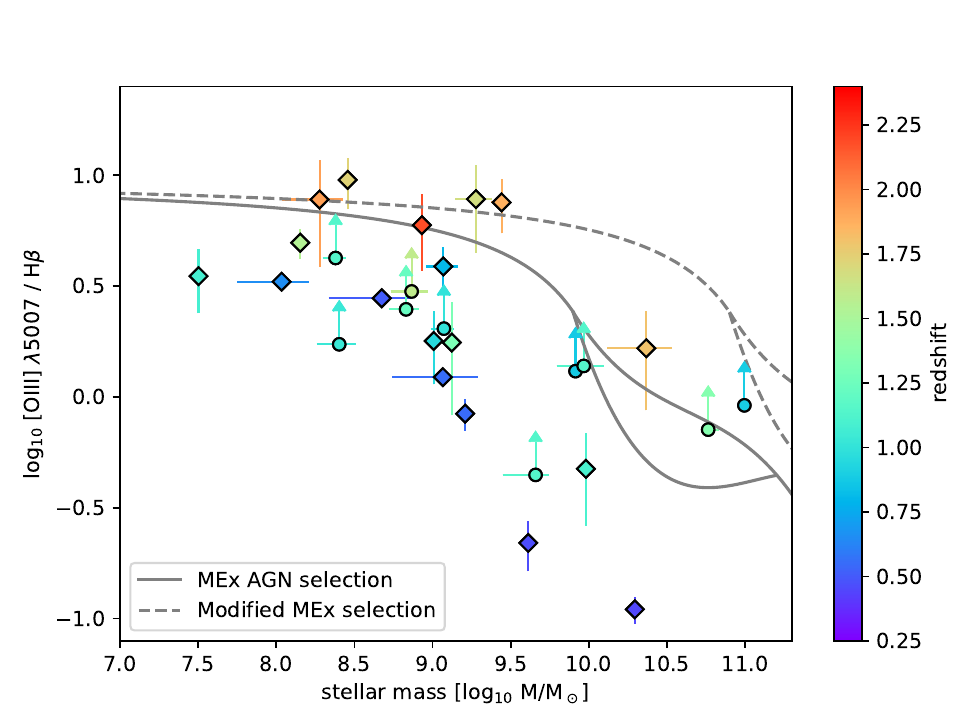}
    \includegraphics[width = \columnwidth]{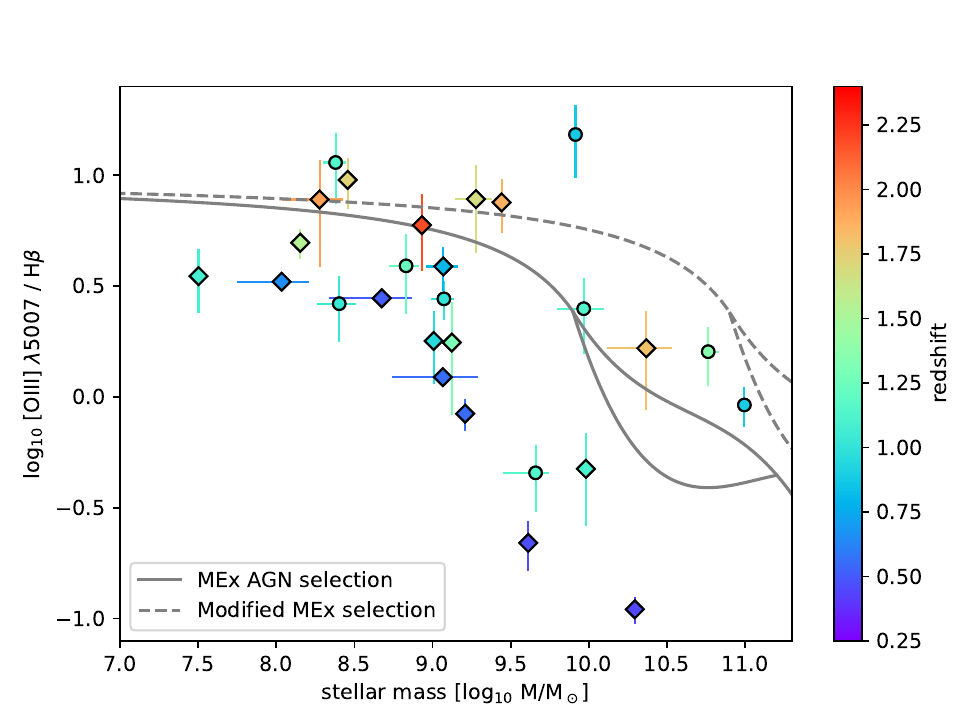}
    \caption{The mass excitation (MEx) diagnostic diagram shows that the ionising radiation field in the majority of our sample is consistent with star formation, rather than AGN activity. Top: galaxies without H$\beta$ detections ($\geq2\sigma$, circles) are treated as $2\sigma$ lower limits on the line flux ratios. Bottom: galaxies without H$\beta$ detections are plotted using $f$(H$\alpha$)/4.07 (consistent with  A(H$\alpha$)= 1 magnitude). }
    \label{fig:MEx}
\end{figure}

The measurements of [S\textsc{ii}] from our WISP WFC3 spectroscopy are often low S/N, and so we want another check on the presence of AGN. We also consider the mass-excitation (MEx) diagnostic proposed by \citet{juneau11}, where the [N\textsc{ii}]/H$\alpha$ ratio of the original BPT diagram is replaced by the stellar mass (on the grounds that the average stellar mass of a galaxy is related to metallicity, from the mass-metallicity relation, see Section \ref{sec:mass_metal}). 
This also allows us to inspect the ionising conditions of the highest redshift galaxies within our sample, where H$\alpha$ no longer fell within the HST/WFC3 WFSS coverage. Plotting the stellar mass (which we describe in Section \ref{sec:stellar_mass}) against [O\textsc{iii}]$\lambda5007$/H$\beta$ allows SFG to be separated from AGN, and we present our MEx diagram in Figure \ref{fig:MEx}, where we plot the low-redshift dividing line, along with another division which may be more appropriate at higher redshift (due to evolution in the mass-metallicity relation e.g., \citealt{Newman14, coil15, Kashino19, Henry21}). We adopt the AGN/star-formation separation in the MEx diagram from \citet{coil15} who consider a sample of MOSDEF galaxies.

Within the MEx diagram, there are 6 galaxies with H$\beta$ detections (diamonds in Figure \ref{fig:MEx}) that lie in the $z\sim0$ AGN region, although each of these lies either inside or is consistent within their uncertainties of the modified $z\sim2$ SFG region. When undetected H$\beta$ flux limits are replaced with the H$\alpha$ flux divided by 4.07 (that consistent with a H$\alpha$ attenuation of 1 magnitude), 5 further galaxies lie above the $z\sim0$ AGN/SFG separation line, although 4 are at least consistent with the SFG region on the modified MEx diagram. Only one galaxy (FORS2 ID \texttt{309\_1\_2}) is clearly an AGN considering both MEx diagrams, this galaxy is also identified as an AGN in the [S\textsc{ii}]-based BPT diagram.

In Table \ref{tab:AGN_candidates} we present the AGN candidates identified by the [S\textsc{ii}]-BPT and MEx diagnostics, and report that only 2 galaxies are identified by both methods. These 2 are also found to be unresolved in the F110W and F160W imaging,
suggesting that these are indeed AGN contaminants of our star forming sample. We remove them from our sample for the remaining analysis. The remaining AGN candidates are identified as being SFG in at least one of the methods with many also being spatially resolved. We conclude that star formation is the dominant ionising source in most of our emission-line-selected galaxies, with only 2 out of 41 galaxies (which have sufficient data to be plotted on at least one of the [S\textsc{ii}]-BPT or MEx diagrams) likely being AGN. For the overwhelming majority of our sample, the H$\alpha$ emission is most likely dominated by a SFG rather than an AGN.  

\begin{table*}
\begin{tabular}{lllll}
FORS2 ID   & Modified [S\textsc{ii}] BPT & MEx  & Modified MEx & imaging/SED \\ \hline
309\_1\_2$^{a}$  & AGN              & AGN  & AGN  & Compact \\
309\_1\_6$^{a}$  & AGN              & AGN  & AGN* & Compact \\
309\_1\_13 & AGN                    & SFG  & SFG  & Resolved \\
309\_2\_13 & AGN*                   & SFG  & SFG  & Resolved  \\
309\_2\_1  & AGN*                   & SFG  & SFG  & Compact \\
64\_2\_10  & AGN*                   & SFG  & SFG  & Resolved   \\
236\_1\_14 & AGN                    & -    & -    & Resolved    \\
309\_1\_3  & SFG                    & AGN* & SFG  & Resolved    \\
309\_2\_4  & -                      & AGN  & AGN* & Compact          \\
309\_2\_5  & -                      & AGN  & AGN* & Resolved           \\
309\_2\_7  & -                      & AGN* & AGN* & Resolved           \\
64\_2\_5   & SFG                    & AGN  & SFG  & Resolved     \\
64\_2\_9   & -                      & AGN* & SFG  & Compact \\
64\_2\_15  & SFG                    & AGN  & SFG  & Resolved       \\
64\_2\_16  & -                      & AGN* & AGN* & Compact    \\
62\_1\_14  & -                      & AGN  & SFG  & Resolved          
\end{tabular}
\caption{Potential AGN candidates meeting various selection criteria. We consider the [S\textsc{ii}]-BPT diagram  and the mass-excitation (MEx) diagram (where we indicate if it meets the AGN threshold at $z\sim0$ and also a modified threshold which may be more appropriate at high redshifts). 
More marginal AGN candidates marked with ($^*$) were consistent with the SFG region within their uncertainties or had limits consistent with the SFG region due to non-detections in the [S\textsc{ii}] emission line. Two galaxies, marked by ($^a$), are consistent with being AGN in both diagnostics and are removed from the sample for the remaining analysis.}
\label{tab:AGN_candidates}
\end{table*}
 
\subsection{Star formation rates and stellar masses of the WISP galaxies}\label{sec:SFR}
The star formation rate and stellar mass are two fundamental galaxy properties that influence the observational characteristics of a galaxy, including the luminosity, comparative broadband colour and nebular emission line strength. Several indicators of the star formation rate have been used (see for example the review by \citealt{Kennicutt12}) for a range of redshifts and galaxy samples, and these have been combined to trace the evolution of the star formation rate density in the Universe \citep[e.g., ][]{Lilly96, Madau96, Hopkins03, Madau14}.
The sample selection from the HST/WFC3 WFSS favours galaxies with high luminosity emission lines, and it is the expectation that these will exhibit high specific star formation rates (sSFR). The reliance of the sample selection on the detection of strong emission lines without the requirement for any continuum detection is expected to allow a large range of stellar masses to be observed, including low mass star-forming systems potentially missed in spectroscopic surveys following up broad-band magnitude-limited samples.
However, we caution that the WISP sample and our FORS2 follow-up is not a stellar-mass selected sample (by design we are targetting the star-forming population over a range of redshifts), and galaxies of any mass with little or no star formation will fall below the emission line flux detection threshold in the WISP slitless spectroscopy, and hence will be missing from our star-forming sample. 
In this Section we compare SFR estimates derived using three measurements: the H$\alpha$ luminosity; the rest-frame UV continuum luminosity; and through SED fitting to the full broadband photometry. We also derived stellar mass estimates from the SED fitting.
The derived properties are shown in Table \ref{tab:derived_properties}.

\begin{table*}
\begin{tabular}{llllllllllll}
ID &  SFR(UV) & SFR(H$\alpha^{\rm{a}}$)& SFR(SED) & M(SED) & Mv & H$\alpha^{\rm{a}}$ EW$_0$ & metal(O32) & metal(R23)  \\ 
  \hline
64\_2\_10 
&1.70$\pm$0.33
&1.49$\pm$0.27
&1.01$\pm$0.29 
&8.03$\pm$0.23 
&-16.85
&300.0$\pm$58.0 
&8.33$\pm$0.02 
&7.88$\pm$0.25 
\\

309\_1\_16 
&24.54$\pm$0.76 
&27.99$\pm$1.34 
&29.27$\pm$2.65
&9.07$\pm$0.11
&-19.69 
&306.2$\pm$9.4 
&8.41$\pm$0.04 
&$<8.44$
\\
...
\end{tabular}
\caption{Derived galaxy properties from emission line measurements and photometric SED analysis, available for the full sample as a machine readable table. Star formation rates are given in $M_\odot/yr$, BEAGLE SED derived masses in log$_{10}$(M$_\star$/$M_\odot$) and rest-frame equivalent widths given in \AA. Metallicity is in $\log_{10}(O/H)$ where solar is 8.69 \citep{Asplund09}.
$^a$ H$\alpha$ corrected for seeing, stellar absorption and [N\textsc{ii}] contribution}
\label{tab:derived_properties}
\end{table*}

\subsubsection{Spectroscopic star formation rate analysis}
H$\alpha$ is one of many nebular emission lines whose luminosity is sensitive to the star formation rate of a galaxy. H$\alpha$ is a recombination line of atomic hydrogen which arises from the integrated stellar light from photons below the Lyman-limit ($\lambda<912$\,\AA) which photo-ionise neutral hydrogen in the ISM.
The H$\alpha$ luminosity therefore provides a probe of the ionising flux and is sensitive to the population of young massive stars, which have hot photospheres and short lifetimes. Taken with a population synthesis model and assumptions of the underlying initial mass function, a calibrated scaling relation can be constructed between the luminosity of the emission line and the star formation rate.

For our sample we derive the SFR using the H$\alpha$ fluxes corrected for [N\textsc{ii}] contribution, stellar absorption and reddening (see Sections \ref{sec:NIIcorr} and \ref{sec:balmer}). The H$\alpha$ fluxes are converted to luminosities using the FORS2 emission line measured redshift or HST/grism emission line redshift when FORS2 had no line detections. 
To determine the SFR, we adopt the SFR - H$\alpha$ luminosity relation from \citet{Kennicutt12} (with a conversion factor = 1.86$\times10^{41}$ erg s$^{-1}$ / M yr$^{-1}$, see also \citealt{Murphy11, Hao11}), which assumes a \citet{Kroupa01} IMF.

\subsubsection{Star Formation Rates inferred from the rest-UV}
\label{sec:SFR-uv}
\begin{figure}
    \centering
    \includegraphics[width=\columnwidth]{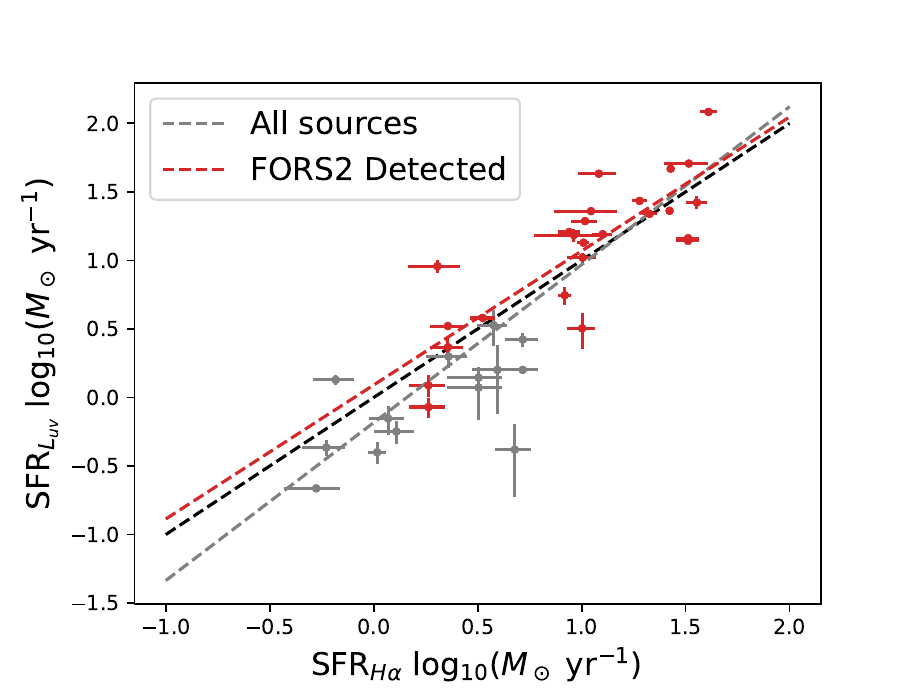}
    \caption{The H$\alpha$-derived SFR compared to the $L_{UV}$ (rest-frame 2800\,\AA) -derived SFR. Targets with FORS2 emission line detections, that can confirm their redshift, are shown in red whilst those which were undetected in FORS2 are in grey. Both SFR measurements are corrected for dust assuming $A_{\rm H\alpha} = 1$\,mag.
    The black line shows the 1:1 line. The best fit model for the whole sample has a log slope of $1.15\pm0.10$ and the sub-sample with FORS2 detections has a log slope of $0.98\pm0.14$. 
    }
    \label{fig:SFR_sFR}
\end{figure}

The luminosity of the non-ionising UV continuum is sensitive to the stellar population of a galaxy and towards shorter wavelengths it is increasingly dominated by the emission from younger stellar populations. The rest-frame UV continuum luminosity can be used as a probe of the underlying star formation rate (e.g., \citealt{Kennicutt12}). We adopt the rest-frame 2800\,\AA\ UV continuum luminosity ($L_{UV}$) as a probe of the star formation rate, available across the majority of our redshift range. We utilise the \citet{Kennicutt12} $L_{UV}$ - SFR scaling relation (see also \citealt{Murphy11, Hao11}), which assumes a constant star forming history (SFH) and a \citet{Kroupa01} IMF.
We note that the non-ionising rest-UV continuum is sensitive to the star formation rate integrated over longer timescales ($\sim$100\,Myr) than that probed by the H$\alpha$ (since the OB stars which dominate the ionising flux have lifetimes of $<$10\,Myr, e.g., \citealt{Sullivan01}). Hence, scatter in the measured $L_{UV}$- and H$\alpha$-derived SFR (see Figure \ref{fig:SFR_sFR}) could be attributable to different recent star formation histories. UV wavelengths are also more sensitive to dust attenuation, which could lead to scatter between the two SFR measurements.

Across our sample, the rest-frame 2800\,\AA\, has broadband filter coverage spanning from the ground-based $r$-band for our lowest redshift galaxies to the HST/WFC3 F110W band for our highest. To estimate the stellar continuum luminosity ($L_{UV}$) from the broadband photometry, we first correct the flux density for any contribution from emission lines (e.g., [O\textsc{ii}]$\lambda3727$) measured in either the HST/WFC3 WFSS or VLT/FORS2 spectroscopy. The flux from any emission lines is subtracted from the filter flux density, weighted by the filter transmission profile at the respective location of each line. For each galaxy in our sample we determine $L_{UV}$ from the broadband magnitude of the filter that contains the rest-frame 2800\,\AA\, based on the spectroscopic redshift. We remove galaxies from this sub-sample if the rest-frame 2800\,\AA\, did not fall within an observed filter or if it lies within 100\,\AA\, of the edge of the filter where the filter response is minimal. The SFR is then computed using the \cite{Kennicutt12} relation for these 41 remaining galaxies out of the 85 observed with FORS2.

Our observed $L_{UV}$ measurements will underestimate the intrinsic luminosity due to attenuation by dust. We correct the 2800\,\AA\, continuum for dust extinction. In Section \ref{sec:balmer} we corrected the nebular emission line flux for reddening by adopting a 1 magnitude attenuation ($A$(H$\alpha)$ = 1) along with a \citet{Calzetti00} extinction law. Like \citet{Wuyts13}, we note the need for extra attenuation in nebular emission lines (e.g., H$\alpha$) compared to the continuum and we adopt the \citet{Calzetti00} parameterisation $E(B-V)_{star} = 0.44 E(B-V)_{gas}$. This parameterisation accounts for the young O/B stars responsible for the UV continuum being within their birth cloud and generates A$_{2800} = 1.29 \rm{A}_{H\alpha}$ as given by Equation 14 of \citet{Calzetti01} (we note that the attenuation at UV wavelengths is much higher than optical wavelengths).

For 39 galaxies, both H$\alpha$-derived and $L_{UV}$-derived star formation rates are available, and we present their comparison in Figure \ref{fig:SFR_sFR}. We find good agreement between the star formation rates derived from each method, 
with a fitted slope of SFR($L_{UV}$) $\propto$ SFR(H$\alpha$) raised to the power of $1.15\pm0.10$, close to a linear relation. We find a scatter in the residuals to be $0.33$dex.
A similar power law slope is identified in \citet{Wuyts13} where, as in this work, a linear relation between $A_{cont}$ and $A_{H\alpha}$ is assumed. 

\subsubsection{Stellar population fits}
\label{sec:stellar_mass}
The available broadband photometry and emission line flux measurements allows us to sample each target's spectral energy distribution (SED) from which we can constrain galaxy properties including the stellar population, stellar mass and SFR. The composite spectrum of a galaxy is dictated by several factors including the galaxy's star formation history, initial mass function (IMF) and dust extinction law. In this section we utilise the Bayesian analysis of Galaxy SEDS (BEAGLE, version 0.24.5, \citealt{Chevallard16}) to model the SED of the sub-set of our sample with sufficient broadband coverage. We therefore do not consider WISP survey field 236 which only has photometry in the HST/WFC3 F140W filter and is unable to place constraints on the stellar templates.

We fix the galaxy redshift to the spectroscopic redshift measured from either the FORS2 or HST/WFC3 emission lines. We assume a star formation history that is comprised of a declining exponential component and a burst. This reflects that the emission line galaxies identified through the HST/WFC3 WFSS are star forming galaxies (typically with high sSFR, indicative of a burst). We assume a \citet{Calzetti00} dust attenuation law and \citet{Chabrier_03} IMF (which finds near identical stellar masses and SFR results to the \citealt{Kroupa01} IMF e.g., \citealt{Chomiuk11}, which we adopted for both the H$\alpha$- and $L_{UV}$-derived SFRs for consistency). We input flux densities and filter transmission curves for each broad-band, along with the emission line fluxes and wavelengths from our spectroscopy. We do not remove the flux contribution from emission lines in the broadband photometry as this is considered as part of the modelling.

\begin{figure}
    \centering
    \includegraphics[width=\columnwidth]{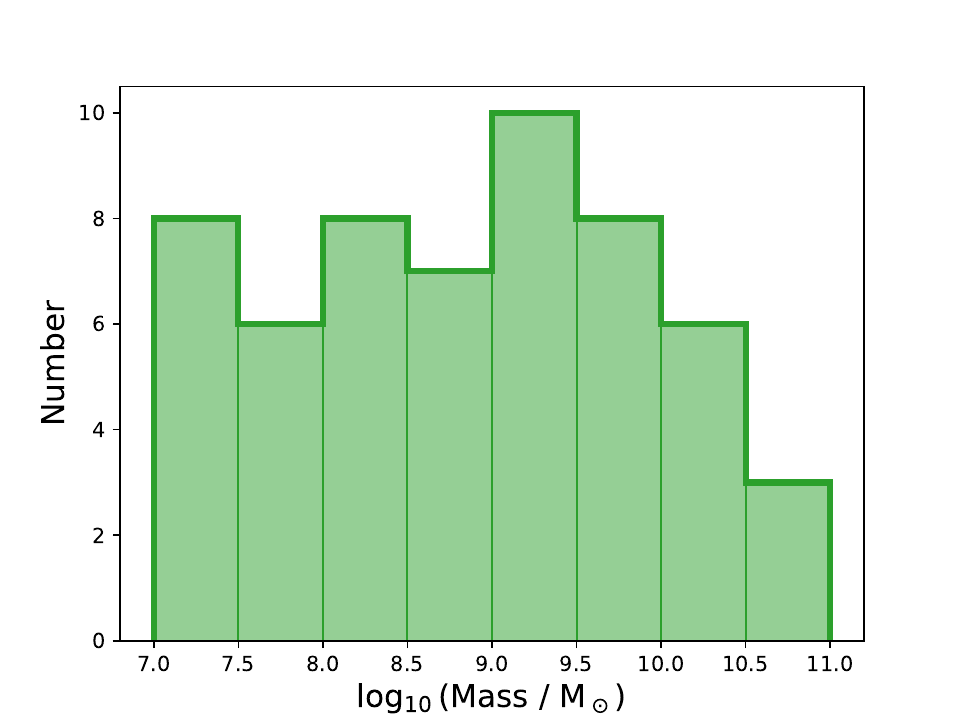}
    \caption{Stellar mass distribution, derived using BEAGLE SED photometric fitting code. This includes all galaxies excluding those from field Par236 where there was insufficient photometric data points to constrain the SED.}
    \label{fig:mass_dist}
\end{figure}

We fit a total of seven parameters in BEAGLE: stellar mass, stellar age, characteristic star formation timescale (the exponential component), SFR (the burst), stellar metallicity, effective dust attenuation, and effective galaxy-wide ionisation parameter. We freely vary the ionisation parameter so the emission line fluxes can better constrain the spectral fits. A burst component is included for physical reasons while also enabling better identification of multi-modal solutions. 
Within the SED fitting, the derived stellar mass acts as a free parameter used to constrain the normalisation of the composite synthetic galaxy stellar population spectrum.
In Figure \ref{fig:mass_dist} the BEAGLE SED derived mass distribution for the 56 galaxies with sufficient broadband photometry to constrain the stellar population is presented (excluding field 236 which only had photometry available in F140W). In particular, the availability of Spitzer/IRAC channel 1 photometry (3.6$\mu$m) for three of our four fields gives us the rest-frame continuum at $>1\mu$m, which is more sensitive to the underlying stellar mass than recent star formation (unlike at shorter wavelengths), and improves our measurement of the stellar mass. We note that of the 56 galaxies covered by Spitzer, 14 are undetected down to $2\sigma$ limits of AB=24.2-24.5mag, and these are consistent with being very low mass systems ($M_{\star}<10^8M_{\odot}$).

With a median mass of $\log_{10}(M_\star/M_\odot)=8.94$ and a dynamic range of $7 < \log_{10}(M_\star/M_\odot) <  11$, our sample reaches lower masses and is typically fainter on average than broadband-magnitude-limited spectroscopic surveys e.g., VVDS (\citealt{Garilli08, fevre05}, see Section \ref{sec:HaEW}). We note that due to the requirement for the galaxies to be detected in the WISP survey direct imaging, to calibrate the wavelength of the emission lines, the lowest mass galaxies may still be missing from our sample. This explains the roll-off of stellar masses below $10^9\,M_{\odot}$ seen in Figure~\ref{fig:mass_dist}. For star-forming galaxies, this potential bias is lessened by having the direct image be taken through a filter which covers most of the wavelength coverage of the grism used, and hence will include the emission line, so strong line-emitting objects will indeed enter the sample through detection in the broad-band image even if the continuum is weak (i.e., low mass objects with high equivalent width line emission will be in the sample).

\begin{figure}
    \centering
    \includegraphics[width=\columnwidth]{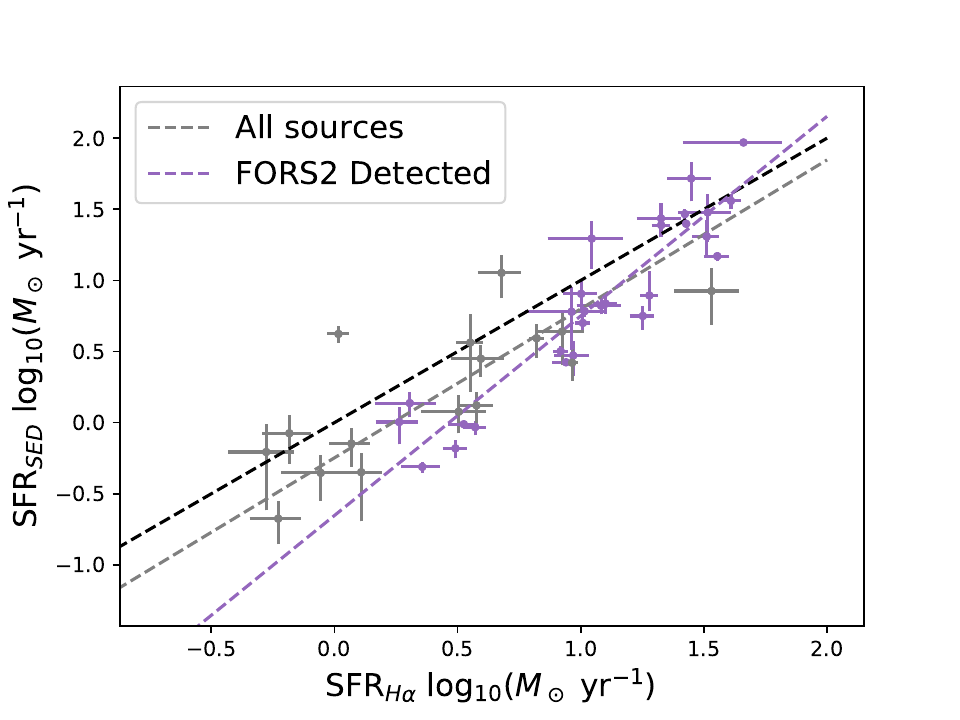}
    \caption{The H$\alpha$-derived SFR compared to the SED-derived SFR. Targets with FORS2 emission line detections, that can confirm their redshift, are shown in purple whilst those which were undetected in FORS2 are in grey.  The black curve shows the 1:1 line. The best fit model for the whole sample has a log slope of $1.05\pm0.08$ with an offset of 0.21dex, while the FORS2 detected sub-sample has a steeper log slope of $1.40\pm0.10$.
    Two galaxies have been removed due to having significantly underestimated SFRs which may indicate AGN contamination, see Section \ref{sec:AGN}}
    \label{fig:SFR_Ha_SED}
\end{figure}

We additionally use our SED fitting to estimate the SFRs of our sample, and in Figure \ref{fig:SFR_Ha_SED} we present their comparison to our H$\alpha$-derived SFRs. We find reasonable agreement between the two methods, with a fitted power law slope of 1.05$\pm$0.08 (1.40$\pm$0.10 when we consider only the FORS2 detected sub-set). We find a scatter in the residuals to be $0.30$dex, similar to that found by \citet[]{Reddy15} (0.34dex) in the comparison of SED- and H$\alpha$-based SFR estimates. We also determine an offset from the 1:1 line of 0.21$\pm$0.05\,dex compared to the H$\alpha$-derived SFRs, which we put down to the difference in underlying assumptions within the SFR calibrations between \citet{Kennicutt12} and BEAGLE (e.g., IMF, mass range, metallicity). 

\subsubsection{Star forming main sequence} 
\begin{figure*}
    \centering
    \includegraphics[width=\textwidth]{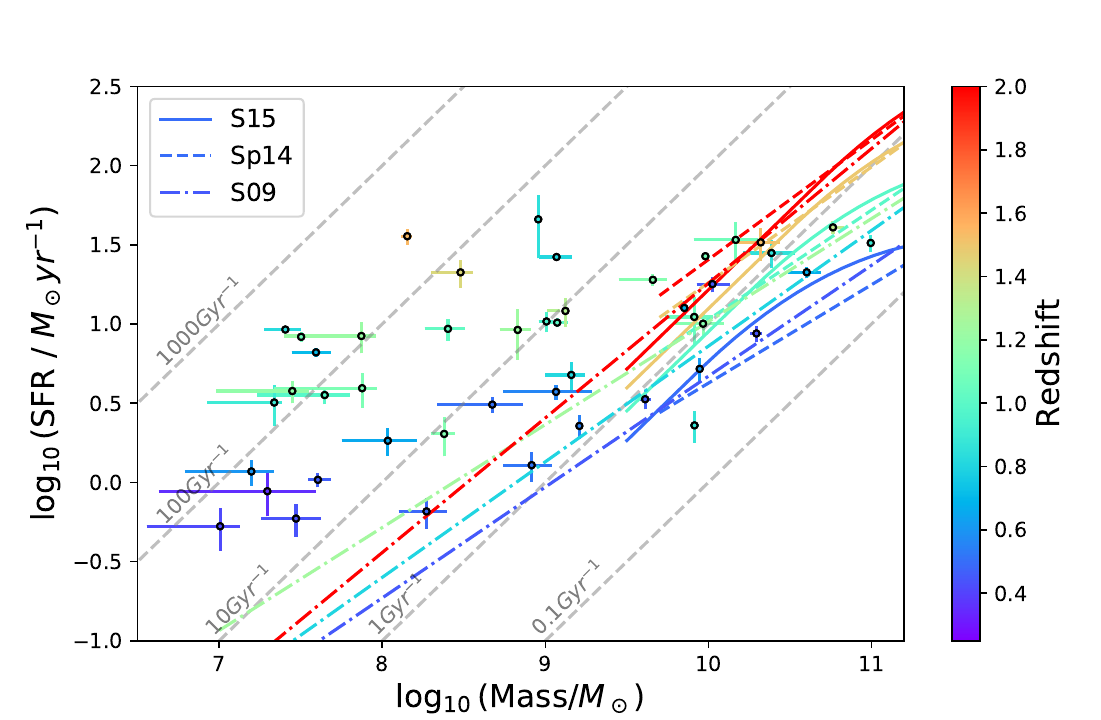}
    \caption{The main sequence, H$\alpha$-derived star formation rate against SED-derived stellar mass. H$\alpha$-derived SFR are adjusted for seeing, [N\textsc{ii}] contribution, stellar absorption and reddening corrections. Grey dashed lines indicated lines of constant sSFR. Over a range of redshifts we over-plot the literature main-sequence from \citet{schreiber15} (solid), \citet{speagle14} (dashed) and \citet{santini09} (dot-dashed). Our emission line selected galaxies, colour coded by redshift, typically lie above the main-sequence. }
    \label{fig:mass_sfr}
\end{figure*}

Galaxies exhibit a relation between their stellar mass and star formation rate, commonly known as the star forming main sequence, which has been studied in great detail from local redshifts out to beyond $z>4$ \citep[e.g.,][]{speagle14, Renzini15, schreiber15, Battisti22}. Most commonly the main-sequence is parameterised with a log-space slope and intercept, with the redshift evolution mainly affecting the normalisation of the relation in line with the increasing SFR density with redshift \citep[e.g., ][]{Madau14}. We present the stellar mass - H$\alpha$ derived SFR \lq main-sequence\rq\ in Figure \ref{fig:mass_sfr} for the sub-set of our sample which have measured H$\alpha$ fluxes. We over-plot the main-sequence from three studies over a redshift range $0<z<2.5$ \citep{speagle14, schreiber15, santini09}.  Galaxies within our sample lie above the main-sequence of the literature curves at the corresponding redshift, matching the expectation that the emission line selected galaxies in the HST/WFC3 WFSS exhibit high sSFR, as one would expect with an increasing SFR, such as a burst. The overall trend of our sample is flatter than the literature main-sequence due to the Eddington bias imposed by our selection on strong H$\alpha$ emitters. At low masses, only those galaxies with the highest SFR produce sufficient H$\alpha$ luminosity to make it into our selection. This selection-driven flattening of the main sequence has been reported for surveys that select on SFR-sensitive proprieties, such as surveys selecting on strong emission lines or UV continuum selections \citep[e.g., ][]{Battisti22, Cochrane18, Rodighiero14, Rodighiero11, erb06b}.      

\subsection{Determination of Metallicity from Emission Lines}
\label{sec:line_diagnostics}
One of the key ISM properties that can be inferred from our observations is the gas-phase metallicity, and in this Section we will derive the gas-phase oxygen abundance from two different line-ratio diagnostics. Using this, we will investigate the relation between the metallicity and the stellar mass of the galaxies in our emission-line selected sample. 

The observed metallicity as measured from the gas-phase oxygen abundance is sensitive to the galaxy's stellar population, star formation history and history of gas inflows and outflows. The chemical enrichment of the ISM reflects the star formation history, with metal-enriched gas returned to the ISM over time through stellar ejecta and explosions. The degree of enrichment of the ISM can be diluted by the flow of cold gas (typically of lower metallicity, and perhaps pristine) from the inter-galactic medium (IGM) or circum-galactic medium (CGM) into the galaxy. Outflows of chemically-enriched gas being ejected from the galaxy driven by supernovae or black hole feedback may also alter the ISM metallicity \citep[e.g.,][]{lilly13}.

\subsubsection{Metallicity Determinations using the R23 and O32 Diagnostics}
\label{sec:metallicity_diganostics}
The wavelength range available from the combined WFC3 and FORS2 spectroscopy provides coverage of metallicity-sensitive nebular emission lines, in particular the [O\textsc{ii}]$\lambda\lambda3727,3729$ and [O\textsc{iii}]$\lambda\lambda4959,5007$ doublets.
In this Section we will utilise these strong oxygen emission lines along with empirical calibrations of the gas-phase oxygen abundance in star-forming galaxies to derive the metallicity of our sample. Alternative approaches include the measurement of stellar metallicity rather than the gas-phase (often using absorption lines, e.g.,  \citealt{Cullen19}) or \lq\lq direct methods" to determine gas-phase metallicity based on electron temperature measurements using auroral emission lines \citep[e.g.,][]{Sanders_2015, Curti23, Laseter24}. However, our spectra from WFC3 and FORS2 are insufficiently deep to achieve good detections of absorption lines or weak auroral emission lines, and hence we can only use strong-line metallicity diagnostics for our sample. 

The flux ratio of recombination emission lines (e.g., H$\alpha$, H$\beta$) to forbidden collisional emission lines (e.g., [O\textsc{iii}]$\lambda5007$, [O\textsc{ii}]$\lambda3727$, [N\textsc{ii}]) is sensitive to the gas-phase metallicity and ionisation conditions of the ISM. Standard nebular line flux ratios R$_2$ = f([O\textsc{ii}]$\lambda3727,3729$)/f(H$\beta$)  and R$_3$ = f([O\textsc{iii}]$\lambda5007$)/f($H\beta$) and the combination of these by \citet{Pagel79} into R$_{23}$ = f([O\textsc{ii}]$\lambda3727,3729$ + [O\textsc{iii}]$\lambda5007, 4959$)/f(H$\beta$) are frequently used as strong-line metallicity indicators. However, these empirically show non-monotonic behaviour (i.e., multiple metallicities can exhibit the same diagnostic value) and, traditionally, a second line diagnostic is introduced to break the degeneracy  (e.g., \citealt{Alloin79, Pagel79,Kewley02,Maiolino_2008, Sanders_2015,Strom_2017}). 
Many works have noted that the excitation conditions in high redshift galaxies are very different from galaxies and HII regions at $z\approx 0$ \citep[e.g.,][]{Masters14, Strom_2017}, and the strong-line metallicity calibrations depend on the ionisation conditions \citep[e.g.,][]{Kewley_13(mean)}. Typically, the metallicity indicator O32 = f([O\textsc{iii}]$\lambda5007$)/ f([O\textsc{ii}]$\lambda\lambda3727,29$) or the f(H$\alpha$)/f([N\textsc{ii}]$\lambda6583$) line flux ratios are used as they are sensitive to the ionisation conditions of the ISM. Since [N\textsc{ii}] is blended into the H$\alpha$ emission line at the spectral resolution of the HST/WFC3 WFSS, and the nearby [S\textsc{ii}]$\lambda\lambda6717,31$ doublet is often at very low signal-to-noise in our WFC3 spectra, we use the O32 diagnostic (rather than [N\textsc{ii}]) to break any degeneracy of the R23 metallicity indicator. 

We utilise the \citet{Curti:2017} empirically-calibrated metallicity diagnostics for R23 and O32 to derive the metallicities for galaxies in our sample. \citet{Curti:2017} use the direct temperature method to measure the metallicity of SDSS galaxies and of individual low-metallicity candidates, from which they construct strong emission line -- metallicity calibrations using both R23 and O32. 

We additionally determine a highest likelihood metallicity. Here, we combine the metallicity estimates with associated Gaussian uncertainties from the \citet{Curti:2017} R23, O32 and R3 metallicity indicators to create a posterior likelihood distribution. The peak of the likelihood and the 68\% confidence intervals of the distribution give a best estimate for the metallicity. 

\begin{figure*}
    \centering
    \includegraphics[width=12cm]{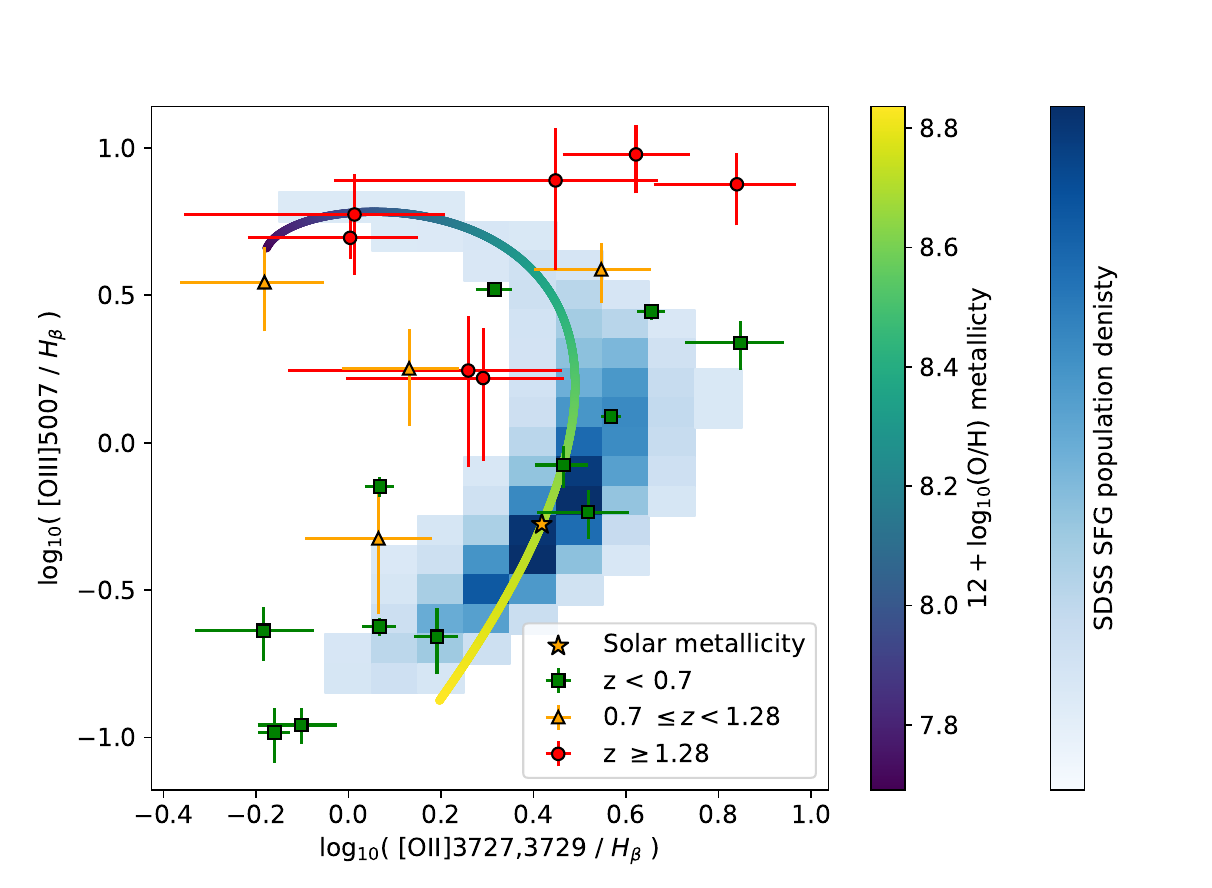}
    \caption{[O\textsc{ii}] vs [O\textsc{iii}] diagnostics normalised by H$\beta$, for the sub-sample of 23 galaxies that have detections in all of [O\textsc{ii}], [O\textsc{iii}] and H$\beta$, split into three redshift bins. The over-plotted \citet{Curti:2017} strong-line diagnostic curve (colour coded by metallicity) indicates that the WISP galaxies generally lie at sub-solar metallicities, with a trend of decreasing metallicity with increasing galaxy redshift. A heat map of SDSS SFG density at $z\sim0$ is under-plotted, taken form \citet{Curti:2017}.}
    \label{fig:OIII/I/Hb}
\end{figure*}

\begin{figure*}
    \centering
    \includegraphics[width=12cm]{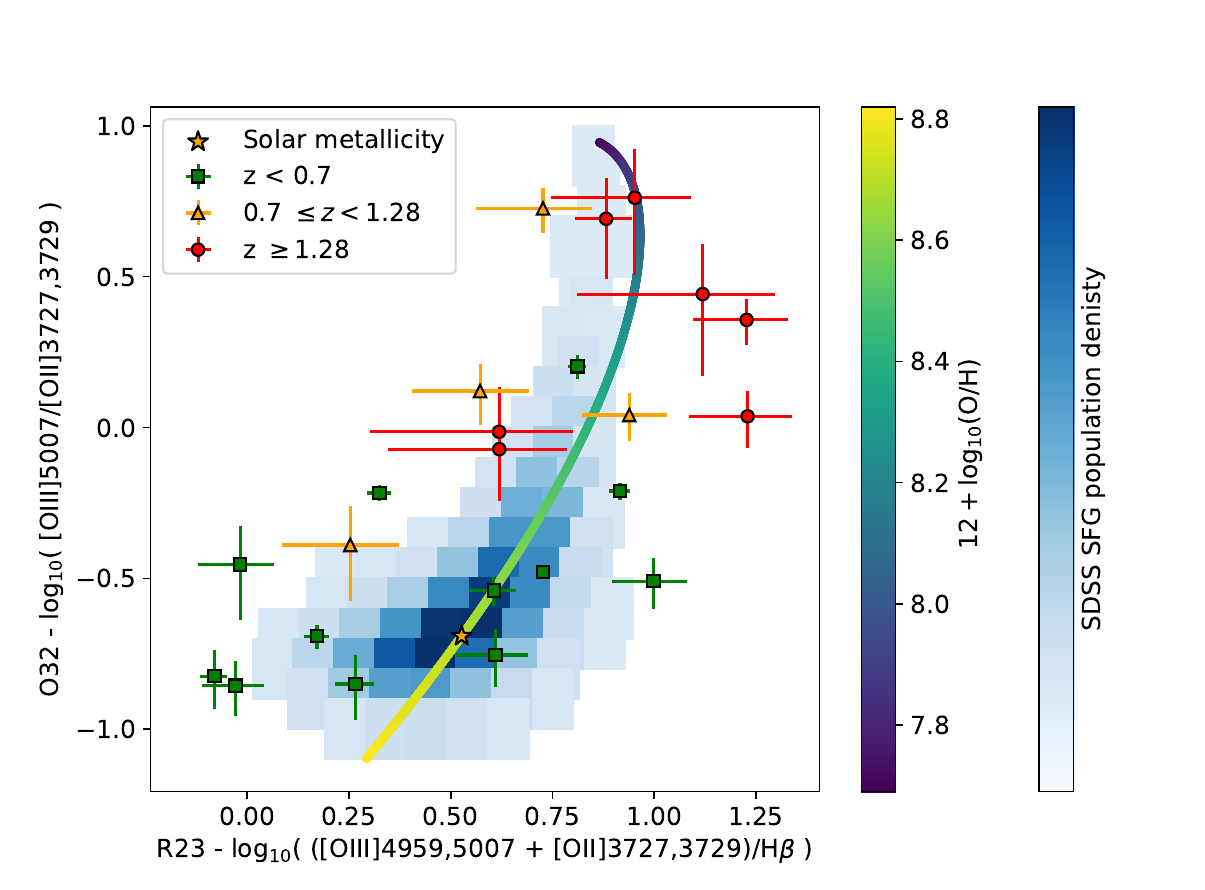}
    \caption{O32 against R23 diagnostic, for the sub-sample of 23 galaxies that have detections in all of [O\textsc{ii}], [O\textsc{iii}] and H$\beta$, split into three redshift bins. The over-plotted \citet{Curti:2017} strong-line diagnostic curve (colour coded by metallicity) indicates most lie at sub-solar metallicities, with a trend of decreasing metallicity with increasing galaxy redshift. }
    \label{fig:O32vsR23}
\end{figure*}

The metallicities for the galaxies in our sample are presented in Table \ref{tab:derived_properties}. We present the line diagnostic R2-R3 in Figure \ref{fig:OIII/I/Hb} and R23-O32 in Figure \ref{fig:O32vsR23} for the individual galaxies within our sample (colour coded by redshift), and compare these with the low redshift results from SDSS, and  the empirical metallicity calibration from \citet{Curti:2017}.
\citet{henry13, Henry21} also analysed the metallicity of a different sub-sample of galaxies from the WISP survey, based on the WFC3 spectra alone and focusing on higher redshifts than in this paper where all the lines [O\textsc{ii}], H$\beta$ and [O\textsc{iii}] are covered by the HST/WFC3 WFSS. Our current paper can explore the metallicities of WISP galaxies down to lower redshift thanks to our VLT/FORS spectroscopic follow-up at shorter wavelengths. 

\begin{figure*}
    \centering
    \includegraphics[width = 12cm]{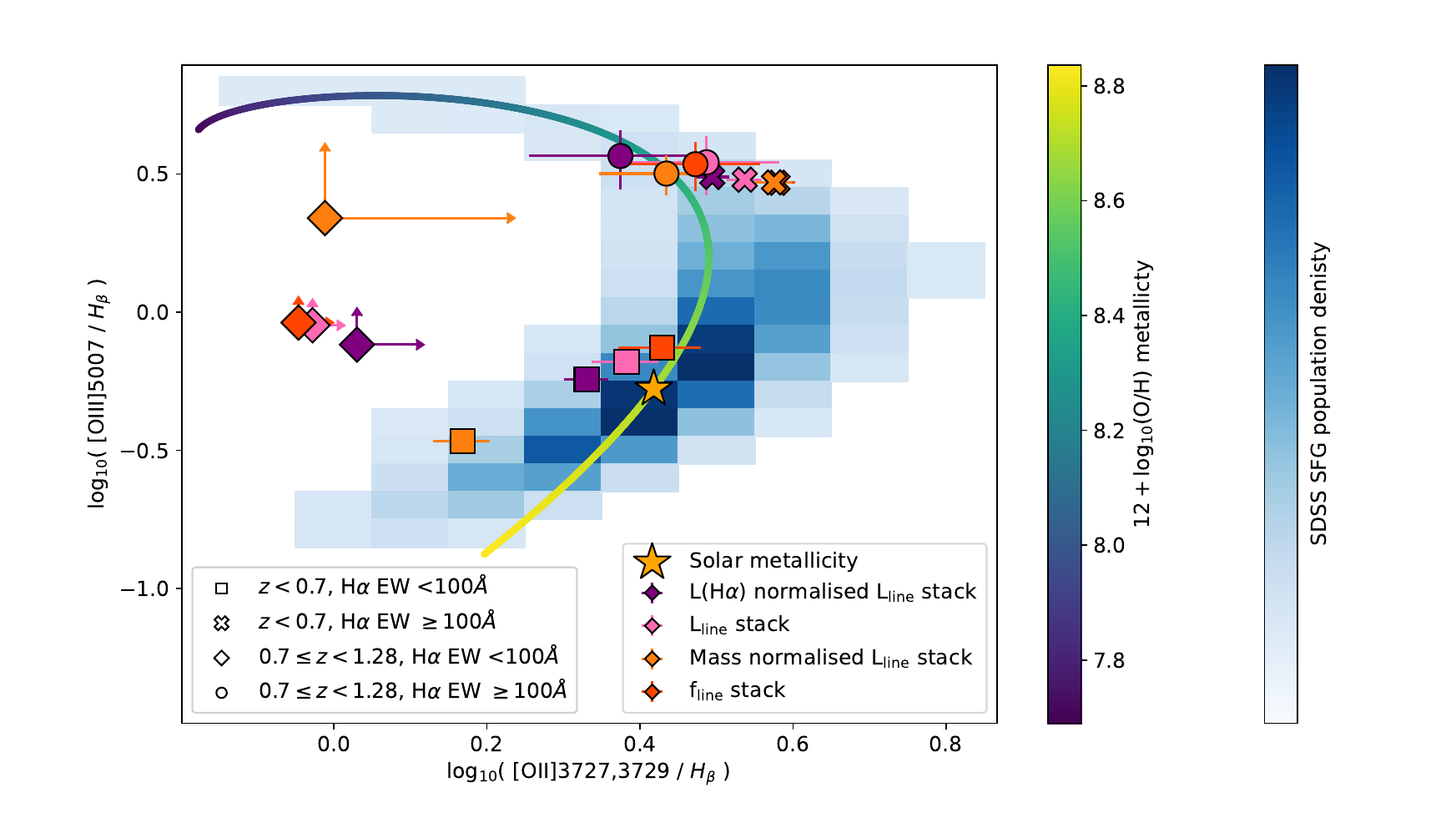}
    \caption{As in Figure \ref{fig:OIII/I/Hb}, the line diagnostic [O\textsc{ii}] vs [O\textsc{iii}] normalised by H$\beta$ plot for sub-samples of galaxies binned by redshift and H$\alpha$ rest-frame EW. Four different weighting schemes are plotted, as in the legend. Since the medium-$z$ low-EW sample of 8 objects had no H$\beta$ detections, even in the stack, we present the H$\beta$ $2\sigma$ limit.}
    \label{fig:OIII/I/Hb_stack}
\end{figure*}

\begin{figure*}
    \centering
    \includegraphics[width = 12cm]{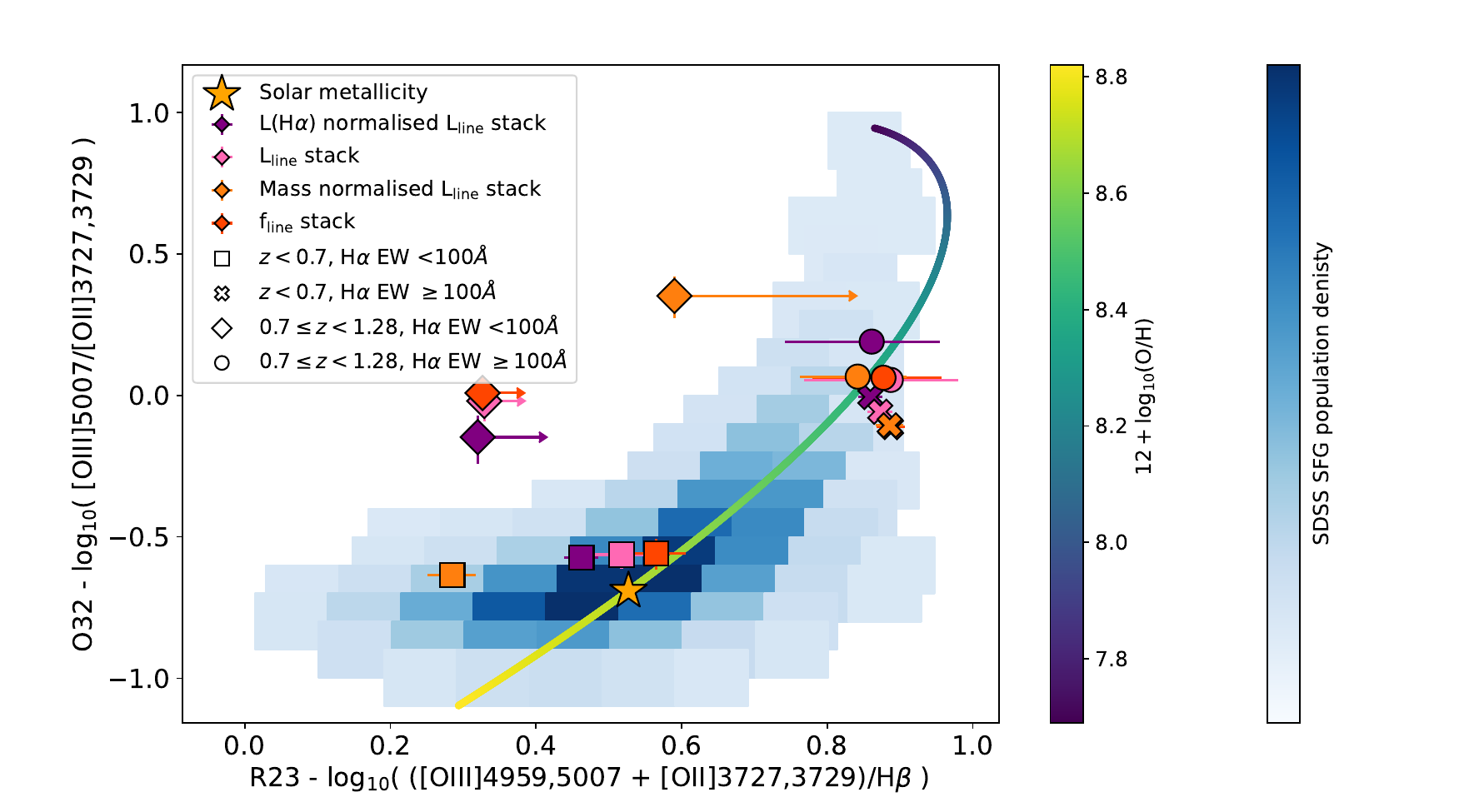}
    \caption{As in Figure \ref{fig:O32vsR23}, the line diagnostic O32 vs R23 plot for sub-samples of redshift and H$\alpha$ rest-frame EW (as in Figure \ref{fig:OIII/I/Hb_stack}).}
    \label{fig:O32vsR23_stack}
\end{figure*}

The diagnostics diagrams in Figures  \ref{fig:OIII/I/Hb} \& \ref{fig:O32vsR23} show measurements from individual galaxies where the [O\textsc{ii}], [O\textsc{iii}] and H$\beta$ emission lines are all detected at $S/N>2$. We note that we have removed one galaxy (FORS2 ID: \texttt{309\_2\_18}) which has a large spatial extent and where the FORS2 slit spectrum would not capture the full flux from the [O\textsc{ii}] emission line, and after removing this one object we have 23 galaxies with individual measurements, of which 18 also have H$\alpha$ flux measurements. We also consider galaxies where individual lines are undetected ($<2\sigma$) by combining the emission line fluxes from several galaxies together in a stacking analysis. We stack our galaxy spectra, so that sources with fainter emission lines are also considered. We generate several sub-samples and scaling schemes, and these are shown in Figures  \ref{fig:OIII/I/Hb_stack} \& \ref{fig:O32vsR23_stack}. First, we perform a straight average of the emission line fluxes (which would typically minimise the noise, but would give equal weight to low-luminosity galaxies at lower redshift and higher luminosity galaxies at greater redshift with the same observed line flux). Secondly, we perform a straight average of the emission line luminosity, so as not to penalise higher redshift sources. Next, we considered a stack where we normalise the line luminosities by the stellar mass before averaging.  Finally,
to ensure that we are not dominated by a small number of very bright sources, we normalise the individual line luminosities before averaging such that the integrated line luminosity from H$\alpha$ was the same (where the H$\alpha$ luminosity is a proxy for the star formation rate, and has been corrected for [N\textsc{ii}], reddening and stellar absorption, as described in Section~\ref{sec:balmer}).  

In this stacking, we only include galaxies with faint emission line measurements ($<2\sigma$) if an accurate redshift is available from the higher-spectral-resolution FORS2 spectrum,
which enables us to accurately locate the wavelength of the emission line in the extracted spectra. 
For each emission line of interest, we sum the line fluxes, luminosities or the luminosities after normalising by the stellar mass or H$\alpha$ luminosity, and take the ratios of the summed fluxes for different samples to plot on the diagnostic diagrams.
For inclusion in the H$\alpha$ luminosity-normalised stacked spectrum we require each galaxy to have H$\alpha$, H$\beta$, [O\textsc{ii}] and [O\textsc{iii}] coverage across our FORS2 or WFC3 spectroscopy, which results in a sample of 30 galaxies, 15 of which have $S/N<2$ in one or more of these emission lines (typically H$\beta$). The other 15 galaxies entering the sample for the stacking analysis are a sub-set of the 23 individual objects plotted in Figures \ref{fig:OIII/I/Hb} and \ref{fig:O32vsR23} with five galaxies excluded as H$\alpha$ was not covered by the WFSS at their redshift and three excluded as they were spatially extended and the [O\textsc{iii}], H$\beta$ \& [O\textsc{ii}] line fluxes determined from FORS2 will be subject to slit losses compared to the H$\alpha$ line measured from the  HST/WFC3 WFSS (see Section~\ref{sec:balmer}).
Our sample covers a broad redshift range and
to study any evolutionary trend we consider sub-samples of objects binned by redshift (low $z<0.7$ and medium $0.7<z<1.28$). We note that there are only two galaxies in the sample at $z>1.28$ that meet our criteria for the stacked sub-samples. 
We additionally split our stacking sub-samples by H$\alpha$ rest-frame equivalent width, which is sensitive to the presence of bursty star formation and young stellar populations. We wish to test if the ISM conditions depend on the H$\alpha$ rest-frame equivalent width, $EW_0$(H$\alpha$), and we create sub-samples\footnotemark\ of galaxies with $EW_0$(H$\alpha$)$< 100$\AA\ and $EW_0$(H$\alpha$)$> 100$\AA\ (corrected for [N\textsc{ii}] and stellar absorption). We note that the high-equivalent-width galaxies identified in the WISP WFSS may have been missed in traditional broad-band selected spectroscopic galaxy surveys, which we will discuss further in Section \ref{sec:HaEW}.  
\footnotetext{There are 7 galaxies in the low-redshift low-H$\alpha$ EW stack, 2 galaxies in the low-redshift high-H$\alpha$ EW stack, 8 galaxies in the medium-redshift low-H$\alpha$ EW stack, and 11 galaxies in the medium-redshift high-H$\alpha$ EW stack }

The medium redshift - low H$\alpha$ equivalent width sub-sample of 8 galaxies exhibited no H$\beta$ detections, even in the stack, so we plot instead $2\sigma$ limits for H$\beta$.
The stacks of these sub-samples are shown in Figures \ref{fig:OIII/I/Hb_stack} and \ref{fig:O32vsR23_stack}, with the results of \citet{Curti:2017} at low redshift shown for comparison. The stacks of our galaxies in each of our various sub-samples all lie on or near the locus of SDSS $z\sim0$ galaxies. We note that each of our sub-samples is consistent with being sub-solar metallicity.  

In recent years concern has been raised over the contribution of line emission from the Diffuse Ionised Gas (DIG) affecting the metallicity inferred from the strong-line diagnostic ratios \citep[e.g.,][]{Sanders17, ValeAsari19}. In local galaxies the DIG can be responsible for 30-60\% of the H$\alpha$ line emission \citep[e.g.,][]{Zurita00}. \citet{Sanders17} present a correction for DIG contamination involving the [N\textsc{ii}] emission line. Since our observations do not recover the [N\textsc{ii}] line emission flux, we do not apply this DIG correction, but we note that at the high redshifts of our WISP galaxies ($z\sim1$) the contamination from the DIG drops to $<$20\% \citep{Sanders17}. 

\subsubsection{Metallicity trend with redshift}

As apparent from Figures \ref{fig:OIII/I/Hb}-\ref{fig:O32vsR23_stack}, we find a trend that metallicity, as measured from the \citet{Curti:2017} calibration, decreases with increasing redshift. This redshift evolution reflects changes in the conditions of HII regions, although a changing contribution of the DIG at different redshifts may also contribute to apparent evolution of the diagnostic ratios. 
This trend with redshift is supported by previous surveys: our low redshift sample lies consistently within the R23 and O32 region occupied by the $0.5<z<1$ sample of \citet{Lilly03} and our medium and high redshift galaxies are consistent with the higher redshift $z\sim 2-3$ sample from \citet{Cullen14}, which exhibit a higher ionisation parameter (higher O32 and R23 values) indicative of a harder ionising radiation field than local galaxies. This supports the conclusions of \citet{Cullen14}, \citet{Hainline09} and \citet{Nakajima13} whose higher-redshift samples ($z\sim2$) are systematically offset to higher O32 than low-redshift SDSS galaxies. \citet{Strom_2017} and \citet{Maiolino_2008} at even higher redshift ($z\sim3$) show offsets to still larger values of R23 and O32. These results at higher redshift are in broad agreement with our highest redshift WISP galaxies. The rise of [O\textsc{iii}]/H$\beta$ with redshift is often attributed to a rise in the ionisation parameter, which could be driven by the changing metallicity, age and geometry of the stellar population in HII regions with redshift \citep[e.g., ][]{Kewley15, Jaskot19}. 

\subsubsection{Metallicity trends with H$\alpha$ Equivalent Width}
We find that galaxies exhibiting a high rest-frame equivalent width of H$\alpha$ ($>100$\AA) typically have higher O32 values (and hence lower metallicities on the \citealt{Curti:2017} calibration) than the low-EW galaxies at similar redshifts, which agrees with previous studies who find a positive trend between the equivalent width of Balmer lines and the O32 line ratio (e.g., \citealt{Kewley15} at $0.2 <z < 0.6$ and \citealt{Reddy18} at $1.6 < z < 2.5$). As can be seen in Figure \ref{fig:OIII/I/Hb_stack}, there is a dramatic difference in the [O\textsc{iii}]/H$\beta$ line flux ratio (and similarly in both the O32 and R23 ratios in Figure \ref{fig:O32vsR23_stack}) going from low-EW of H$\alpha$ to high-EW in the low-$z$ sub-samples, with the higher-ionisation [O\textsc{iii}] line becoming more dominant compared to [O\textsc{ii}] for the galaxies with high EW H$\alpha$. This trend is driven by a higher ionisation parameter at higher EWs, indicative of a young stellar population and high sSFR due to the relative abundance of young ionising stars to older non-ionising stars \citep{Mingozzi20, Kaasinen18, Kewley15} 

\subsubsection{Metallicity trends with stellar mass}
\label{sec:mass_metal}
The metallicity of star forming galaxies is typically found to increase with their stellar mass. This trend, the \lq\lq mass-metallicity relation", has been well studied observationally over a broad redshift range (e.g., \citealt{Tremonti04} at $z\sim0.1$;  \citealt{Yabe14} at $z\sim1.4$; \citealt{Revalski24} at $z\sim1-2$; \citealt{henry13} at $z\sim2$; \citealt{Sanders18} at $z\sim2.3$; \citealt{Maiolino_2008} at $z\sim3.5$; \citealt{Curti24} at $z\sim3-10$). 
The mass-metallicity relation can be motivated theoretically and may reflect different levels of star forming chemical enrichment and regulation driven by secular processes including gas inflows and outflows, such as in the bathtub model of \citet{lilly13}, with outflows more efficiently removing material at lower galaxy stellar masses (\citealt{Kobayashi07}, see also models by \citealt{Forbes14}). At a fixed mass, galaxies at higher redshifts are typically observed to exhibit lower metallicities (see for example the review by \citealt{Maiolino19}). Within this redshfit evolution the observed shape of the mass-metallicity relation remains roughly constant. These observational results have also been recovered in theoretical chemical evolution models \citep[e.g.,][]{Tinsley80} and in high resolution hydro-dynamic simulations which trace the evolution of chemical enrichment \citep[e.g.,][]{Taylor16, Torrey18, DeRossi17}. 

\begin{figure*}
    \centering
    \includegraphics[width=\textwidth]{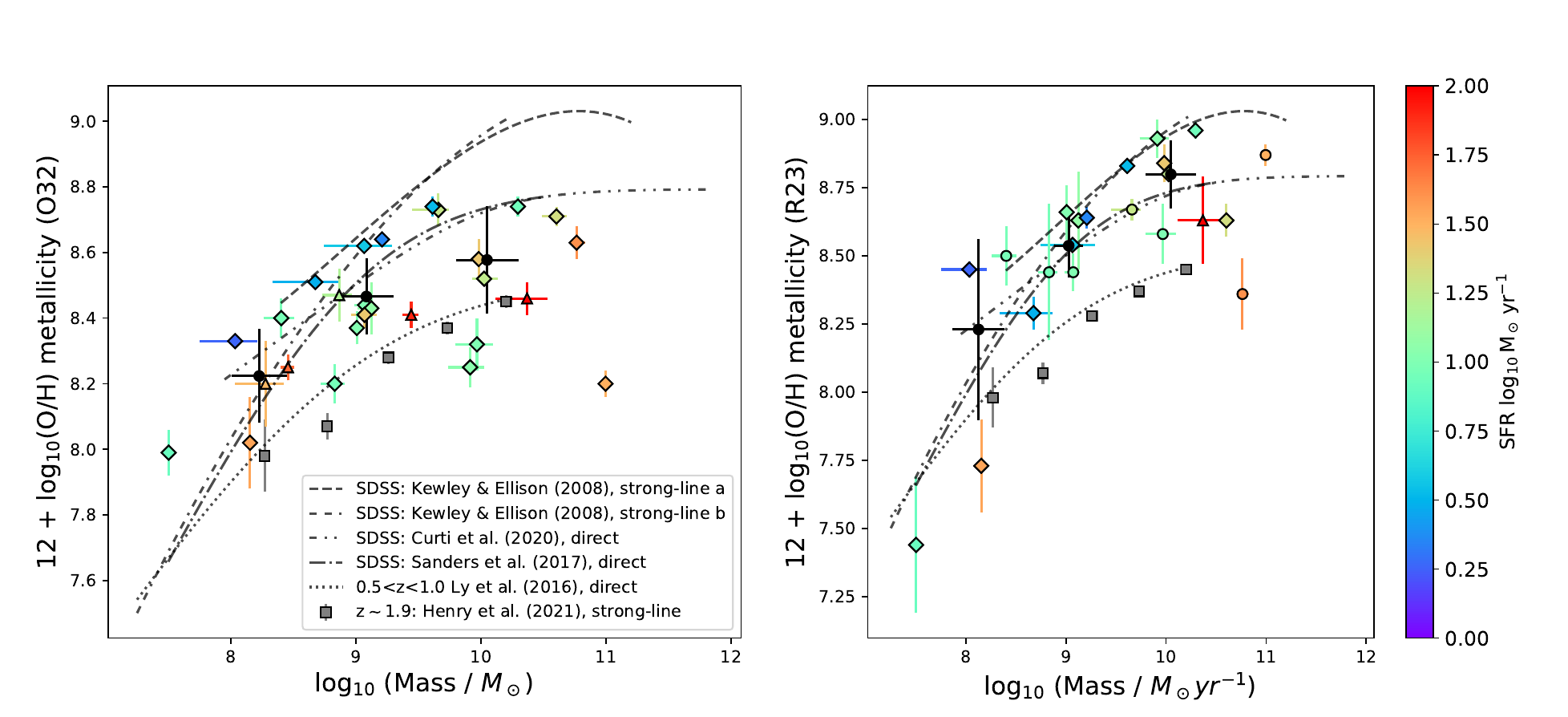}
    \caption{Mass metallicity relation for O32 (left panel) and R23 (right panel) derived metallicities, colour coded by star formation rate (H$\alpha$, unless unavailable then SFR-UV is used, these are shown as triangles). Circles are the locations of galaxies with undetected H$\beta$ where we instead infer the H$\beta$ flux from the measured H$\alpha$ flux, as in Figures \ref{fig:BPT} and \ref{fig:MEx}. Black circles show the data binned by mass for $\log_{10}(M_\star/M_\odot)$ = 7.5-8.5, 8.5-9.5 and  9.5-10.5.}
    \label{fig:mass_metal}
\end{figure*}

\begin{figure*}
    \centering
    \includegraphics[width=\textwidth]{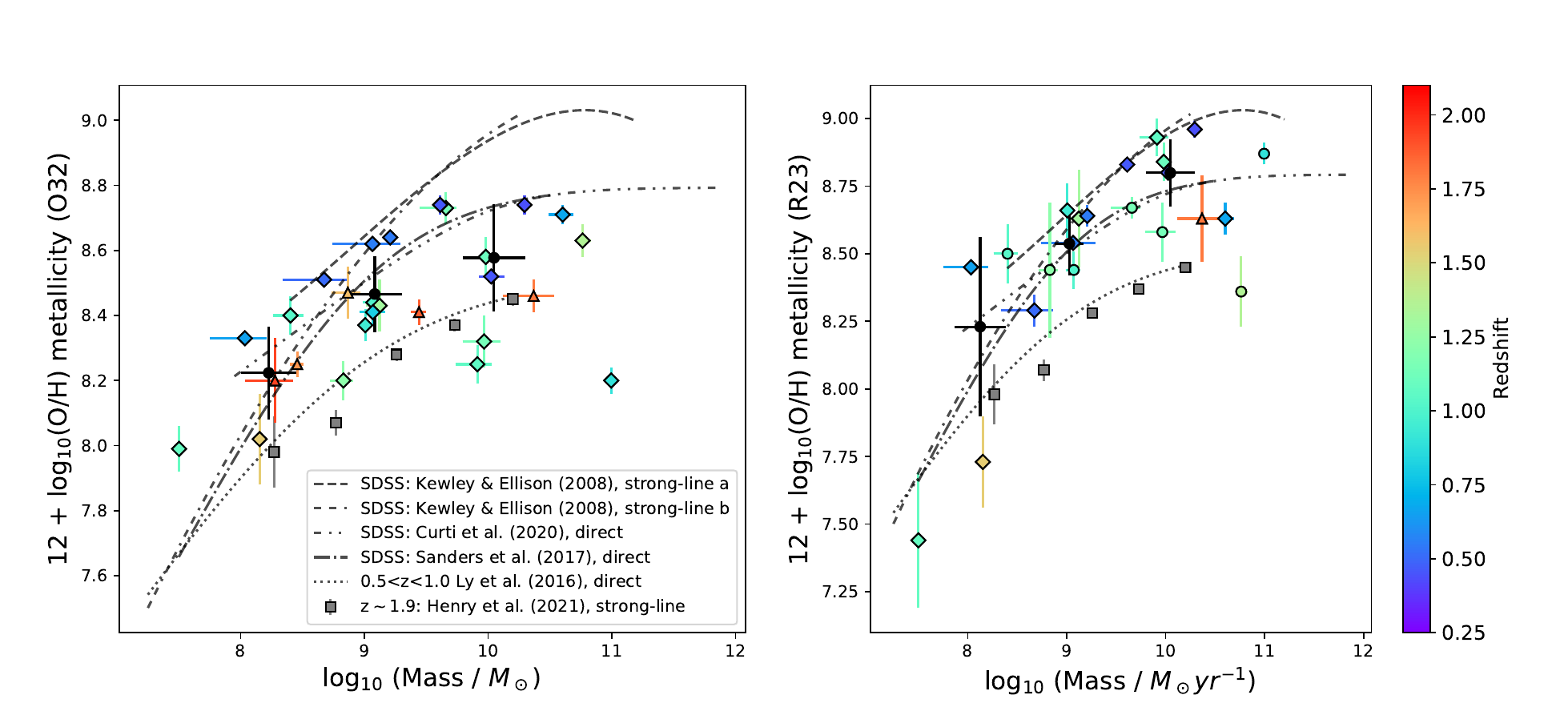}
    \caption{Mass metallicity relation for O32 (left panel) and R23 (right panel) derived metallicities, colour coded by redshift. Circles are the locations of galaxies with undetected H$\beta$ where we instead infer the H$\beta$ flux from the measured H$\alpha$ flux, as in Figures \ref{fig:BPT} and \ref{fig:MEx}. }
    \label{fig:mass_metal_z}
\end{figure*}

\begin{figure*}
    \centering
    \includegraphics[width=\textwidth]{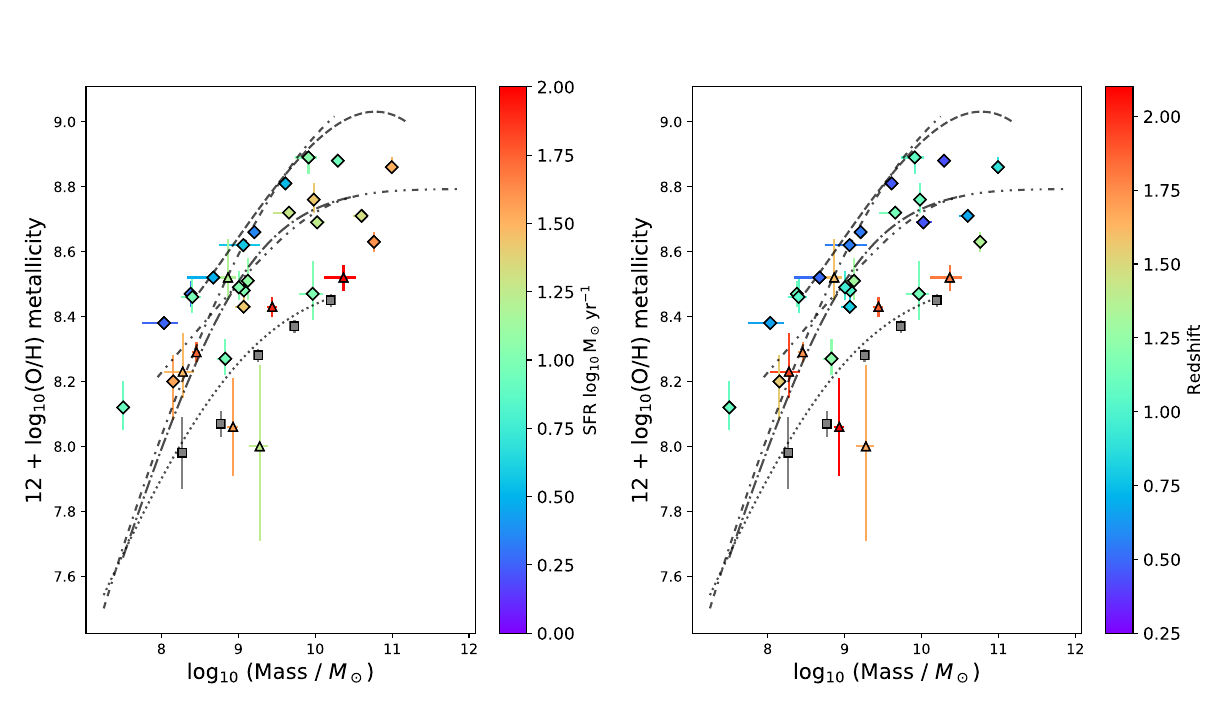}
    \caption{Mass metallicity relation for highest likelihood metallicity, colour coded by redshift (left) and star formation rate (right), as in Figures \ref{fig:mass_metal} and \ref{fig:mass_metal_z}.}
    \label{fig:henry_mzr}
\end{figure*}

Here we compare the derived stellar masses (see Section \ref{sec:stellar_mass}) to the metallicity of the galaxies within our sample. We present the mass-metallicity relation for the two gas-phase metallicities derived using the R23 and O32 line diagnostics, shown in Figure \ref{fig:mass_metal} colour coded by the derived star formation rate, and in Figure \ref{fig:mass_metal_z} colour coded by the redshift. In Figure \ref{fig:henry_mzr}, the highest likelihood metallicity MZR is presented with redshift and SFR colour coding. We over-plot three local observed relations derived from SDSS samples, first by \citet{Tremonti04} who derived metallicities using strong-line ratios, second by \citet{Curti20} who derived metallicities using direct temperature methods and third by \citet{Sanders17} who also use a direct temperature method and adjust their relation to correct for DIG. Our metallicities come from the O32 and R23 strong-line method which are calibrated against the direct temperature method by \citet{Curti:2017, Curti20}. Additionally we over-plot two higher redshift samples; \citet{ly16} at $0.5 < z < 1.0$ who use direct temperature methods to derive their metallicities, and the binned data of \cite{Henry21} at $z\sim1.9$ who use strong-line derived metallicities (and find consistency with \citealt{Sanders18} at $z\sim2.3$ who use direct temperatures methods).  

The mass-metallicity relation in our sample becomes very apparent when we bin our galaxies by stellar mass (Figures \ref{fig:mass_metal} and \ref{fig:mass_metal_z}), and individually we find that our data points lie typically in between the $z\sim0.1$ and $z\sim1.9$ mass-metallicity trends of \citet{Tremonti04} and the high-$z$ WISP sample of \cite{Henry21}. Within the scatter of our data we find a trend that higher redshift galaxies exhibit a lower metallicity at a given mass, as shown by the colour-coding of the individual data points in Figure \ref{fig:mass_metal_z}. This trend is found in both the O32 and R23 diagnostic derived metallicity values.   

Many studies have probed dependencies on the scatter within the mass-metallicty relation, which led to the development of the Fundamental Metallicity Relation (FMR, \citealt{Mannucci10}) where the scatter is thought to be driven by the SFR of the galaxies (see also, \citealt{Tremonti04}, \citealt{Lara-lapez10}, \citealt{Hunt12}). However,  observationally the FMR is not consistently detected within galaxy samples (e.g., \citealt{Sanchez17, Barrera-Ballesteros17}, c.f., \citealt{Cresci19}) and some work has claimed that the FMR correlations may arise from observational systematics, for example calibration uncertainties in nebular emission line metallicity diagnostics \citep{Telford16}, or slitloss effects from fibre/slitmask spectroscopic surveys compared to integral field spectrograph results \citep{Barrera-Ballesteros17}. In high resolution hydro-dynamic simulations \citep[e.g.,][]{DeRossi17, Torrey19} agreement with the empirical form of the FMR has been found. 
The anti-correlation between metallicity and SFR at a fixed stellar mass in the FMR may arise from the joint dependency of the SFR and metallicity on the gas fraction $M_{HI}/M_{\star}$ \citep[e.g.,][]{Bothwell13}.
A galaxy with a plentiful reservoir of pristine gas (a high gas fraction) may fuel a large SFR and has a corresponding low metallicity. After a period of star formation the SFR will decrease as the reservoir becomes depleted (a decreasing gas fraction) whereas chemical enrichment from the preceding star forming episode will increase the metallicity. The similarity of the time-scale of the SFR and metallicity variability drives the observable anti-correlation in the two galaxy properties \citep{Torrey18}.  

To investigate the possible existence of the FMR,
in Figure \ref{fig:mass_metal} we colour-code by the derived SFR (similar to \citealt{Maiolino19} figure 22), using our H$\alpha$-derived SFRs for galaxies with detected H$\alpha$ and the $L_{\rm{UV}}$-derived SFR for those at higher redshifts where H$\alpha$ was not available (which we have shown to be comparable methods in Figure \ref{fig:SFR_sFR}). In the mass-metallicity diagram based on the O32 diagnostic we see an obvious trend in the mass-metallicity relation, in that low-SFR galaxies have higher metallicities than high-SFR galaxies of the same stellar mass. The trend is less obvious when we use R23-based metallicities, because our H$\beta$ detections (which go into the R23 diagnostic) are often low S/N. 

\begin{figure*}
    \centering
    \includegraphics[width=\textwidth]{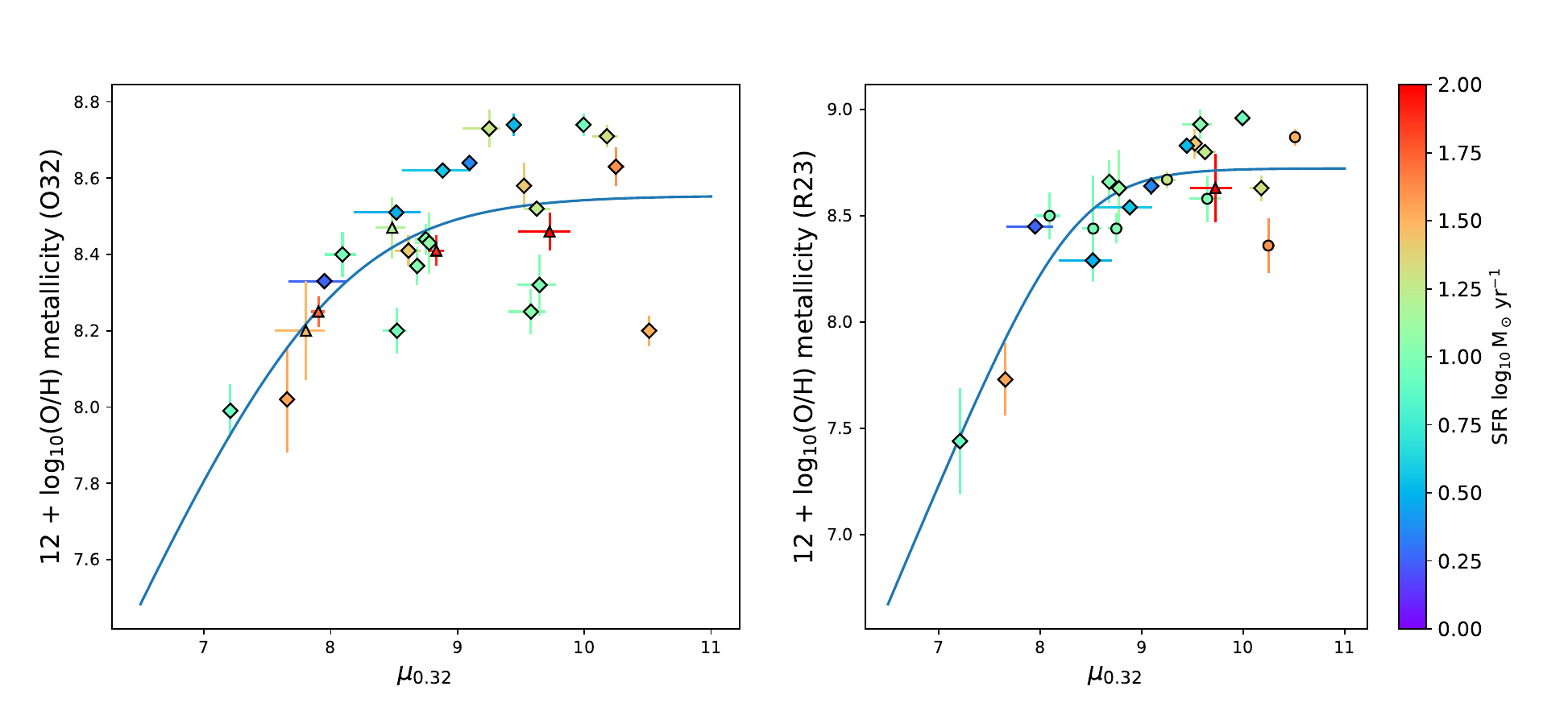}
    \caption{Modified mass-metallicity diagram, with $\mu$=log$_{10}$(M)-$\alpha$log$_{10}$(SFR) and $\alpha=0.32$ from \citet{Mannucci10}. Left panel: metallicity derived from O32. Right panel: metallicity derived from R23. In both panels the data are colour coded by the SFR, as in Figure \ref{fig:mass_metal_z}. The best fit parameterisation is given by the blue curve.}
    \label{fig:modified_mass_metal}
\end{figure*}

The FMR defines a surface that galaxies occupy in the 3D space of stellar mass, metallicity and SFR. Hence, the scatter in the mass-metallicity relation can be reduced by scaling the mass as a function of the star formation rate. We adopt the \citet{Mannucci10} modified mass-metallicity scaling parameter $\mu = \log_{10}(M) - \alpha\,\log_{10}(SFR)$ where $\alpha = 0.32$ to inspect whether our scatter decreases.
We determine the scatter by considering the RMS of the residuals after fitting our data with a standard mass-metal relation parameterisation \citep[e.g., ][]{Sanders17} 
\begin{equation}
12+\log_{10}\left ( O/H \right ) = 12+\log_{10}\left ( O/H \right )_{asym} - \log_{10}\left (1 + \left (\frac{M_{TO}}{M_\star} \right)^\gamma \right)
\end{equation}
where each term defines the asymptotic gas-phase metallicity, the turn-over mass ($M_{TO}$), and a power slope ($\gamma$). Plotting metallicity against $\mu_{0.32}$ in Figure \ref{fig:modified_mass_metal} shows a slightly tighter relation than seen when plotted against mass, with an RMS scatter of 0.146 compared to 0.157 when plotted against mass for the O32 metallicity (and 0.172 and 0.199 for $\mu_{0.32}$ and mass using R23). This effect is subtle but we also note that we find that $\alpha=0.32$ tightens the scatter to a greater extent than the best fit $\alpha=0.17$ measured by \citet{Henry21}. The best fit asymptotic metallicity, turn over mass, and $\gamma$ are given in Table \ref{tab:best_fit}.  

We have found a mass metallicity relation using the metallicities derived from the O32 and R23 strong line diagnostics. We also find that the scatter in these reduced when we consider the SFR as a third dependent, as part of the FMR. However, our best fit models for the O32 and R23-derived work are not identical, this disparity has also been found by \citet{Topping21} who find that the evolution of the MZR depends on the strong line diagnostic used to determine the metallicity. 

\begin{table}
\begin{tabular}{llllll}
Diagnostic & $asym^{\rm{a}}$ & $M_{TO}^{\rm{b}}$ & $\gamma$    \\ \hline
Mass-metallicity (O32)        & 8.55$\pm$0.07        & 8.14$\pm0.31$     & 0.67$\pm0.30$      \\
Mass-metallicity (R23)    & 8.74$\pm0.08$        & 8.65$\pm0.28$     & 0.95$\pm0.26$      \\
$\mu_{0.32}$-metallicity (O32)        & 8.55$\pm0.06$        & 7.84$\pm0.27$     & 0.74$\pm0.30$      \\
$\mu_{0.32}$-metallicity (R23)       & 8.72$\pm0.06$        & 8.26$\pm0.21$     & 1.13$\pm0.26$    
\end{tabular}
\caption{The best fit parameters ($^a$ asymptotic metallicity, $^b$ turnover $\log_{10}$ mass) for the mass-metallicity and modified $\mu_{0.32}$-metallicity diagnostics.}
\label{tab:best_fit}
\end{table}

\subsection{High equivalent width H$\alpha$ sample}
\label{sec:HaEW}
The WFSS from the WISP survey selects galaxies on the basis of their line emission. This is in contrast to the usual broad-band magnitude-limited surveys, which select on the continuum emission. Hence WISP may miss galaxies which have very low SFR (and hence weak line emission) but we are more sensitive to galaxies with high equivalent width emission lines and weak stellar continuum - these low-mass but actively star-forming systems would typically not enter a broad-band selection. To examine this we consider deep spectroscopic surveys targetting $z\sim1$ galaxies. These typically extend to optical magnitudes of $AB=22.5$\,mag over wide fields and $AB=24$\,mag over smaller regions, for example: VVDS-wide with $I_{AB}<22.5$ \citep{Garilli08}, VVDS-deep with $I_{AB}<24.0$ \citep{fevre05}; DEEP2 with $R_{AB}<24.1$ \citep{Newman13}; zCOSMOS-wide $I_{AB}<22.5$, zCOSMOS-deep with $I_{AB}<24.0$ \citep{Lilly07}; VUDS $I_{AB}<25.0$ \citep{fevre15}. Many of our WISP sources with reliable detections of emission lines have $H_{AB}\sim24-25$ (see Figure \ref{fig:Hmag_dist}). A typical star forming galaxy with moderate extinction $E(B-V)=1$ at $z\approx 1$ has a colour of $(I-H)_{AB}\approx 1$\,mag (e.g., figure 1 of \citealt{Doherty05}, where $(I-H)_{AB}\approx (I-H)_{\rm{Vega}}-0.9$). About half (51$\%$) of the emission line galaxies we detect in WISP, and spectroscopically confirm in this paper, are fainter than $H_{AB}=24$\,mag ($I_{AB}\sim25$) and will have escaped previous spectroscopic surveys, and 30$\%$ of our WISP galaxies are fainter than $H_{AB} > 25$\,mag. 

\begin{figure}
    \centering
    \includegraphics[width=\columnwidth]{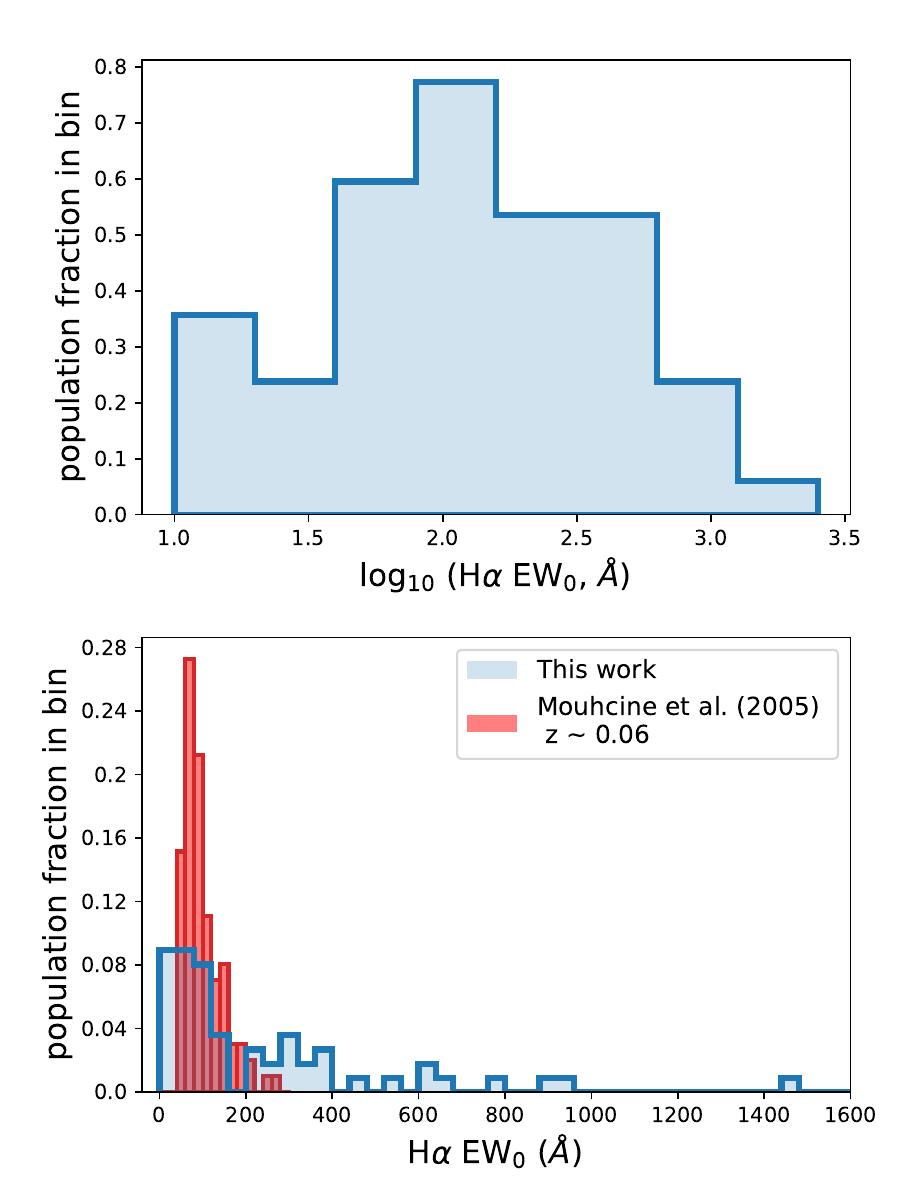}
    \caption{Upper panel: Logarithmic H$\alpha$ rest-frame equivalent width distribution of the WISP selected galaxies with follow-up FORS2 observations. H$\alpha$ rest-frame equivalent widths are corrected for seeing, [N\textsc{ii}] contribution and stellar absorption. We remove two AGN candidates (see Section \ref{sec:AGN}). Lower panel: we compare our EW distribution (blue histogram) with that determined from the broad-band selected survey of \citet{mouchine05} in red, and note the tail of high-EW galaxies which is missed in the broad-band selection. The \citealt{mouchine05} sample has selected on $EW_0(H\beta)>0$\,\AA, which results in the sharp decline in the red histogram at $EW_0(H\alpha)<40$\AA.}
    \label{fig:Ha_EW}
\end{figure}

We determine the H$\alpha$ equivalent width of our sources through the ratio of the H$\alpha$ emission line flux, corrected for [N\textsc{ii}] contribution and stellar absorption (see Sections \ref{sec:NIIcorr} and \ref{sec:balmer}), and the continuum flux density determined from the broadband magnitude of the photometric filter the line fell in. 
We remove the contribution of the emission line to the flux in the broad-band filter when calculating the stellar continuum, using the transmission profile of the filter and assuming a spectral slope of $f_{\lambda}\propto \lambda^{\beta}$ with $\beta = -2$ (a slope flat in $f_{\nu}$ appropriate for ongoing star formation in the absence of significant reddening, e.g., \citealt{Wilkins11}).
We convert all the observed-frame equivalent widths to the rest-frame values ($EW_0$) using $EW_0 = EW_{\rm obs} / (1+z)$.
For 11 galaxies we can not provide a H$\alpha$ EW since we don't have H$\alpha$ detections (due to the wavelength coverage at higher redshifts) and for a further 16 we do not have the required HST photometry used to determine the stellar continuum in the broadband filter containing the H$\alpha$ emission, and we are unable to determine an EW. We additionally remove the two potential AGN\footnote{309\_1\_2 and 309\_1\_6, see section \ref{sec:AGN}.} from this analysis.
The distribution of the H$\alpha$ rest-frame equivalent widths is shown in the top panel of Figure \ref{fig:Ha_EW}, and we additionally compare our distribution with the H$\alpha$ rest-frame equivalent widths distribution from the \citet{mouchine05} broad-band-selected survey (bottom panel). We note that we have a tail of high-$EW_0$ galaxies picked up in the WFSS that is missing from the broad-band selection. The high-$EW_0$ tail will also be enhanced compared to the $z\sim0.06$ \citet{mouchine05} sample due to the evolution of H$\alpha$ EW with redshift.  

\begin{figure*}
    \centering
    \includegraphics[width = \textwidth]{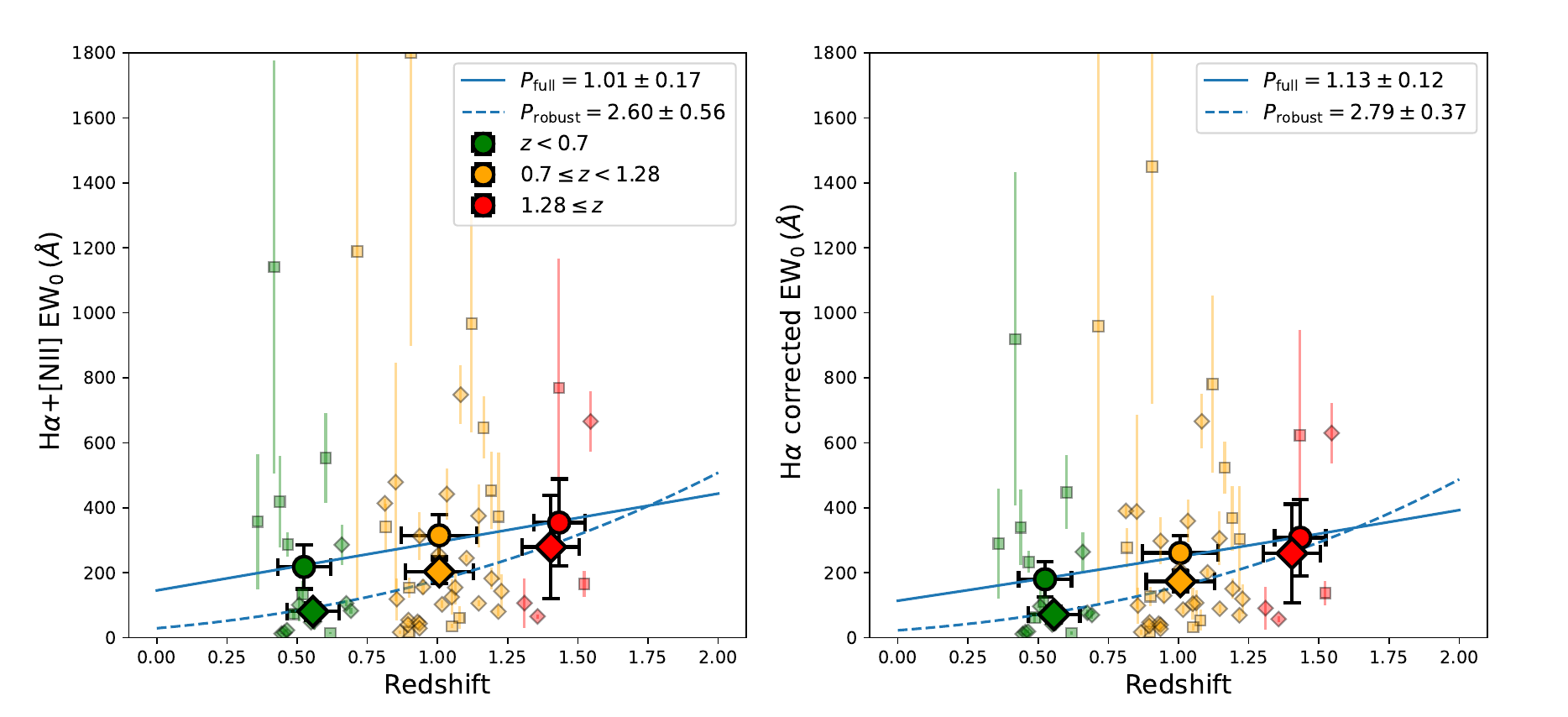}
    \caption{H$\alpha$ rest-frame equivalent widths against for the low, medium and high redshift samples. We present  H$\alpha$ EWs with and without [N\textsc{ii}] and stellar absorption correction in the right and left panel, respectively. Galaxies with FORS2 emission line detections are presented as small Diamonds, and those without as small Squares. We determine the mean EW for each redshift bin, where we consider our whole sample (large Circles, black outline) and 
    only those with robust redshifts from FORS2 detections or having multiple emission lines (large Diamonds, black outline).
    We overlay the best fit power laws (blue) to the mean EW of each redshift bin with the form $EW_0 \propto (1+z)^P$.
    }
    \label{fig:EW_z}
\end{figure*}

While our WISP survey is not a stellar-mass-selected (or even magnitude-limited) sample, by design it samples star forming galaxies over a range of redshifts. As can be seen in figure 5 of \citet{Atek10}, the 5-$\sigma$ emission line sensitivity of WISP is $\approx 4\times 10^{-17}$\,erg s$^{-1}$ cm$^{-2}$ at the middle wavelength of both the G102 and G141 gratings, the sensitivity declines rapidly at shorter wavelengths, meaning that the line luminosity limit to H$\alpha$ is $\approx 10^{41}$\,erg s$^{-1}$ over a fairly wide range of redshifts, so the WISP sample selects star-forming galaxies with SFRs exceeding $\sim 1-3\,M_{\odot}$\,yr$^{-1}$ (depending on reddening corrections and SFR conversion factors, see also Figure 17 in \citealt{Battisti24}). Hence we can consider the evolution in properties of star forming galaxies across cosmic time from our sample. 
We examine the distribution of rest-frame H$\alpha$ equivalent widths across our three redshift bins and present the EWs binned by redshift in Figure \ref{fig:EW_z}. 
The median (and mean) rest-frame H$\alpha$ equivalent widths for the low ($z < 0.7$), medium ($0.7 \leq z < 1.28$) and high ($z \geq 1.28$) redshift bins are 
77\,\AA\ (mean $180 \pm 55$\,\AA ), 
129\,\AA\ (mean $261 \pm 52$\,\AA ) and 
138\,\AA\ (mean $308 \pm 117$\,\AA ) 
respectively, from bins of 17, 34 and 5 galaxies
with H$\alpha$ observations, where we quote the standard error on the mean. 

If we only consider the sub-set of galaxies with robust redshifts, either having FORS2 emission line detections or having multiple emission lines from WISPS (shown in Diamonds in \ref{fig:EW_z}), we find a lower average EW in the low and medium redshift bin. 
This sub-sample also exhibits a steeper redshift evolution compared to our full sample.  As discussed in Section \ref{sec:non-detect}, a sub-set of the galaxies in our sample without FORS2 detections despite the wavelength coverage had an expected S/N that would have been sufficient for detection, suggesting therefore that their WISPS single-line detections may have been spurious in nature. Some of these likely spurious lines were attributed to very faint galaxies, and hence high EWs were attributed.  Eliminating the single-line measurements from WISPS (without corroborating FORS\,2 spectroscopic confirmations) leads to ha decrease in the average EW when considering only the sub-set of galaxies with robust redshifts.   
We parameterise the growth of rest-frame EW with redshift as 
\begin{equation}
EW_0 \propto (1+z)^P
\label{eq:Ha_EW}
\end{equation}
where we determine the best-fit value of $P$ to be $1.13\pm0.12$
for our full sample and $2.79\pm0.37$
for our robust sub-sample.
 
Our data shows a trend of increasing H$\alpha$ rest-frame EW with redshift\footnote{We note that our sensitivity to lines {\em vs.\ } continuum (i.e.\ EW) is boosted at high redshift due to a fixed observed-frame EW limit dropping in rest-frame EW$_0$ by $EW_{obs}/(1+z)$, but most of our sources are well above the EW threshold for line identification in WFSS, and the main selection effect is the H$\alpha$ S/N.}, in line with the picture of increasing sSFR (e.g. \citealt{Marmol16}). Similar results on the redshift evolution of H$\alpha$ equivalent width are obtained by the 3DHST survey \citep{Fumagalli12} which also uses WFSS with HST/WFC\,3 and determine the evolution of rest-frame H$\alpha$ equivalent width\footnotemark of $P=1.79\pm0.18$ for H$\alpha$ detected sources at $10<\log_{10}(M_{\star}/M_\odot) < 10.5$ using the same parameterization as Equation \ref{eq:Ha_EW}, which lies between our determination of $P=1.13$
for our full sample, and $P=2.79$
when we restrict to galaxies with robust redshifts. \footnotetext{In the HST Grism the H$\alpha$ emission line is blended with the [N\textsc{ii}] doublet and \citet{Fumagalli12} do not correct for the contribution from [N\textsc{ii}]. The value of the power law evolution will remain comparable to our results unless there is significant evolution in the [N\textsc{ii}]/H$\alpha$ ratio. As can be in Figure \ref{fig:EW_z} where we determine consistent power laws whether we correct for [N\textsc{ii}] contribution or not.}
However, we note that our results are not directly comparable to \citet{Fumagalli12}, who select a sample based on multi-colour imaging in the well-studied CANDELS fields and hence also include galaxies with undetected line emission (i.e. low SFR, below the sensitivity of WFC\,3 spectroscopy). Such galaxies are not included in our emission-line-selected sample, where we prioritise spectroscopic follow-up of high equivalent width sources, and hence the average SFR in our sample are higher than the average of the whole sample in \citet{Fumagalli12}, but are comparable to the sub-set of star forming galaxies (SFGs) shown in their figure~2 (but we note that they plot the combined equivalent width of H$\alpha$+[N\textsc{ii}]). 

We note that our strong evolution of the H$\alpha$ EW  echoes a similarly strong redshift dependence on the fraction of galaxies which are Extreme Emission Line Galaxies (EELG), with very high equivalent widths of the [OIII]$\lambda$5007 and/or H$\alpha$ emission lines. \cite{Boyett22A} report the fraction of galaxies exceeding $\sim 750$\,\AA\ evolves as $(1+z)^P$ with $P=2.5\pm0.3$, using a sample at $1.7<z<2.3$ combined with literature results at $z>5$. More recently, \cite{Boyett24arXiv} have used the {\em JWST} Deep Extragalactic Survey over a wide range of redshift to determine $P=2.6\pm0.1$ for the evolution of the EELG fraction. While the EELG fraction determined in these papers is not exactly the same measure as the average equivalent width of the star forming population (reported in this paper), they both trace the change in the distribution of equivalent widths with redshift and show strong evolution.

We can use our spectroscopic follow-up to analyse any differences with H$\alpha$ equivalent width in the physical properties of galaxies. In Section \ref{sec:metallicity_diganostics} we split the population approximately in half by H$\alpha$ equivalent width, and showed that galaxies in our sample with $EW_0>100$\AA\ had higher [O\textsc{iii}]/H$\beta$, indicating lower metallicity and/or higher ionisation parameters than galaxies with $EW_0<100$\AA\ (Figures \ref{fig:OIII/I/Hb_stack} $\&$ \ref{fig:O32vsR23_stack}). This supports our contention that the population of galaxies captured by broadband photometric selection criteria do not represent the full variety of star forming galaxy properties, and may miss low stellar mass galaxies with high specific star formation rates. 

\section{Conclusions}

In this paper we have followed up galaxies identified through line emission in the near-infrared in the WFSS HST/WFC3 WISP survey, using VLT/FORS2 optical spectroscopy to confirm the redshifts and to study the reddening and metallicity of this emission line population. Half of the WISP galaxies in our sample are fainter than $H_{AB}=24$\,mag, and would not have been included in many well known spectroscopic surveys based on broad-band magnitude selection. Over 4 WISP survey fields, we targetted 85 out of 138 line emission objects identified in the WFC3 WFSS spectra over a redshift range $0.4<z<2$. We confirm 95\% of the initial WFC3 grism redshifts in the 38 cases where we detect lines in the FORS2 spectra. Spectroscopic confirmation from FORS2 is important because in many cases the WFC3 spectroscopy detected only a single emission line, usually assumed to be H$\alpha$. 
For the 43 single-line WFC3 sources targetted with FORS2 at $z<1.28$ (where we expted line emission to fall in the FORS2 wavelength coverage), we confirm the initial WISP-based redshifts in 15 out of 17 galaxies where emission lines were detected in FORS2. However, there were a further 26 single-line WFC3 sources where we expected to see line emission in FORS2 which were not detected, most likely indicating that some of the single-line low-S/N WFC3 sources were spurious. From this, we determine a maximum potential false-detection rate of WFC3 WFSS single emission line galaxies of 65$\%$ (28/43 at $z<1.28$) 
which improves to 33\% (1/3) using the latest public v6.2 WISP emission line catalogue. 

We measure the Balmer decrement (the H$\alpha$/H$\beta$ flux ratio, after correcting for stellar absorption and for [N\textsc{ii}] emission blended with H$\alpha$ at the resolution of WFC3 grism), and find that the extinction of the WISP galaxies is consistent with $A($H$\alpha)=1$\,mag with some evidence for an increase in reddening with H$\alpha$ luminosity, and less reddening than seen in $z\sim 0$ samples. After correction for reddening, we find good agreement between star formation rates derived from the H$\alpha$ emission line and those from the rest-frame UV continuum ($L_{UV}$) with a near-linear relation of $SFR(L_{UV}) = [SFR(H\alpha)]^{1.15\pm0.10}$. We obtain comparable star formation rates to those from the dust-corrected rest-UV from the SED-fits of the stellar population with BEAGLE on multi-band photometry (including Spitzer $3.6\,\mu$m). We find that stellar masses derived from this fitting have a median mass of $\log_{10}(M_{\star}/M_\odot)=8.94$, and our emission-line-selected galaxies tend to lie above the star-forming main sequence when plotting SFR against stellar mass (i.e.\ they have higher sSFR than the comparison samples at similar redshifts). 

We use emission line ratios to identify two likely AGN ($\sim$5\% of the sample) from the [S\textsc{ii}]-BPT diagram (modified to use [S\textsc{ii}] rather than [N\textsc{ii}]) and the mass-excitation (MEx) diagnostic. For the star-forming galaxies, we use the [O\textsc{iii}], [O\textsc{ii}] and H$\beta$ lines to derive gas-phase metallicities using the strong-line O32 and R23 diagnostics and the calibrations from \citet{Curti:2017}. Individual galaxies detected in these lines lie within the region of the diagrams occupied by the low-$z$ SDSS galaxies, but our galaxies typically have sub-solar metallicity, with metallicity decreasing with redshift. 

Our WISP galaxies have a mass--metallicity relation, and lie below the $z=0$ relation (i.e.\ the WISP galaxies have lower metallicity) and our results are consistent with the evolution seen in other high redshift studies. We see a trend that for galaxies of the same mass, those with higher SFR have lower metallicity, consistent with the existence of a Fundamental Metallicity Relation between SFR, stellar mass and metallicity. 

Finally, we consider the evolution of the H$\alpha$ rest-frame equivalent width of galaxies in our sample, and find a strong dependence on redshift of $EW_0(H\alpha)\propto (1+z)^{1.13\pm0.12}$ in our full sample (with a steeper power law of $P=2.79\pm0.37$ in our robust redshift sub-sample). When we split our sample by H$\alpha$ equivalent width at $EW_0(H\alpha)=100$\,\AA , we find that higher EW galaxies have a larger [O\textsc{iii}]/H$\beta$ and O32 ratio on average, suggesting a lower metallicity or higher ionisation parameter in these extreme emission line galaxies.

Looking forward, we emphasise the importance of securing multiple line detections in future WFSS to obtain robust redshifts and avoid single emission line ambiguity. 
The next generation of WFSS surveys on {\em JWST} and Euclid will provide rich samples of star forming galaxies below traditional photometric broadband thresholds, and to higher redshifts than possible with HST/WFC3 WFSS. Towards higher redshifts, where galaxies typically show higher sSFR, WFSS is even better positioned to capture more complete samples down to fainter magnitudes, crucial for the study of galaxy evolution and the most extreme emission line galaxies.
We note that while high purity (low false-detection rates) can be achieved in these future WFSS surveys with high S/N thresholds, it is important that follow up spectroscopy at complimentary wavelengths is still utilised to confirm single-emission line galaxies below these S/N thresholds that would otherwise be discarded from such pure samples. As shown in this paper, multiple emission lines from such follow-up spectroscopy can help characterise the physical conditions of these galaxies.

\section*{Acknowledgements}
We would like to thank Prof. Michele Cappellari for detailed discussion of PPXF, and Prof. Chiaki Kobayashi for useful discussion on the MZR.  
% K. Boyett & A. Battisti
K.~B., A.~B. acknowledge support from the Australian Research Council Centre of Excellence for All Sky Astrophysics in 3 Dimensions (ASTRO 3D), through project number CE170100013. 
% A. J. Bunker & J. Chevallard
A.~J.~B., J.~C, K.~B. acknowledge funding from the \lq\lq First Galaxies" Advanced Grant from the European Research Council (ERC) under the European Union's Horizon 2020 research and innovation programme (Grant agreement No. 789056).
% A. L. Henry - None
% S. Wilkins
S.~W. thanks STFC for support through ST/X001040/1.
% M. A. Malkan - None
% J. Caruana
%J. C. acknowledges support from the University of Malta Research Support
%Services Directorate (PHYRP19-20 and PHYRP19-21). J
% H. Atek
H.~A. acknowledges support from CNES, focused on the JWST mission, and the Programme National Cosmology and Galaxies (PNCG) of CNRS/INSU with INP and IN2P3, co-funded by CEA and CNES.
% I. Baronchelli - None
% J. Colbert - None
% J. P. Gardner - None
% M. Rafelski - None
% C. Scarlata - None
% H.I. Teplitz - None
% X. Wang
X.~W. is supported by the Fundamental Research Funds for the Central Universities, and the CAS Project for Young Scientists in Basic Research, Grant No. YSBR-062.

This work is based on data obtained from the Hubble Space Telescope, as part of G0 14178, 13517, 13352, 12902, 12568, 12283 and 11696. Support for these programs was provided by NASA through a grant from the Space Telescope Science Institute, which is operated by the Association of Universities for Research in Astronomy, Incorporated, under NASA contract NAS5-26555.

%%%%%%%%%%%%%%%%%%%%%%%%%%%%%%%%%%%%%%%%%%%%%%%%%%
\section*{Data Availability}
The HST/WFC3 observations conducted and reduced by the WISP Survey are publicly available on MAST at \url{https://archive.stsci.edu/prepds/wisp/}. The latest  v6.2 emission line catalogues and catalogues of all the WISP photometry \citep{Battisti24} are publicly available at the same MAST page. 

In this paper we report in Tables \ref{tab:identification}, \ref{tab:photometry}, \ref{tab:emission_flux} and \ref{tab:derived_properties} the id, photometry, emission line fluxes and derived measurements from the galaxies that we target and we provide a electronic machine readable table at \url{https://github.com/Kitboyett/Boyett_et_al_2024}.

The data products produced as part of this paper are available upon reasonable request. 

%%%%%%%%%%%%%%%%%%%%%%%%%%%%%%%%%%%%%%%%%%%%%%%%%%

%%%%%%%%%%%%%%%%%%%% REFERENCES %%%%%%%%%%%%%%%%%%

% The best way to enter references is to use BibTeX:

\bibliographystyle{mnras}
\bibliography{references} % if your bibtex file is called example.bib

%%%%%%%%%%%%%%%%%%%%%%%%%%%%%%%%%%%%%%%%%%%%%%%%%%

%%%%%%%%%%%%%%%%% APPENDICES %%%%%%%%%%%%%%%%%%%%%

\appendix
\section{Emission line catalogue redshift comparison}\label{App:catalogue_redshift}
\begin{figure}
    \centering
    \includegraphics[width=\columnwidth]{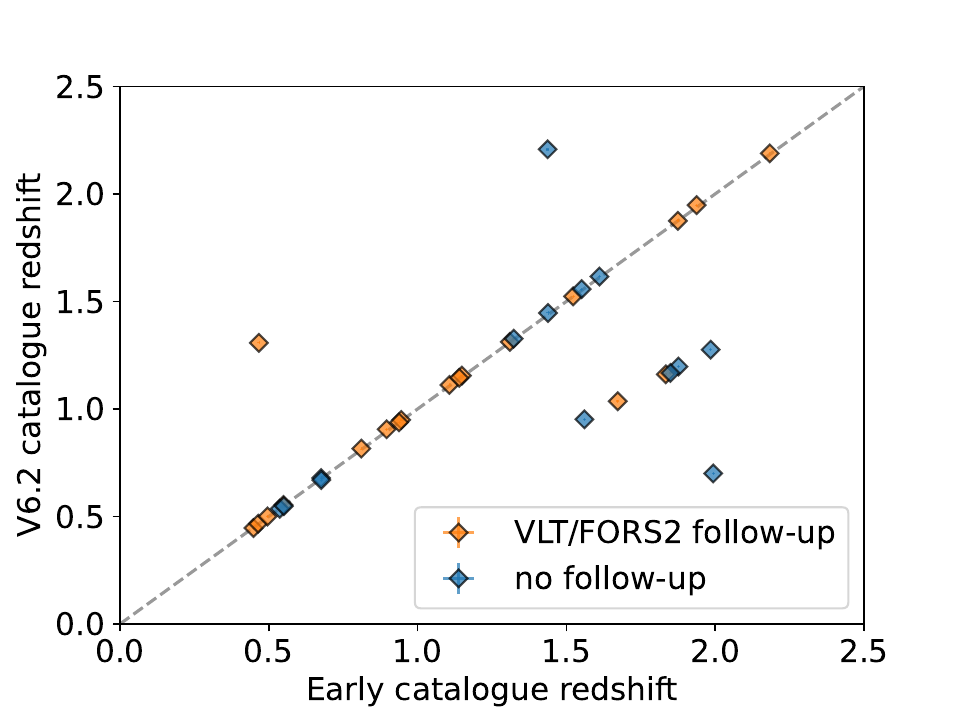}
    \caption{Redshift comparison of 33 galaxies in common between the early WFSS emission line catalogue, from which the slitmasks were designed, and the public v6.2 data release catalogue from \citet{Battisti24}. Galaxies targetted as part of the VLT/FORS2 observations are shown are in orange.}
    \label{fig:catalogue_redshift}
\end{figure}

When matched, 33 of 114 candidates in the original WISPS emission line catalogue (from which the slitmask design was based) appear in the v6.2 catalogue. We do not consider Par62 when matching as this field was not included in v6.2. In Figure \ref{fig:catalogue_redshift} we present the reported redshifts for this sub-set of candidates given in the two iterations of the emission line catalogues. The galaxies shown in orange are those followed up in our VLT/FORS2 observations. 

Consistent redshifts between the catalogues are found for 24/33 (73$\%$) candidates. We find 9 galaxies which have changed redshift between the two catalogues. For 6 galaxies (see Figure \ref{fig:catalogue_redshift}), the early catalogues reported detections of the \Hb and \OIIIdoublet lines, however in the v6.2 catalogue the \OIIIdoublet feature is re-identified as the \Ha line. This has resulted in the updated redshift for these galaxies being lower, demonstrating how there is still ambiguity in distinguishing \Ha and \OIIIdoublet emission lines at our spectral resolution.  

The remaining 3 cases of changing redshifts in Figure \ref{fig:catalogue_redshift} present further re-identifications between the catalogue iterations. We refer to these galaxies by their early WISP catalogue id (see Table \ref{tab:identification}).  
For \texttt{236\_121} ($z_{\rm{initial}}=2.0, z_{\rm{v6.2}}=0.7$), \OII and \OIIIdoublet lines had been reported in the early catalogue however, v6.2 re-identifies \OII as \Ha and the \OIIIdoublet feature to be spurious. 
For \texttt{309\_182} ($z_{\rm{initial}}=0.5, z_{\rm{v6.2}}=1.3$), a single \Ha line had been reported in the early catalogue however, v6.2 identifies \OIIIdoublet and \Ha from two separate features. There is no clear line identification for the emission line features reported in the early catalogue using the redshift solution from v6.2.
This galaxy was observed with FORS2 (ID \texttt{309\_1\_12}) but no emission features were detected. This supports the higher redshift solution of the v6.2 catalogue, as no lines would be expected to fall in the FORS2 wavelength coverage.
Finally, for \texttt{64\_70} ($z_{\rm{initial}}=1.4, z_{\rm{v6.2}}=2.2$), \OII and \Ha were claimed in the early catalogue, however v6.2 re-identify \Ha as \OIIIdoublet resulting in a higher redshift solution in v6.2. We note that the v6.2 catalogue contains multiple entries for this galaxy, since this object happened to be observed in the overlap region between WISPS fields 63, 64 and 66. The other two entries for the same galaxy, which were reduced and analysed independently, support the original redshift solution ($z=1.4$) from the early catalogue. Clearly, even when independent observations and analysis of the same source is made, there is still ambiguity over the reported \Ha and \OIIIdoublet degeneracy.  However, we do not detect any emission in FORS2, so the redshift should still be considered tentative. 

\section{VLT/FORS2 and HST/WFC3 WFSS spectroscopic redshift comparison}\label{App:comparison_redshift}

\begin{figure}
    \centering
    \includegraphics[width=\columnwidth]{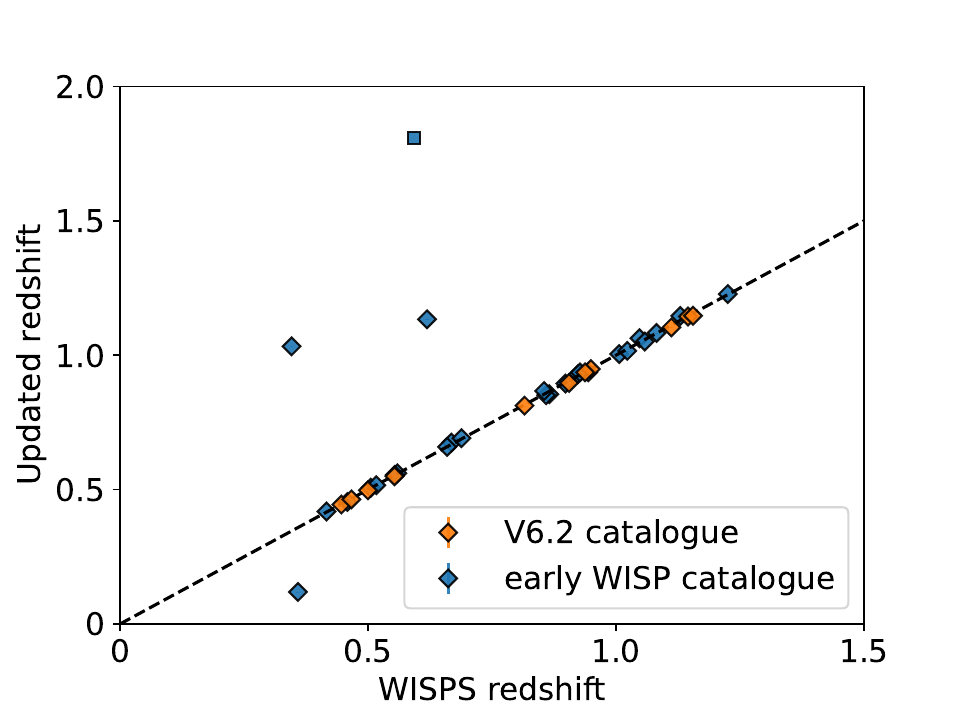}
    \caption{VLT/FORS2 and HST/WFC3 Grism spectroscopic redshift comparison for the 38 galaxies with FORS2 spectroscopic redshifts (Diamonds). We additional include one galaxy (Square) which was not detected in FORS2, but our re-analysis of the reported grism emission lines allowed us to update the redshift.
    Here the WISP redshift is either provided as the input early catalogue redshift (blue) or when available the \citet{Battisti24} v6.2 catalogue redshift (orange). In the 11 cases when we use the v6.2 redshift, we confirm that these were consistent with the early catalogue redshift.}
    \label{fig:spectroscopic_redshift}
\end{figure}

We compare the measured redshift from the FORS2 spectroscopy, for the 38 galaxies with detected line emission, with their WISP grism measured redshift. There is good agreement with 36 out of 38 showing consistent measurements between the two, with marginal redshift differences being due to the relative spectral resolution of the two instruments. There are two galaxies where the determined redshift from FORS2 spectroscopy was different from the input catalogue (and one further case which had multiple redshift entries in the initial catalogue). This can be seen in Figure \ref{fig:spectroscopic_redshift}, where we additionally include a galaxy (\texttt{62\_1\_14}, shown as a circle) which had no FORS2 detections but we were able to re-examine the WFC3 emission features to update the redshift. 

\texttt{236\_2\_12} was expected to be a redshift $z\sim0.36$ galaxy based on a single WISPS emission line, assumed to be $H\alpha$, however FORS2 detected $H\alpha, H\beta, [O\textsc{ii}]$ and $[O\textsc{iii}]$ at a redshift $z\sim0.12$. No standard emission lines lie at the WISPS emission line $0.89\mu m$ wavelength, leading us to believe this must have been an erroneous detection due to a cosmic ray or CCD effect coincident with the dispersed slitless spectrum of the galaxy.

\texttt{62\_1\_13} had a single WISPS detection assumed to be H$\alpha$ at $z=0.62$ which is contradicted by a FORS2 detection of [O\textsc{ii}] at $z=1.13$. Reviewing these two detected lines, the [O\textsc{ii}] from FORS2 is deemed to be real from its doublet line profile and without any standard lines matching the WISPS wavelength, the detection is determined to be a mis-identified cosmic ray/CCD effect.

One galaxy we targetted, \texttt{309\_1\_13}, had two discrepant redshifts reported in the early catalogue, with up to three emission lines detected. This early catalogue reported a faint H$\alpha$ detection at $z=0.35$ and additionally a different (brighter) H$\alpha$ and [O\textsc{iii}] detection at $z=1.03$. In FORS2 we detect an [O\textsc{ii}] emission line which confirms the $z\sim1.03$ solution, the mis-identified `H$\alpha$' emission line does match H$\gamma$+[O\textsc{iii}]4365 at the correct redshift. Although we note this is a faint line and therefore this is more likely a spurious feature.
We regard this as a successful confirmation of a WISP redshift with FORS2.

Finally we note galaxy \texttt{62\_1\_14}, which had 3 contradictory emission line detections reported in the early catalogue, each claiming a different redshift. H$\alpha$ at $z=0.59$, H$\beta$ at $z=1.81$ and another H$\alpha$ at $z=1.14$. 
Re-examining the observed wavelengths of these 3 emission line features it is clear that these are in fact \OII, \Hb and \OIIIdoublet at $z=1.81$. No emission lines were detected in FORS2 which supports the $z=1.81$ result, as the other two redshift solutions would predict multiple lines to fall in the FORS2 wavelength coverage.

\section{Example spectra}\label{App:example_spectra}
\begin{figure*}
    \centering
    \includegraphics[width=\textwidth]{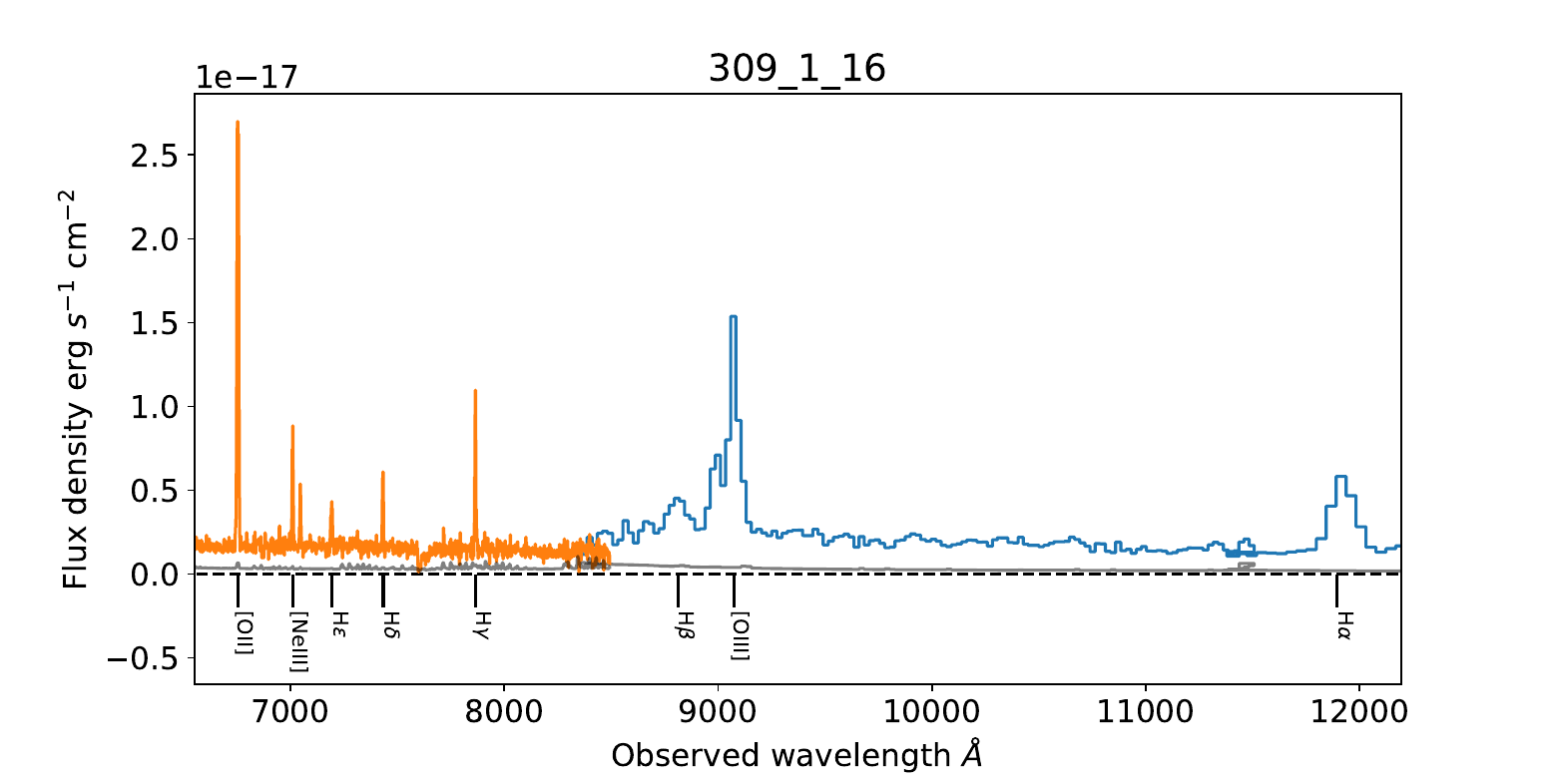}
    \includegraphics[width=\textwidth]{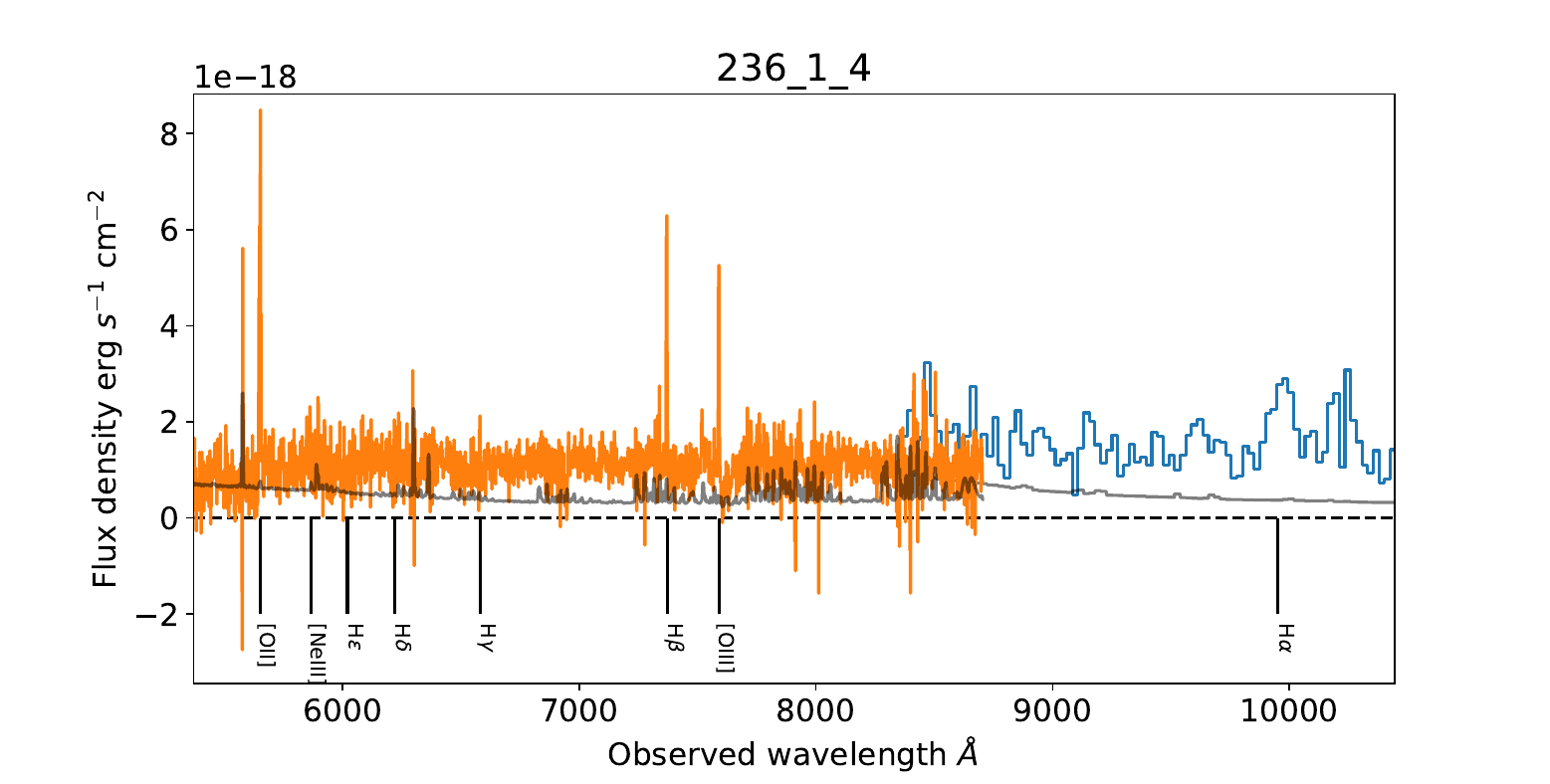}
    \caption{VLT/FORS2 and HST/WFC3 Grism spectroscopy of two example galaxies in our sample with detected line emission. 
    Top: galaxy id \texttt{309\_1\_16} which had multiple emission lines detected in WISPS (\Hb, \OIII and \Ha). We confirm the redshift of this object at $z=0.812$ with detection of multiple lines in FORS2 ([O\textsc{ii}], [Ne\textsc{iii}], H$\gamma$, H$\delta$).
    Bottom: galaxy id \texttt{236\_1\_4} which had only a single emission line detected in WISPS, assumed to be H$\alpha$. This galaxy did not make the threshold for inclusion in the \citet{Battisti24} v6.2 emission line catalogue, yet we confirm the redshift of this object at $z=0.516$ with detection of multiple lines in FORS2 ([O\textsc{ii}], H$\beta$, [O\textsc{iii}]).
    The spectra shown in orange and blue present the VLT/FORS2 and HST/WFC3 grism spectroscopy, respectively. In grey we present the associated noise spectrum.}
    \label{fig:example_spectra}
\end{figure*}

To provide examples of the VLT/FORS2 and HST/WFC3 grism\footnote{The reduced 1d grism spectra presented here are publicly available on the WISP Survey MAST page. The associated IDs for these two spectra are Par309 object 40 and Par236 object 59.} spectroscopy we present in Figure \ref{fig:example_spectra} the spectra from two galaxies with FORS2 detected emission lines. 
The first is \texttt{309\_1\_16} which had multiple emission lines detected in WISPS (\Hb, \OIII and \Ha). We confirm the redshift of this object at $z=0.812$ with detection of multiple lines in FORS2 ([O\textsc{ii}], [Ne\textsc{iii}], H$\gamma$, H$\delta$).
The second is \texttt{236\_1\_4} which had only a single emission line detected in WISPS, assumed to be H$\alpha$. This galaxy did not make the threshold for inclusion in the \citet{Battisti24} v6.2 emission line catalogue, yet we confirm the redshift of this object at $z=0.516$ with detection of multiple lines in FORS2 ([O\textsc{ii}], H$\beta$, [O\textsc{iii}]).
From the spectra, it is immediately clear that the VLT/FORS2 spectroscopy has a higher spectral resolution, as seen by the narrower line emission profiles. 

%%%%%%%%%%%%%%%%%%%%%%%%%%%%%%%%%%%%%%%%%%%%%%%%%

% Don't change these lines
\bsp	% typesetting comment
\label{lastpage}
\end{document}